\documentclass[b5paper,oneside]{book}

\usepackage[round]{natbib}

\usepackage{physics}
\usepackage{type1cm}         
\usepackage{enumerate}
\usepackage{makeidx}         
\usepackage{graphicx}        
\usepackage{multicol}        
\usepackage[bottom]{footmisc}

\usepackage[T1]{fontenc}
\usepackage{newtxmath}       

\usepackage{soul} 
\usepackage{tikz}

\usepackage[
	colorlinks=true,
	hypertexnames=false
			]{hyperref}
\PassOptionsToPackage{linktocpage}{hyperref} 

\hypersetup{
  colorlinks   = true, 
  urlcolor     = magenta, 
  linkcolor    = black, 
  citecolor   = black!65
}

\renewcommand{\cite}{\citep} 

\makeindex             

\begin{document}

\author{Dominik Hangleiter, Jacques Carolan, and Karim Th\'{e}bault}
\title{Analogue Quantum Simulation: \\A New Instrument for Scientific Understanding}
\date{(Draft, March 2021)}
\maketitle

\frontmatter

%
%

\chapter{Preface}

Philosophy of science, like science, is an amorphous and adaptable human endeavour. The shape it takes at any particular time is conditioned both by prevailing endogenous intellectual currents and a multitude of exogenous influences, drawn from science and other areas of human discourse. Moreover, like science, the philosophy of science is a inherently collaborative enterprise: one cannot conduct research in philosophy of science in isolation from either scientists and other philosophers. We take the case of analogue quantum simulation to provide a particularly vivid embodiment of these observations. Just as the science in question is the product of recent, revolutionary changes at a theoretical, experimental and technological level, so the philosophical methodology we have sought to formulate and apply in what follows has been iteratively hewn from the collaborative project of grappling with an array of methodological and epistemological questions, both discussed in the more general philosophical literature, and specific to the case of analogue quantum simulation. We hope that the reader will share our wonderment and excitement in exploring the rich and productive interplay between contemporary science and philosophy of science that we take to have been embodied in this short book.   

Moreover, we hope in our discussion the reader will find well illustrated the extent to which the philosophy of science may be adapted for useful scientific ends.
In particular, in the context of cutting edge scientific practice that is methodologically innovative, we hope to have successfully demonstrated that the application of the philosophical conceptual toolkit can help clarify, inform and even refine scientific practice. Conversely, we hope to convince the reader that philosophical methodology can and should be refashioned anew by the demands of a new scientific context. To the extent that our philosophical methodology is to combine detailed case studies with interpretative and generalising analysis, we have very much followed a tried and tested philosophical approach. However, this more traditional approach has been deployed in a modified form, focused upon constructing a normative framework to refine as much as interpret scientific methodology. In particular, our principal goal has been not just to articulate the implicit nascent epistemological framework that scientists deploy in analogue quantum simulation, but to also contribute to a dialogue about what form that framework should take.       

This has been an deeply collaborative project, drawing upon our varied expertise spread across many different areas of philosophy and both theoretical and experimental science. One of us is an experimental physicist, with expertise in optical quantum simulation, one of us is a theoretical physicist with expertise in the mathematics and computer science of quantum simulations, and one of us is a philosopher of science, with expertise in the foundations of classical and quantum gravity. We all have deep interests in each others areas of practice and in the collective enterprise of deepening the store of human knowledge through interdisciplinary collaboration. For scientists and philosophers alike, we hope that his work will go some small way towards encouraging others to follow our approach.

This short book owes it origin in chance meetings and conversations in Munich and Bristol and to a spirit of collaborative engagement and interdisciplinary enthusiasm involving conversations with a number of philosophers and scientists. We are extremely grateful in particular to Radin Dardashti, Pete Evans, Sam Fletcher, Stephan Hartmann, Maxime Jacquet, James Nguyen, Hugh Reynolds, and Paul Teller for helpful comments on various stages of draft material, and to Cameron Beebe for early insightful conversations. Stephan Hartmann deserves particular thanks for a comprehensive and detailed read through of the entire draft manuscript leading to a great number of changes that have improved the coherence and precision of the manuscript. We are also extremely grateful to two anonymous reviewers who provided insightful constructive feedback that allowed us to improve the manuscript in a number of important respects. DH thanks Ulrich Schneider and Immanuel Bloch for enlightening in-depth conversations about their perspective on the epistemic goals of cold-atom quantum simulators, and Lukas Schwarz on the physics of the Higgs mode. JC thanks Dirk Englund, Seth Lloyd and Anthony Laing for their constant encouragement and thoughtful insights.  KT is particularly grateful to Radin Dardashti, Pete Evans, Sean Gryb, Stephan Hartmann, Patricia Palacios, and Eric Winsberg for invaluable insights gained from earlier collaborative projects that were instrumental towards constructing the framework presented here. All errors, omissions and oversights found within the present text we take full credit for. 

Furthermore, we are grateful to audiences at the annual conference of the BSPS in Edinburgh in July and ECAP9 in Munich in August 2017 for comments on earlier versions of this work. We also are particularly grateful to participants at the workshop `Analogue Experimentation' held in Bristol in 2018. KT is appreciative to audience members at the online workshop `The Aesthetics Nature of Scientific Experiments' (Cambridge 2021) for a number of valuable comments and suggestions. During the course of this work DH has been supported by the Templeton foundation, JC has been supported by the European Union's Horizon 2020 research and innovation programme under the Marie Sklodowska-Curie grant agreement No.\ 751016 and KT has been supported by an Arts and Humanities Research Council (UK) grant No.\ AH/P004415/1. We are also gratefully to the Munich Center for Mathemitcal Philosophy and the Bristol Centre for Science and Philosophy, and the Bristol Institute for Advanced Studies for supporting various stages of this project. We are greatly appreciative of the support provided by us by the Springer Editorial team, in particular Lucy Fleet and Svetlana Kleiner, during the course of this book's publication. Finally, KT is profoundly grateful to Ana, for everything.





\begin{flushright}\noindent
Berlin, London, Bristol\hfill {\it Dominik Hangleiter}\\
March, 2021\hfill {\it Jacques Carolan}\\
\hfill {\it Karim Th\'ebault}
\end{flushright}

\tableofcontents

\mainmatter

\chapter{Introduction: A New Instrument of Science?}
\label{intro}



Over the past few decades experimenters have learned to manipulate quantum systems at regimes that were hitherto unimaginable.
The radius of the proton has now been measured to within $1$ part in $10^{17}$ \cite{bezginov2019measurement} and quantum states of light have been used to observe gravitational waves due to black hole mergers billions of light years away \cite{tse2019quantum}.
It is not just the precision with which quantum systems can be probed that enables new frontiers for scientific knowledge, but also the \textit{scale}.
%
State of the art experiments can now coherently control thousands of ultra-cold atoms using arrays of laser beams \cite{Jaksch1998,greiner2002quantum}, or electrically trap dozens of individual ions that can be individually addressed \cite{blatt:2012}.
The precision with which these systems can be controlled, enables these \textit{quantum technologies} to serve as a window into fundamentally new science, from the observation of emergent many-body properties \cite{Bakr-Science-2010} to performing computation outside of the reach of conventional computers \cite{arute_quantum_2019}.

While the long term vision is to scale up the number of quantum systems (qubits) to thousands or even millions, such that these processors can run quantum algorithms, there may be many practical applications before we get to that regime.  
These applications are likely to rely on intrinsic properties of the quantum systems themselves.
As Aaronson and Arkhipov note:
\begin{quote}
    \textit{``One might suspect that proving a quantum system's computational power by having it factor integers is a bit like proving a dolphin's intelligence by teaching it to solve arithmetic problems. Yes, with heroic effort, we can probably do this, and perhaps we have good reasons to. However, if we just watched the dolphin in its natural habitat, then we might see it display equal intelligence with no special training at all.''} \hfill \cite{aaronson_computational_2010}
\end{quote}
In other words, if we just look a bit harder at the quantum systems we currently have, perhaps they are already doing something useful. 



\textit{Analogue quantum simulation} is a new and powerful inferential tool whereby scientists manipulate and probe the dynamics of a well-controlled quantum system in the lab to learn about features of another quantum system they do not have direct access to.
%
Much like how a mechanical orrery simulates planetary motion, analogue quantum simulators replicate key features of quantum systems.  
The mechanical orrery can do this because planetary motion is well approximated by Newtonian physics.
However, in the case of quantum systems there are unique properties that are not well described by classical physics: e.g. entanglement, superposition, wave-particle duality. 
Therefore to accurately and efficiently simulate a quantum system of interest, one must use another well controlled quantum system.
One of the first proposals to formalise this notion is due to Richard \citet{feynman:1982}. 
Feynman pondered the problem of how to efficiently simulate quantum-mechanical systems given that at that point in time computing devices were understood to be entirely classical in nature.
With his characteristic clarity, he noted: 
\begin{quote}
	\emph{``I therefore believe it's true that with a suitable class of quantum machines you could imitate any quantum system.''} \hfill \cite[p.~475]{feynman:1982}
\end{quote}

Feynman envisioned a \textit{universal} quantum simulator, a single device that is capable of simulating \textit{arbitrary} quantum systems.  In contrast, \textit{analogue} quantum simulators are typically `bespoke' devices capable of simulating a much more \textit{restricted} class of quantum systems.
Specifically, analogue quantum simulators establish a mapping between the continuous dynamics of the simulator in the lab and the quantum dynamics of a `target' physical system of interest.
Much as a model airplane in a wind tunnel is used to make precise predictions about the aerodynamics of full-scale air craft, an analogue quantum simulator is used to make predictions about the dynamics of particular quantum systems.
In principle it is possible to build a full computer model of the fluid dynamics of the airplane, and this system would also be able to simulate various other systems governed by the Navier-Stokes equations (blood flow, heat transfer, weather).
However, such models are intensely computationally demanding and wind tunnel simulations remain a critical tool for the aeronautics industry --- both in terms of making predictions but also for verifying computer models.



Whereas classical analogue simulators have largely been superseded by digital classical computer simulation, quantum analogue simulators provide the most plausible near-term device for efficiently simulating a wide range of quantum systems. 
While more limited in the range of problems to which they can be applied, analogue quantum simulators have proved more achievable in practice than universal quantum simulators. 
Quantum computers are, at the present date, fairly far from reaching a regime in which they are able to outperform the best available classical computers on practical problems \cite{haner_factoring_2017,ogorman_quantum_2017,gheorghiu_benchmarking_2019,gidney:2019}. 
While, recently, there have now been first proof-of-principle demonstrations of a quantum speed up \cite{arute_quantum_2019,wu_strong_2021,zhong_quantum_2020,zhong_phase-programmable_2021}, arguably, bespoke analogue quantum simulators have already surpassed this `quantum speed up' milestone for the past $10$ years on computations of quantum-mechanical properties of physical models.\footnote{Examples of experiments which have been argued to outperform classical computers on such tasks are particularly prominent in cold-atom platforms \cite{BlochEisertRelaxation,choi_exploring_2016,bernien_probing_2017}. This is because, here, a large number of atoms evolves coherently, making classical simulations particularly challenging.}

But analogue quantum simulators may not only be used to learn about properties of systems that could not be \emph{computed} otherwise. 
Rather, analogue quantum simulation also promises to shed light on \textit{empirical questions} in a manner inferentially analogous to conventional experimentation. More uncontroversially, analogue quantum simulations have the clear potential to both guide future experiments on the target system and provide heuristic insights for theory building. More contestably, and most exciting from a fundamental physics perspective, analogue quantum simulations are a potential new means to experimentally probe by proxy, physics which is not accessible in our earthly laboratories nor even via astronomical observation. Most vividly, it has been argued that experiments on analogue models of black holes, implemented variously via fluids, Bose-Einstein condensates, or fiber optic platforms, can \emph{confirm} the existence of astrophysical Hawking radiation, an effect that is in practice
impossible to observe via conventional means.\footnote{We will return to this exacting and controversial claim in Chapter \S5. See \cite{dardashti2015confirmation,Dardashti:2019,Crowther2019,evans:2019} for arguments along this line.} 

Analogue quantum simulation may thus allow us to extend our \textit{understanding} beyond the horizons that limit us today, be it in terms of computational power, or the observability of phenomena that are not accessible via conventional experiments or observations. 
However, it remains to be seen how well the conventional framework for thinking about scientific inferences is fitted to the evaluation of analogue quantum simulation as a novel scientific practice. What precisely does it mean for `a quantum system to simulate another'? 
Is analogue quantum simulation a single unitary scientific practice or does the simple label lump together importantly distinct types of inference and experiment? 
What, precisely, are the scientific goals undertaken under the label `analogue quantum simulation'? 
Is analogue quantum simulation really a novel means of inference or just the `same old stew'?\footnote{For an argument along these lines targeting claims of philosophical novelty relating to the philosophy of conventional computer simulations see \cite{frigg:2009}.} 
Just another form of computation on a different platform, just a new variant of analogical argument, or just a particular type of well controlled  experiment? 
Is analogue quantum simulation really a new tool of science of just an adaptation of an old one?   

In this short book, we will attempt to answer these questions via an interdisciplinary analysis that cuts between physics, computer science, and the philosophy of science. 
Our philosophical analysis will be built upon a range of detailed case studies in contemporary experimental practice.
We will examine a range of examples of analogue quantum simulation with regards to the extent to which they provide \emph{scientific understanding}~\cite{Regt-Synthese-2005,grimm2012,grimm2016,khalifa2017,de_regt_understanding_2017}. This focus allows us to identify the value of analogue quantum simulation as a form of scientific inference, and to highlight the differences to other forms of scientific inference. 
Furthermore, our focus on understanding as a goal of science will  allows us to draw a multifaceted `methodological map' \cite{galison1996} of modern science on which analogue quantum simulation can be situated. 


In answering these questions we hope to clarify what is meant by `analogue simulation' and the function it serves in scientific practice. 
Furthermore, in addition to these descriptive aims, our project will involve the formulation of prescriptive guidance that we hope will be useful for practising scientists in the field of quantum simulation. 
In particular, we will argue that scientists should distinguish between two important epistemic goals within analogue quantum simulation: obtaining \textit{how-possibly} scientific understanding and obtaining \textit{how-actually} scientific understanding. 
We will analyse the conditions under which these distinct aims are achieved in practice and assess the importantly different conditions under which they obtain.

The foundations for our argument are formed by a set of distinctions that will refine precision of our language in discussing `analogue quantum simulation' (Chapter~\ref{Distinctions}). 
The key distinction that we will make is that between analogue quantum \textit{computation} and analogue quantum \textit{emulation} as two faces of analogue simulation. 
The epistemic goal of analogue computation is to learn about a mathematical model, the goal of analogue emulation is to learn about a concrete physical system. In practice, a single experiment can be, and often is, conducted with the view to performing both functions. It is thus not surprising that they are often conflated. However, the difference in epistemic goals has important implications for the difference in validation procedures and forms of understanding that can be achieved, and is thus ultimately a scientifically important one in practice. The case studies (Chapters~\ref{ch:cold atoms}--\ref{ch:hawking}) and subsequent detailed philosophical analysis (Chapters~\ref{ch:understanding}--\ref{ch:method mapping}) are all orientated around establishing this point.  

We will provide detailed analysis of four case studies in total, two examples of analogue quantum computation and two examples of analogue quantum emulation. In Chapter~\ref{ch:cold atoms} we will outline two experiments using an ultracold-atom platform performing analogue quantum computation for the existence of a Higgs mode in two dimensions and the existence of many-body localisation (MBL) in two dimensions, respectively. In Chapter~\ref{ch:enaqt}, we will then consider an analogue quantum emulation in quantum biology, which has as its target individual photons of a quantum mechanism potentially involved in photosynthesis, so called environment-assisted quantum transport (ENAQT). Finally, in Chapter~\ref{ch:hawking} we will discuss analogue quantum emulation of astrophysical Hawking radiation with a particular focus upon dispersive optical media platforms. 
In each of the case studies, we will provide an accessible summary of the key features that will be important for the philosophical argument later in the book, as well as a detailed discussion of the physical details for the interested reader. 

In Chapter~\ref{ch:understanding} we will move our focus on the topic of scientific understanding. We will extend the framework for understanding via models developed by \citet{Strevens2008,Strevens2013understanding} and \citet{ToyModels} to further scientific activities, including computation, analytical derivations, and experiments as well as different types of targets to be understood, formal properties of a mathematical model and physical phenomena. Our arguments will be substantiated with a range of examples from the sciences. We will argue that the goal of analogue quantum computation and emulation is to obtain understanding of their respective target -- abstract mathematical models and concrete physical systems, respectively. In Chapter~\ref{ch:assessing case studies}, we will then bolster our argument, drawing upon the detailed examples of analogue simulation in practice provided by the case studies. We will demonstrate how in these specific experiments the framework for understanding in analogue quantum computation and emulation is not only applicable, but importantly clarificatory. 

In Chapter~\ref{ch:norms}, we draw back to take a wider perspective on analogue quantum simulation. In particular, we will look back at both the abstract framework for understanding via analogue simulation and its application to our case studies in order to isolate `norms' for scientific practice. Our aim is that these norms will constitute useful guidance both for researchers working on analogue simulations and those who want to assess the claims made, with respect to the epistemic goal of understanding. Finally, in Chapter~\ref{ch:method mapping}, we will situate analogue quantum computation and emulation on the methodological map of modern science, again, with respect to the goal to obtain understanding of the respective target phenomenon at hand.

To complete this introductory chapter let us foreshadow our principal conclusions. 
Our analysis is built around the combination of two distinctions. The first  distinction is between how-actually and how-possibly understanding. This is a distinction between understanding that is vertical (or truth-like) and understanding that is modally weaker and not required to be vertical. The second distinction is between the two relevant objects of understanding: formal features of a target model and physical target phenomena. This difference analogue quantum computation and analogue quantum emulation can be characterised by the respective target; while in analogue quantum computation, we aim to learn about a \emph{formal property of a target model}, in analogue quantum emulation, we aim to learn about a \emph{concrete physical phenomenon}. 

\begin{figure}[t]
\centering
\includegraphics[width = \textwidth]{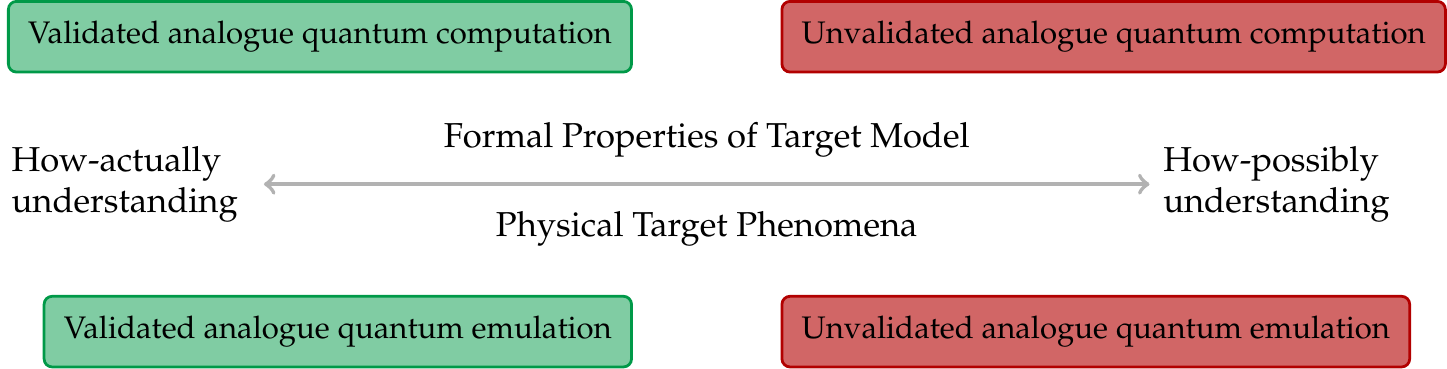}
\caption{Methodological map with respect to the type of understanding aimed at and the object that is to be understood. Whether or not these aims are achievable depends on further features of analogue quantum emulation and computation as discussed in this chapter and the next. 
} 
\label{fig:methodological map1}
\end{figure}

We will argue that \textit{veridicality conditions} can be  established in a validation procedure that involves both mathematical verification of correspondences between different models in certain parameter regimes and experimental certification of models as empirically adequate models of physical phenomena. This argument leads us to situate analogue computation and emulation on the \textit{methodological map} given in Figure~\ref{fig:methodological map1}, depending on whether or not they are validated. In particular, by considering the target and the model strength of the understanding we can usefully compare different forms of analogue quantum simulation with more traditional scientific methodologies (full discussion will be provided in Chapter~\ref{ch:method mapping}).   

Analogue quantum computations yield understanding of formal properties of a target model. 
If an analogue computation is validated we can assert that it yields how-actually understanding of its target formal property, while an unvalidated analogue quantum computation yields the modally weaker how-possibly understanding. 
This is because without validation, we cannot assert the truthfulness of the results obtained in an analogue simulation. 
We will argue that heuristic and exact computer simulations are co-situated with unvalidated and validated analogue quantum computations, respectively, as in the latter case we have a guarantee on obtaining the true formal property, while in the former case we do not. 

In an analogue quantum emulation scientists aim to obtain understanding of physical target phenomena. 
Again, validation of the relevant source and target models promotes the modally weaker form of how-possibly understanding to how-actually understanding of the target phenomena. 
We will argue that unvalidated analogue quantum emulations are similar to material analogies in this respect, while validated analogue quantum emulations may play a role resembling that of conventional experiments. 

The significance of this analysis is that it allows us to isolate the sense in which analogue quantum simulation is both methodologically and epistemically peculiar. It is precisely \textit{because} analogue quantum simulations function in some contexts like material analogies, in other contexts like computer simulations, and in still other like experiments, that looking for a unified and unequivocal analysis of what we can and cannot learn from them is simply inappropriate. If this book has a single underlying story to tell it will be that the diversity of practice within the label `analogue quantum simulation' must be recognised before a reliable appraisal of the relevant epistemological issues can be undertaken.  Given such a fine grained analysis, both the power and the limitations of this new an exciting tool of scientific understanding can be recognised.

\chapter{Distinctions with a Difference}
\label{Distinctions}

The term `analogue quantum simulation' is widely used in the context of quantum information science and carries with it a number of subtle and interrelated connotations. Following a standard philosophical approach, we will start our analysis of analogue quantum simulation in contemporary scientific practice with a number of terminological and methodological distinctions. Together these distinctions will prove crucial to achieving the requisite level of linguistic precision in the analysis that follows. 

\section{Analogue and Analog} 
\label{sub:analogue_vs_analog}
 The first distinction is between the two senses of `analogue' that are at play in the term `analogue quantum simulation'. The first sense relates to the idea that a physical system is similar to, comparable with, or has parts or functions in common with another physical system. This sense of analogue brings to mind the idea of an `analogue model' or `argument by analogy', which are the subject of a fascinating literature in the philosophy of science.\footnote{See in particular \citet{keynes:1921,hesse:1964,hesse:1966,bailer:2009,bartha:2010,bartha:2013,sep-models-science,frigg:2018}.} The examples of analogue quantum simulation will be `analogue' in broadly this sense, since they all involve two systems or models of systems that exhibit features `analogous' to features of certain other systems. 
 Analogue quantum simulation is built upon the exploitation of these common features. 
 
 The importance of this analogue aspect should not be overstated, however. Almost everything is analogous to almost everything else in \textit{some} sense. As we shall see in our examples, the scientific practice of analogue quantum simulation is about the exploitation of a specific type of controllable formal relationship between dynamics of two systems or models. For this reason, the `simulation' aspect is just as  significant, if not more significant, than the `analogue' aspect of analogue quantum simulation.   
 
 The second sense of analogue in analogue quantum  simulation relates to the idea that the simulation in question is a non-digital or continuous computation in one of the senses discussed shortly. Note that in the rest of this chapter, focused on definitions and distinctions, for clarity, we will consistently use the British spelling `analogue' for the `similar to' sense and American spelling analog for the `not digital' sense. 
 In the remainder of the book we will typically revert to the British spelling, with both senses of the term implicit, but reintroduce the distinction when strictly necessary.  
 

\section{Analog and Digital Computational Models} 
\label{sub:analog_vs_digital}
The second distinction that we would like to make is between analog and digital computation. We will consider this distinction in the context of the mathematical formalism that describes the computation, that is, the digital and analog \textit{models} of computation.\footnote{This focus on analog vs. digital models of computations means that the philosophical discussions of analog and digital representations are largely tangential to our own. See \cite{goodman:1968,lewis:1971,trenholme:1994,maley:2011} for discussion of the latter distinction.} In a digital model of computation, the mathematical model that underpins the computation is based upon a discretisation of both the in and output encoding of the computation as well as the control parameters of the computation itself. In contrast, in an analog computation, the mathematical model that underpins the computation is based upon continuous  input and output encoding together with continuous control parameters.

A further distinction can be made between a discretisation of the spatial and temporal components of the computational model. The spatial component of the computational model relates to the state space over which the computation is defined, the temporal component to the successive computational states in computational time. Clearly there is no principled reason to require that if one of these is analog or discrete then the other must be also. 
Thus, strictly speaking, analogue computation is an ambiguous term that could refer to one (or all) of computations that are: 
spatially analog and temporally discrete; spatially discrete and temporally analog; or both spatially and temporally analog.  

Let us provide some examples of analog and digital computational models.\footnote{Bournez and Pouly survey a number of other analog models of computation in a recent review \citep{bournez2018survey} and the reader is referred there for detailed discussion of the full range of models.} First, as a contrast case, we have entirely discrete models of computation. 
The most famous example is the Turing machine. The Turing machine is a mathematical model of computation that is used to prove fundamental limits on computation and is defined such that it is discrete in its input/output space and in computational time. At every one of the discrete time steps, it can either read or write symbols according to one of a finite number of rules. 
A second, even more vivid, example is computation via cellular automata such as  Conway's famous \textit{Game of Life}. There the state space in question is a specification of `on' or `off' for each square over a square discrete grid which implies the the space itself is discrete. Moreover, the computational time is discretised via the the application of the rules of the game at each time step.

In comparison to these models that are discrete in space and time, we can consider the general purpose analog computer (GPAC) introduced by \citet{shannon1941mathematical}. 
This is a very general model of analog computation 
involving mechanical elements that are able to implement certain families of circuits. 
This model can be shown to be computationally equivalent to a Turing machine in certain circumstances \cite{bournez:2007}. 
For our purposes the most important feature of the GPAC is that it is continuous in terms of both time and space. 

Finally, as already indicated, it is possible to have models of computation that are discrete in space but continuous in time, or discrete in time but continuous in space. 
We will make the standard assumption that an analog computational model is one in which the space or time of the computation is continuous.   
One example of an analog model of computation which is continuous in space, but discrete in time are neural networks, which have become a dominant trend in computing over recent years \cite{lecun2015deep}. 
The input to the computation is typically a vector of real numbers, such as grey scale pixel values of an image. The computation itself then comprises a sequence of vector-matrix multiplications (parameterised by some weights) and nonlinear activation functions. The output to the computation is also a vector of real numbers, and the classification of whether, say, an image is a `cat' or a `fish' may be given by whichever vector element has the largest value.  The goal is to train the weights of the matrices to realise a particular input-output relation over some large data set. The spatial component of a neural network is continuous (real numbers) and the temporal component is discrete (in terms of a discrete sequence of operations). Neural networks have achieved exceptional performance for a range of applications including natural language processing \cite{Wu:2016wt}, particle physics \cite{Radovic:2018iz}  and cancer diagnosis \cite{Capper:2018dy}. In part, these advances have been enabled by the introduction of dedicated hardware to accelerate this particular computational task \cite{sze2017efficient}.

This points to the following question: since spaces are typically conceived of as continuous entities, is not every computation ultimately analog \emph{qua} being a physical process in the first place? 
For example, even bits on a digital computer are encoded as analogue voltage or current signals; 
logical gates are based on transistors whose output current continuously depends on the input current. 
Conversely, some continuous models of computation are implemented in hardware which are best described discretely, such as neural networks implemented on Google's TensorFlow processing unit (TPU) \cite{jouppi2017datacenter}. 

To fully appreciate the distinction between analog and digital computation, it is thus crucial to distinguish the physical processes underlying a computation on the level of the hardware and the computational model itself. 
Adding this level of analysis complicates matters considerably. 
The hardware itself can be assumed to be governed by the physical laws at the relevant spatio-temporal scale, that is, the micro to milli scale. At these scales the laws take the form of differential equations as implied by classical or quantum field theories, or at least their limits. We can therefore understand the laws governing the hardware as being continuous in both time and space. However, the analog physical processes underlying the hardware may be very accurately described by a digital model of computation. Indeed, in practice we find that many discrete models of computation are implemented in a continuous way on the hardware level. Take the example of a transistor as used in electronic computing hardware, where it acts essentially as a switch: only if a small control current is above a certain threshold, the transistor permits a (much larger) gate current to flow. 
While strictly speaking the transition between full gate current and no gate current as a function of the (much smaller) control current is a continuous one, it is also an extremely sharp one that can be well approximated within a digital model.


An important consequence of -- one might also use this as a criterion for -- a digital computational model for analog physical hardware, is that any errors occurring throughout a computation can also be described within the digital model.
For example, an erroneous increase in current across a transistor may cause a bit flip in the output of the computation.
A key distinguishing feature between analogue and digital computational models is the possibility to correct for those erorrs. 
In digital computational models, the detection and correction of (certain) errors is possible using so-called error correction codes. 
This is not the case in analogue models where we do not know of a way to error correct them, and errors tend to accumulate. 

The distinction between the underlying physical process of a computation and the computational model describing it is a particularly important and subtle distinction for quantum computation. 
Here, a physical qubit may be controlled by, say, the intensity or duration of an incident laser field, but it may also take on an arbitrary state on the Bloch sphere and is thus a continuous object. 
We will say more on this point later, but for the moment we opt for a working definition which distinguishes between analog and digital computation via the \emph{abstract mathematical model which describes the computation.}

\section{Reprogrammable and Bespoke Simulators} 
\label{sub:reprogrammable_vs_bespoke}

On the level of the computational model we can also introduce the distinction between simulators that are reprogrammable and simulators that are bespoke. Reprogrammable simulators can be reconfigured to implement distinct computational tasks. 
In contrast, bespoke simulators are designed to implement a specific task or small sets of tasks for a certain range of input parameters.

Let us focus our attention on analog simulators -- that is devices designed to implement analog models of computation. An early example of a reprogrammable analog simulator is the differential analyzer built by Bush in 1931 at MIT to solve differential equations \citep{sep-computing-history}. In contrast, we can consider classical examples of bespoke analog simulators such as an orrery (a mechanical device that simulates planetary motion) or a Phillips-Newlyn machine (a hydraulic machine that simulates macro economic relationships).\footnote{There is an important difference between bespoke analog simulators such as an orrery or a Phillips-Newlyn machine and the analog simulators that are the focus of our analysis. This difference is that the former, but not the latter, are \textit{artifacts} which may be deployed by a user towards understanding theirs targets (i.e. the solar system or the Guatemalan economy) without mediation through a source model (i.e. model of the clock work mechanics of the orrery or hydrodynamics of the Philips-Newman machine). Such artifactual analog simulators are thus rather different from the experimental quantum simulation devices that are our focus here. We refer to them only to give the reader a simple intuitive connection. For an extensive discussion of the Phillips-Newlyn machine in the contexts of models and representation see the excellent discussion of \cite{frigg:2018}.} The contrast is between the level of controllable specificity of the device. The reprogrammable simulator is designed to be able to be reconfigured such that it implements very different models. In contrast, the bespoke simulator is designed with the intention of implementing a single model or small class of very similar models with the possibility to tune the model parameters within a certain range. 

In reality the distinction between reprogrammable and bespoke analog simulators can often be blurry. 
For example, Ising solvers estimate the ground state of a classical Ising Hamiltonian. 
As well as reprogramming the physical parameters of the Hamiltonian, many classes of important problems in combinatorial optimisation can be mapped onto finding the ground state of an Ising Hamiltonian \cite{10.3389/fphy.2014.00005}.\footnote{We will discuss one example of such a combinatorial optimisation problem, the travelling salesman problem, as well as a strategy for its solution via a mapping to an Ising Hamiltonian, in Section \ref{sec:understanding_formal_properties}. }
Typically, solving these problems is incredibly time consuming using conventional, classical computers, motivating the development of bespoke machines that implement such solvers in laser networks \cite{mcmahon2016fully} and quantum annealers \cite{boixo_evidence_2014}. 
Ising solvers are therefore bespoke in the sense that they solve a specific class of problems, but reprogrammable in the sense that many different problems can be reduced to fidning the ground state of an Ising model.

Finally, we note that the same distinction is possible, though increasingly unusual, in the case of digital simulators. That is, we can have devices that are built to implement a specific discrete computational algorithm, such as running a particular chess algorithm.
However, in most cases of interest digital simulators that are reprogrammable (unlike the chess computer) are computationally universal. 

\section{Classical and Quantum Simulators} 
\label{sub:classical_vs_quantum}

Our next distinction is that between classical and quantum devices. 
As we are focusing on computational devices, we draw the line between classical and quantum with regards to the computational abilities of the device. 
As for the analog/digital distinction, these are determined by the effective model that accurately describes the computation. 
If that computational model is a quantum model, that is, expressed as the unitary transformation of a normalised vector on the complex sphere followed by a quantum measurement according to Born's rule, then we say that a device is `quantum'. In practice, the set of all non-quantum devices can then be taken to be made up of `classical devices'.\footnote{We here neglect the theoretical possibility of super-quantum or post-classical computational devices.} 
A computation carried out on a classical device can be described in terms of a probabilistic mixture of deterministic time evolution, as determined, for instance, by classical mechanics, or the rules of classical logic. 

Notice that we do not frame the distinction in terms of whether the device itself is classical or quantum -- since ultimately we expect all matter to be quantum in nature. 
Take a desktop computer whose logic gates are implemented using transistors. To explain and quantitatively describe a modern CMOS transistor used in microprocessors requires the concept of electronic bands and therefore quantum theory. However, the function it performs in the computation is that of a switch -- an entirely classical concept. 
In other words, the physical phenomena that are exploited for the purpose of implementing the computational model are not necessarily related to that model. 
Conversely, the computational power of a device depends entirely on the computational model and not the physics underlying a specific implementation of that model. 
This is why we focus on the effective computational model of the simulator system when distinguishing between classical and quantum systems for the purposes of discussing simulators. 

We say that a computation is `efficient' if the resources required for the computation in terms of space or time scale at most polynomially in the input size of the problem. Importantly, there is strong evidence that not all computations that can be efficiently performed on a quantum device can be efficiently reproduced by a classical device. 
This evidence is in large part constituted by quantum algorithms with an exponential speedup over the best known classical algorithms such as Shor's algorithm for prime factorisation \cite{shor_algorithms_1994}, the simulation of quantum mechanical systems \cite{lloyd_universal_1996} and quantum algorithms that perform certain random sampling tasks \cite{shepherd_temporally_2009,aaronson_computational_2010,arute_quantum_2019,zhong_quantum_2020}. 
As of today, we do not have a definitive answer to the question why quantum devices are more powerful than classical devices \cite{sep-qt-quantcomp}. 
Na\"ively, the exponentially large dimension of the state space might be considered a reason why dynamics in this space are hard to simulate classically. 
However, the state space is exponentially large in classical computation as well ($n$ classical bits may take $2^n$ configurations). 
Indeed, more intricate mechanisms are actually at play.
Such mechanisms include so-called `quantum parallelism', that is, the possibility to perform computations in superposition \cite{nielsen:2010}, destructive interference giving rise to a complex entanglement structure \cite{hangleiter_sampling_2020}, and contextuality \cite{howard_contextuality_2014,bermejo-vega_contextuality_2017}. 

In our terminology, a desktop computer is a reprogrammable classical digital simulator that implements computations using classical logic gates such as \texttt{AND}, \texttt{OR} and \texttt{NOT} and is therefore a classical device. 
In particular, if a so-called \emph{universal gate set} can be implemented on a reprogrammable computer, it is possible to implement any possible program that can be run on a Turing machine on this device. 
In contrast, a reprogrammable quantum digital simulator is quantum since it can perform coherent computations on a $2^n$-dimensional quantum state space. 
Such a computer can be realised using, for instance, the quantised flux or charge of superconducting LC circuits \cite{Clarke:2008gi} or hyperfine states of trapped ions \cite{blatt:2012} as its qubits.
Unitary transformations of those qubits, i.e., quantum logic gates, can be performed by letting individual qubits interact coherently. 


Within the category of quantum simulators we can make the distinction between digital quantum simulators and analog quantum simulators. The digital quantum device in question is constructed out of quantum logic gates and, given that the algorithm implemented by these gates can be changed by reordering their connections, constitutes a programmable `quantum computer'. A quantum computer that comprises a universal gate set can efficiently implement any quantum algorithm and is called a `universal quantum computer'.  
It can be proven that such a machine would provide us with the capacity to perform a digital simulation of any physical quantum system that features local interactions \cite{lloyd_universal_1996}. 

Intrinsic to the nature of quantum computation is that the coherence of quantum states is extremely fragile. 
One of the major breakthroughs of the field came with the advent of error correction \cite{aharonov_fault-tolerant_1997,aharonov_fault-tolerant_2008}, which allows one to perform arbitrarily long error-free computations, provided that the error incurred in the application of a single gate is below some threshold \cite{Devitt:2013hb}.
Besides the high accuracy required for quantum error correction, it also involves a large overhead in terms of the size of the computing device. 
The experimental challenge is thus to engineer quantum computing systems with extremely high fidelity and on a sufficiently large-scale --
a monumental challenge.

This is where analog quantum simulators become particularly significant. Given the immense difficulties in constructing a large-scale fault-tolerant quantum computer, intermediate-scale non-universal quantum devices that can solve specific tasks have become promising candidates to put the power of quantum devices to use already today. 
An analog quantum simulator is not constructed out of quantum logic gates. Rather, it is a well controlled quantum system that can be continuously manipulated to implement certain continuous dynamics of a system of interest, in particular, that of an \textit{analogue} system which the experimenter does not have direct epistemic access to. 
Due to the continuous nature of the computation there is no known way to error correct these systems, which in turn makes arguments for a scalable speedup problematic \cite{Hauke:2012dq}. Nonetheless, it is expected that there exist simulation regimes wherein errors are not yet overwhelming and yet classical computers fail \cite{aaronson_computational_2010,bremner_average-case_2016,arute_quantum_2019}, or that errors in the simulator properly correspond to errors in the physical system being simulated \cite{Cubitt:2017ti}.  

The focus of this book will be upon devices that are exploitable analogues of different systems we are interested in learning about, that implement analog computational models, that are bespoke, and whose quantum properties are exploited. 
These are analogue quantum simulators.

\section{Source and Target Phenomena}

All experiments involve physical phenomena, that is, quantitative or qualitative properties of a system. 
It is standard in the philosophy of experimental science to distinguish between the `source system' that is manipulated in the lab and the `target system' about which the experimenter wishes to gain knowledge \cite{hacking:1983,galison:1987,franklin:1989,sep-physics-experiment}. 
With this in mind, we can think of the `source physical phenomena' as the quantitative or qualitative properties of the system manipulated in the lab and the `target physical phenomena' as the quantitative or qualitative properties of the system about which the experimenter is hoping to learn.\footnote{We are using qualitative or quantitative phenomena here to indicate the difference between broad patterns of behaviour as opposed to precise numerical and functional relationships. Later we will also talk about qualitative or quantitative properties of solutions to a set of equations or formal model. One can thus have qualitative or quantitative physical phenomena and qualitative or quantitative formal properties. The qualitative vs. quantitative distinction is therefore tangential to both the physical vs. formal distinction and the source vs. target distinction.} 

In the context of analogue quantum simulation source physical phenomena can be isolated in precisely the same way as in a conventional experiment. 
Consider as an example the dynamical evolution of a quantum system after preparing it in a certain initial state as measured by local particle densities or two-point correlation functions (see Chapter~\ref{ch:cold atoms}).
In this context, the source physical phenomena are both the (potentially cleaned) measurement data, involving prior knowledge of the apparatus, and the equilibration behaviour observed in this experiment, whose interpretation is obviously theory or model mediated. 

What role, if any, we should attribute to target physical phenomena within analogue quantum simulation is far more subtle. 
In fact, it is precisely with regard to the role of target physical phenomena that we shall draw our crucial distinction between analogue quantum computation and emulation. We shall return to this point shortly after some significant preliminary discussion of a distinction between two different types of model. 

\section{System and Simulation Models}

In general, we can always describe a source or target system at different levels, using models with differing degrees of idealisation. Speaking of \emph{the} source or target model is misleading as there typically will in practice be several descriptions of the system at hand that play different roles in the simulation process. 
For a physical system involved in a simulation let us therefore distinguish between the \emph{system model} and the \emph{simulation model}.\footnote{There is a suggestive, but inexact, connection between our distinction here and the hierarchy of models account of  scientific experimentation \cite{suppes:1966,mayo:1996}. 

The system model in our account functions something like the notion of a `model of the data' within the hierarchy account. That said, nothing we will say will depend upon the strength of this connection and our account is very much specialised to the function of models within analogue quantum simulation. In what follows we will, as much as is possible, refrain from engaging with more general considerations regarding models in scientific experimental practice since the structure of relationship between models in the context of analogue quantum simulation requires its own specialised treatment. For a fascinating recent discussion of the hierarchy of models account of scientific experimentation in the context of high-energy physics experiments see \cite{karaca:2018}. For discussion in the context of classical computer simulation see \cite{winsberg:1999b}.} 

The system model provides the best available theoretical description of the particular system at hand in terms of quantitative predictions.
It is typically a very complex model used to represent a given system and involves a low degree of idealisation. 
This model usually takes the form of a Hamiltonian, Liouvillian or Lagrangian.  
In particular, the system model of the source system is key to all types of analogue quantum simulation as it forms the basis of the exploitation of a particular physical source system. 
This is the best available description of the experimental system (or class of experimental systems) which is actually manipulated in the lab. 
This description may include the peculiarities of the particular experimental settings as well as all known experimental imperfections such as fabrication errors and noise processes. Constructing this model also involves factoring in how the experimental controls correspond to changes in the model parameters, which is key to conducting a successful experiment. 
In this process the experimental system is also modified so as to remove imperfections as much as possible.\footnote{There are some interesting connections in this regard to a recent discussion of concrete models due to \cite{pincock:2020}.} 
To emphasise this connection with experimental practice we might also call the highly accurate system model obtained in the process of characterizing the experimental system at hand the `lab model' or `lab Hamiltonian'. Indeed, in some cases the lab model is precisely what the source system model often is: it is a particular Hamiltonian that varies (potentially significantly) from one laboratory implementing a particular system, say a linear-optical system or a cold-atom system, to another laboratory implementing the same system.   

In contrast there is the simulation model. 
This is typically a highly idealised and simplified model that takes on the role of what we labelled `computational model' for classical computations in the case of (analogue) simulations. This model may also take the form of a Hamiltonian, Liouvillian or Lagrangian. The simulation model is typically generic, in the sense that it  describes the \emph{type} of system at hand. 
Simulation models are thus often the standard by which we judge what is meant by `the same system'. Having such a highly idealised description of the system allows for a `clean' theoretical study of a range of concrete systems which does not include specific peculiarities of the concrete system but only the most important interactions and features of the system at hand. It is important to note that while the system model invariably will itself also be idealised, it will be \textit{less idealised} than the simulation model. 
This is because a higher degree of idealisation makes it possible to find formal connections between the source simulation model and the simulation model of other physical systems, and in particular the target system, in the first place. 

The core common feature of analogue quantum simulations is the exploitation of a formal relationship between a target simulation model and source simulation model via the exploitation of the particular experimental source system.
The distinction between the source system and simulation models is  then of great importance for the evaluation of the success of analogue quantum simulation. This is because it is only in virtue of the `validation' of the relationship between the source system model and the source simulation model that the simulation can be used to make reliable inferences. We will return to this crucial point at length in Chapter~\ref{ch:norms}. Before then the distinction between source and simulation models will be grounded in the careful analysis of real experimental practice. 

The key point for the time being is that both `source model' and `target model' can be understood as triples of the respective system model, simulation model, and a description of the formal relationships (typically limiting relations in specific parameter regimes) between the two. Significantly, it is the formal relationship \textit{between models} that is the main focus of the book. This is not, of course, to relegate the physical source system to a level of secondary importance. Without the actual simulator system there can be no analogue quantum simulation. The point of significance is that the system source model is an experimental model and thus although our focus is on the relationship between models, one of the models in question is always essentially linked to the observation and manipulation of a physical source system. This feature will prove crucial to our account of the scientific understanding that analogue quantum simulation can provide.

\begin{figure}[t]
\centering
    \includegraphics[height=.4\textheight]{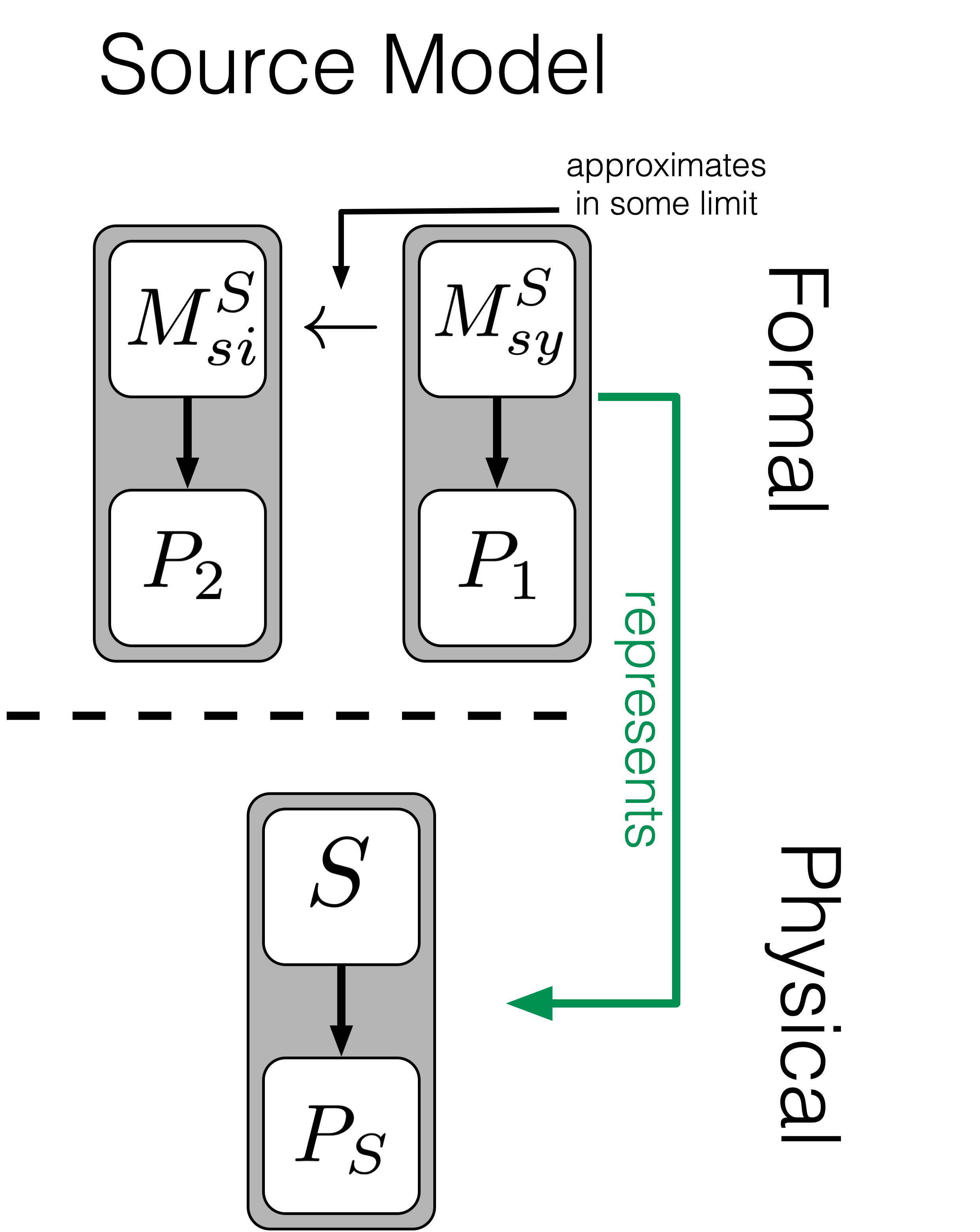}
\caption{
    \label{fig: source_model} 
Schematic representation of the structure of the source model. The three parts are the \textit{system model} ($M^S_\text{sy}$, the complex experimental model), the \textit{simulation model} ($M^S_\text{si}$, the simple idealised model), and the limiting relations between the two models. The system models stands in a representation relationship with the concrete physical system, $S$, which displays some set of phenomena, $P_S$. Our analysis will focus on the formal structure of the relations between models and not require any particular formalisation of the representation relation. However, since $M^S_\text{sy}$ is an \textit{experimental} model, the physical source system plays an essential role in the relevant inferences. $P_1$ and $P_2$ here can be understood as more and less idealised formal representations of the phenomena, $P_S$.     
    }
\end{figure}

As well as the formal relationship between models we can also, schematically, include a `representation relation' between the models and the physical systems that they represent. There is no requirement within our treatment for such relations to be formalised or characterised in any specific way.\footnote{ \label{repnot} Whilst on some account representation is a partial isomorphism, on others it is built around similarity, and still others it involves ideas such as `denotation'. Nothing we will assume about representation in the remainder of the book will rely on any particular account of the concept, and in this sense we will assume our analysis of analogue quantum simulation to be philosophically neutral with regard to representation. For an excellent recent introduction to the scientific representation and the analysis of models see \cite{frigg:2020}. For discussion specifically relating to representation via material models see \citep{frigg:2018}. Further accounts, of representation in science are \citep{hughes:1997,giere:1999,suarez:2004,contessa:2007,bailer:2009,weisberg:2012}. A good overview of various connected issues is provided in \cite[\S2]{gelfert:2016}.} Moreover, for our purposes, what will be significant is the empirical adequacy of the relevant models with regard to the class of phenomena of interest, thus the correspondence between models and the world that our account will be anchored in is a principally empiricist one. That said, representation will be important to our account in the somewhat different context of the `mental representations' that we take to provide scientists with understanding. We defer discussion of this idea to Chapter~\ref{ch:understanding}.

We can represent these first steps in our analysis via the diagrammatic language given in Fig. \ref{fig: source_model}. We will expand and refine such diagrams throughout the book.

\section{Analogue Quantum Computation and Emulation}
\label{sec:computtaionempulation}

We hope that it is now clear what we are talking about when we talk about analogue quantum simulators. In making the distinctions above we have been following the fairly standard scientific usage. Our next and most significant distinction, by contrast, is a prescription towards a new terminological refinement. That is, we would like to encourage scientists to distinguish between two different types of activity that are both currently described under the title `analogue quantum simulation'. 
These are analogue quantum \textit{computation} and analogue quantum \textit{emulation}. 

The distinction that we are enjoining scientists to make is driven by the differing scientific goals that scientists might have in performing an analogue quantum simulation, in particular, the differing conceptions of the `target' about which they aim to learn by performing the simulation. 
Later, we will defend a view in which the relevant scientific goal is one of understanding and differentiate different forms of understanding. 
For the purpose of this preliminary analysis of the distinction between analogue quantum computation and analogue quantum emulation we will leave the fleshing out of what we mean by `understanding' to the reader's intuitions. For the time being, the crucial question when performing an analogue quantum simulation is: What is the `target' about which a scientist aims to gain understanding?

In analogue quantum \textit{emulation} a scientist is interested in gaining understanding of \textit{physical target phenomena}, that is, quantitative or qualitative properties of a target system which can in principle be constrained by conventional forms of inductive reasoning based upon experimental and observational evidence. This notion of `physical phenomena' is purposefully extremely broad. It includes  `observable phenomena' which are directly discernible via the senses (such as the phases of Venus), `manipulable phenomena', which we have mediated causal access to (such as the spin of an electron), accessible phenomena which we can receive signals from but may not be able to manipulate (such as gravitational waves from binary black hole mergers), and inaccessible phenomena which we cannot in fact access but possible experiments could (such as radiation from stars outside our cosmological particle horizon).\footnote{
We should note that there is debate in the philosophy of science regarding the relative epistemic status of these different types of phenomena. At one extreme, constructive empiricists only accept that scientists can make reliable knowledge claims regarding observable target phenomena \cite{vanFraassen:1980,sep-constructive-empiricism}. More reasonably, \citet{massimi:2007}, building on the original work on data and phenomena due to \citet{bogen:1988}, has argued that scientific knowledge regarding unobservable but manipulable phenomena can reliably be drawn from experimental practice in modern particle physics. Finally, \citet{evans:2019} argue that contemporary practice in astrophysics in fact includes numerous uncontroversial examples of reliable knowledge regarding both in practice and in principle inaccessible phenomena. We will consider the relevant aspects of these arguments later.
}

Often, we have a theory or model regarding some set of physical phenomena that we wish to test, but are unable \textit{in practice} to gain experimental access to the phenomena within the full relevant parameter regime. 
This is the situation in which quantum emulation constitutes a new inferential tool. 
In a conventional experiment scientists typically gain understanding of physical target phenomena by probing such phenomena in a \textit{token} source system of the relevant \textit{type} (e.g. gaining understanding about electron spin in general by probing the spin of particular electrons).\footnote{For details on the distinction between types and tokens in the philosophical literature, see the overview by \citet{wetzel_types_2018}.} 
Scientific understanding drawn from experimentation is typically built upon (theory or model mediated) source-target inference towards general properties of a system type based upon an experiment on a token of the relevant type. 
The crucial difference in an analogue quantum emulation is that the system manipulated is of a different type to the system about which the scientist wishes to gain understanding. That is, the physical phenomenon probed in the experiment is displayed by a source system of a different type to the target system of interest to the experimenter. Whilst the source and target systems are, by construction, taken to be similar in the sense that they both display the relevant isomorphic phenomena, they will, in general, be very different in terms of the relevant wider set of phenomena and structure of physical laws -- as different as black holes and fibre optic cables or sulphur bacteria and photonic circuits.\footnote{What we mean by `different types' here is thus made abundantly clear by concrete examples and certainly does not require any metaphysically strong commitment to type-token distinctions. Indeed, in respect of the isomorphic phenomena the source and target system in an analogue quantum emulation will be, in a certain sense, of the `same type'. However, in the examples of analogue quantum emulation we will consider there are always ample well-established senses in which they will be of different types, and thus also in which the analogue quantum emulation is unambiguously different from a conventional experiment.} 

Scientific understanding drawn from analogue quantum emulation is built upon (theory or model mediated) source-target inference towards general \emph{properties of a physical system} of one type based upon an experiment on a physical system of another type. 
The extent to which such inferences can be justified, and the forms of understanding they can bring, will be one of the abiding questions of the book.\footnote{It is important to note that in both practice and principle the limits on what we can observe, manipulate, and access is dependent on time, technology, theory, and models. As time goes on we gain more data about the world around us by simple accumulation, most trivially in the sense that the cosmological horizon of bodies about which we have received information expands. Furthermore, as technology changes we can gain in practice access to phenomena which where hitherto thought to be inaccessible, often simply through new techniques of measurement or simple greater precision. Finally, the scientific concepts of observation, manipulation and access are heavily theory dependent, and as such liable to evolve as our theories change. These considerations are particularly relevant to the context of quantum emulations of in practice inaccessible target phenomena, since precisely such evolution in accessibility can lead to post hoc validation of emulations.}  
Crucial to answering these questions will be a careful analysis of the precise relationship between source and target as mediated by the system and simulation models of both systems. 
 
In analogue quantum \textit{computation} a scientist aims to gain understanding of certain \textit{formal properties} of a target mathematical model via the manipulation of a source physical system. That is, given a target mathematical model, the goal is to understand properties of the solutions to this model via experimenting on a source system. Depending on the modelling framework the mathematical form that  the solution may take will vary. In quantum mechanics it might be the ground state of a Hamiltonian as represented by a ray in a Hilbert space, in hydrodynamics it may be the solution to a set of partial differential equations given some initial condition. 
Properties of these solutions can be quantitative, like the expected value of some Hermitian operator, or qualitative, such as the occurrence of turbulences in a hydrodynamical system. As with physical phenomena, the understanding of these formal properties is essentially theory or model mediated. 

It is important to note that our definition of analogue quantum computation depends upon scientists' goal of understanding formal features of a mathematical model, not the relationship (or not) of the mathematical model with physical phenomena. In many cases of interest, the target mathematical model will be expected to correspond to a physical target system.\footnote{Clearly in such cases, the target system would need to be of a different type to  the source system, else the analogue quantum computation in question would reduce to a conventional experiment.} The properties of the solutions will thus be related to some physical phenomenon occurring in that system. However, in analogue quantum computation such physical phenomena are not the focus of the understanding that scientists wish to gain via analogue quantum computation.\footnote{For example, the modeller may have no physical target system in mind, but rather only an idealisation or abstraction thereof. Prime examples of this are toy models~\cite{sep-models-science,ToyModels}.} 
Furthermore, it may even be the case that the target mathematical model is known to \textit{not} putatively represent any physically possible system, yet this would not bar a scientist from gaining understanding of the relevant formal properties via analogue quantum computation. Thus a scientist can seek to gain understanding via an analogue quantum computation even when no distinct target system exists (or is even physically possible) for a relevant  target mathematical model. They are seeking to gain understanding of mathematical properties of an abstract model, and the role of that model in putatively representing things in the world is not necessarily relevant to that understanding.   

The goal of analogue quantum computation is to understand formal properties of a mathematical target model. This is particularly relevant and interesting in situations in which the target model is not solvable by either numerical simulation algorithms implemented on a classical computer or analytical calculations. 
In this case, although the solutions deductively follow from the specification of the mathematical model and the corresponding modelling framework, one does not know whether or not the solutions to this model display certain properties or not. 
Analogue quantum computations may often solve this problem as they are able to exploit the superior computational power offered by quantum computational models. 
This is particularly so when the formal model involves quantum mechanical equations as already observed by \citet{feynman:1982}. 

Let us consider a simple example. Consider the phenomenon of frustration in models of magnetic spin networks as described in the simplest case by an Ising model. 
In the model a network of spins are connected via anti-ferromagnetic interactions, which attempt to orient neighbouring spins towards opposite directions. 
Under certain lattice geometries, for example three spins arranged in a triangle, no spin configuration can satisfy all interactions so that one pair of spins will inevitably be aligned. 
The target model is the magnetic spin networks model and the property is frustration. We can gain understanding of this formal property of the target model by numerical modelling on a classical computer. 
Similarly, we can gain the relevant understanding by performing an analogue quantum computation on a second source system that is analogous in the relevant respects \cite{Kim:2010ib}. We might also wish to gain understanding of the actual physical phenomena of frustration as displayed by a particular physical target system. For example, the phenomenon of frustration has been shown to be central in the understanding of protein folding \cite{Bryngelson:1987uu} and we may wish to understand better how frustration functions in such a physical context. It is when scientists have such goals that we talk about analogue quantum emulation. The goal is understanding quantitative or qualitative physical properties of a physical target system. 

The distinction between analogue emulation and computation is not always an exclusive one. In some situations, for example, one might want to compute the solution to a mathematical model of a physical target phenomenon in order to better understand that phenomenon. 
Vice versa, in analogue quantum emulation understanding the target phenomenon will often be mediated via a model of the target system. Thus, in an analogue quantum emulation one is implicitly also performing a computation. We take it, however, that while deeply intertwined, most analogue quantum simulation experiments focus on and are tailored towards achieving one of those goals and remain largely silent about the other. 
As such, they may be judged by whether or not they achieve this goal oblivious to whether or not they achieve the other goal, too.
In particular, both methodological practice and the relevant evaluative criteria for success differ between the two contexts. 
Our distinction thus engenders a normative difference. The principal argument of this book will be that this distinction between analogue quantum computation and analogue quantum emulation is one that matters.

Finally, before we move on to our detailed consideration of the case studies it is worth noting  the relevance of the distinction between system and simulation models in the context of the target model used in analogue quantum computation and emulation. In the context of analogue quantum emulation it makes sense to distinguish between target \textit{system} model and target \textit{simulation} model because, here, the target is a concrete physical system. 
The target system model is the less idealised model which we can reasonably take to correspond to or represent the complex and detailed physical process that the target system displays according to our current best physical understanding. 
The target simulation model is the more idealised and simple model that is taken to stand in the relevant formal correspondence (usually a partial isomorphism) with the source simulation model.

Of course, in such cases it is typically not the case that the target system model is conceived of as a `lab' model, since the system itself is not usually manipulable in the lab within the relevant parameter regime. However, the target system model may still be parameterised based upon experiment or observational data. In contrast, in an analogue quantum computation, scientists have to goal of gaining understanding of formal properties of a mathematical target model. Hence, there is no target `system model' as there is no physical system involved, but only a target simulation model. 

\begin{figure}[t]
\centering
    \includegraphics[height=0.4\textheight]{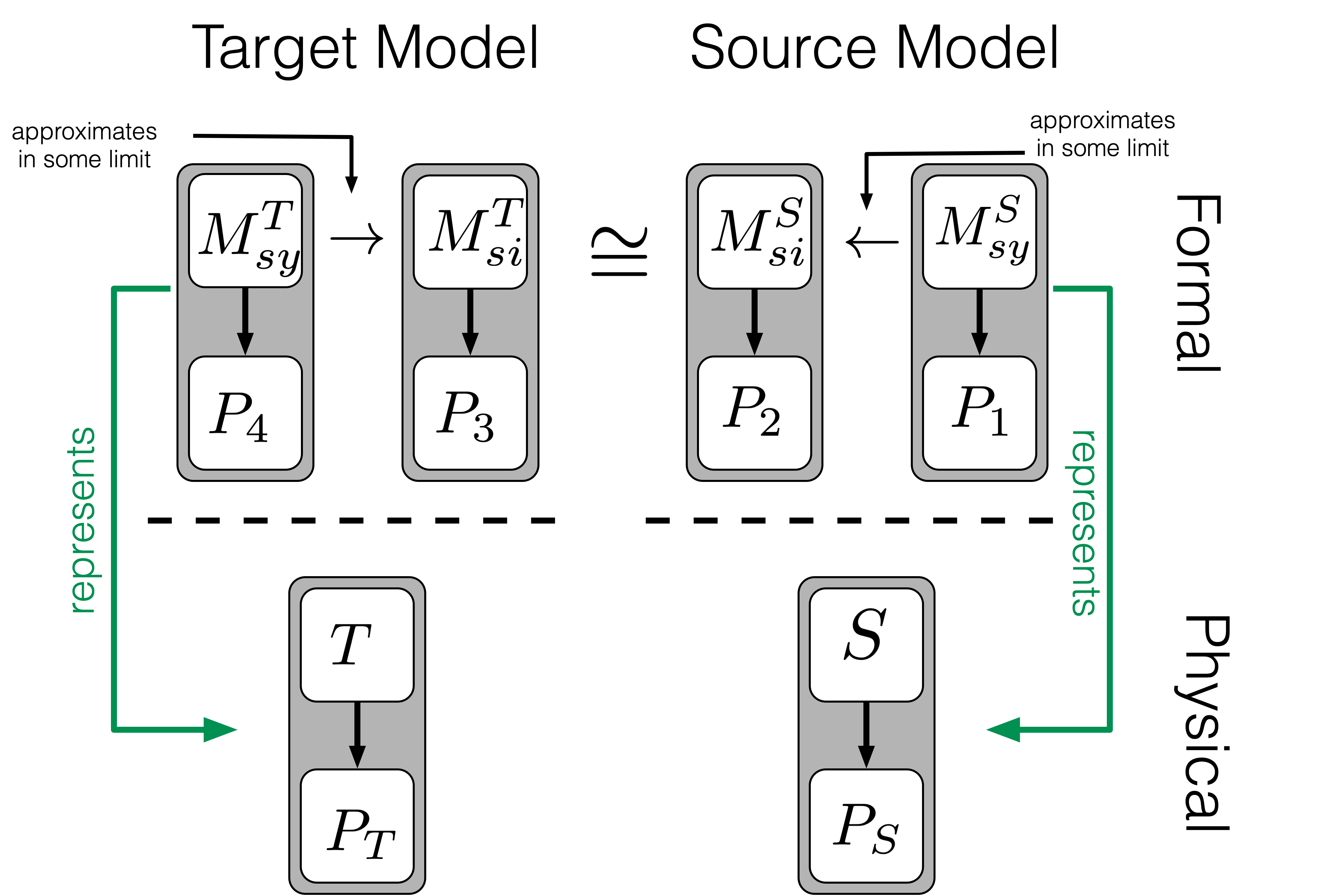}
\caption{
    \label{fig: ae_model} 
Schematic representation of the structure of an analogue quantum emulation. Four models are involved: the source system model ($M^S_\text{sy}$, the complex experimental model of the source system), the source simulation model ($M^S_\text{si}$, the simple idealised model of the source system), the target system model ($M^T_\text{sy}$, the complex experimental model of the target system), the target simulation model ($M^T_\text{si}$, the simple idealised model of the target system). Each of the system and simulation models are related by limiting relations  which we will denote as $M_1\rightarrow M_2$ where $M_1$ is the more general model that approximates the less general model $M_2$ in some limit. The two simulation models are typically related by a partial isomorphism  which we will denote as $\cong$.}
\end{figure}

\begin{figure}[b]
\centering
    
    \includegraphics[height=0.4\textheight]{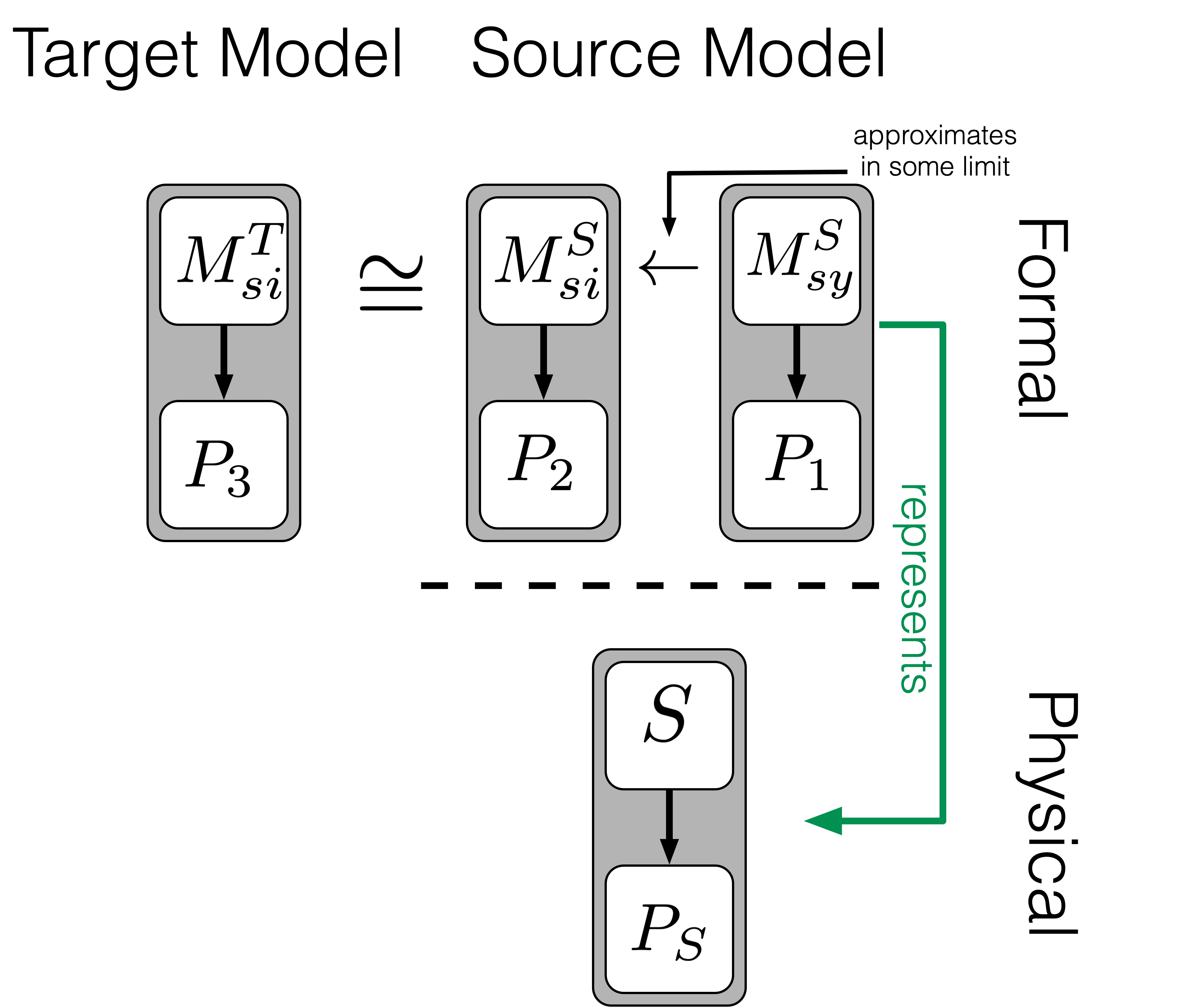}
\caption{
    \label{fig: ac_model} Schematic representation of the structure of an analogue quantum computation
Three models are involved: the source system model ($M^S_\text{sy}$, the complex experimental model of the source system), the source simulation model ($M^S_\text{si}$, the simple idealised model of the source system), the simulation target model ($M^T_\text{si}$, the simple idealised model of the target system). The source system and source simulation models are related by limiting relations which we will denote as $M_1\rightarrow M_2$ where $M_1$ is the more general model that approximates the less general model $M_2$ in some limit. The two simulation models are typically related by a partial isomorphism which we will denote as $\cong$.}
\end{figure}

It is instructive to represent the full structure of the various models and their relations in the context of both analogue quantum emulation and analogue quantum computation using our diagrammatic language, see Fig. \ref{fig: ae_model} and Fig. \ref{fig: ac_model}. 
In the next three chapters, we will provide concrete illustrations of the various system and simulator models and how they are related in cases of analogue quantum computation and emulation. In each case we will fill in the various details applying the same diagrammatic language.  This careful and explicit analysis of real scientific practice will then, in turn, provide the keys to understanding the forms of scientific understanding that analogue quantum computations and emulations can bring.

%
%
%

%


\chapter[Cold atom computation]{Cold Atom Computation: From Many-Body Localisation to the Higgs Mode}
\label{ch:cold atoms}


A powerful and highly flexible platform for analogue quantum simulation is a lattice of cold atoms which can be created in the lab via laser beams and spatially varying magnetic fields. The experimental implementation of the cold atom simulator is such that the system can be tuned to mimic condensed matter phenomena such as many-body localisation and the Higgs mode. Detailed analysis of these case studies shows them to exemplify the notion of analogue quantum computation. That is, the type of analogue quantum simulation in which source phenomena is being appealed to for the specific purpose of gaining understanding of formal properties of a target simulation model.

\section{Science Summary}

Cold atoms in optical lattices are one of the most important platforms for quantum simulations \cite{OptLatQS}. 
In such systems an artificial lattice potential is created using counter-propagating laser beams: a crystal of light. 
The resulting intensity pattern acts as a space-dependent lattice potential for certain atoms via the dipole-dipole coupling between the light field and the
dipole moment of these atoms. Such optical-lattice potentials can be combined with so-called magneto-optical traps (MOT) with which one can create a low-temperature state of a confined atomic cloud consisting of atoms such as $^{87}$Rb (a bosonic atom) or $^{40}$K (a fermionic one). 
One can thus realise a system in which hundreds to thousands of atoms evolve coherently while interacting among themselves and propagating through the lattice. These systems bear strong similarities to real solid-state systems, where one encounters the same lattice structure for the potential of electrons hopping between atoms\footnote{Although for our purpose, the specific experiments at focus aren't considered to be representing an actual concrete system.}. Strikingly, an idealised Hamiltonian that has been found to accurately describe the dynamics of cold atoms in optical lattices is the Hubbard Hamiltonian \cite{Jaksch1998}. 
In its fermionic variant, the Hubbard Hamiltonian is the simplest model describing interacting fermions that features Coulomb repulsion, a nontrivial band structure, and incorporates the Pauli Principle, and thus serves as a simple model of electronic solids \cite{Hubbard-1963}. 
But also its bosonic variant describes a plethora of physical solid-state systems\footnote{See \citet{bruder_bose-hubbard_2005} for a review.} including granular high-temperature superconductors \cite{muller_discovery_1987,muller_flux_1987}, Helium films \cite{mcqueeney_surface_1984,reppy_4he_1984,finotello_sharp_1988} and Josephson junction arrays \cite{bruder_quantum_1992}. 
While the experimental realisation of the Fermi-Hubbard model at low temperatures remains an outstanding challenge, however, the bosonic variant is being implemnted with enormous precisoin in many laboratories across the world today \cite{Bloch-RMP-2008}. 
What makes cold atoms in optical lattices so well suited for the simulation of condensed-matter phenomena is the possibility to both manipulate and probe these systems with high precision. 

\begin{figure}[t]
  \includegraphics[width = \textwidth]{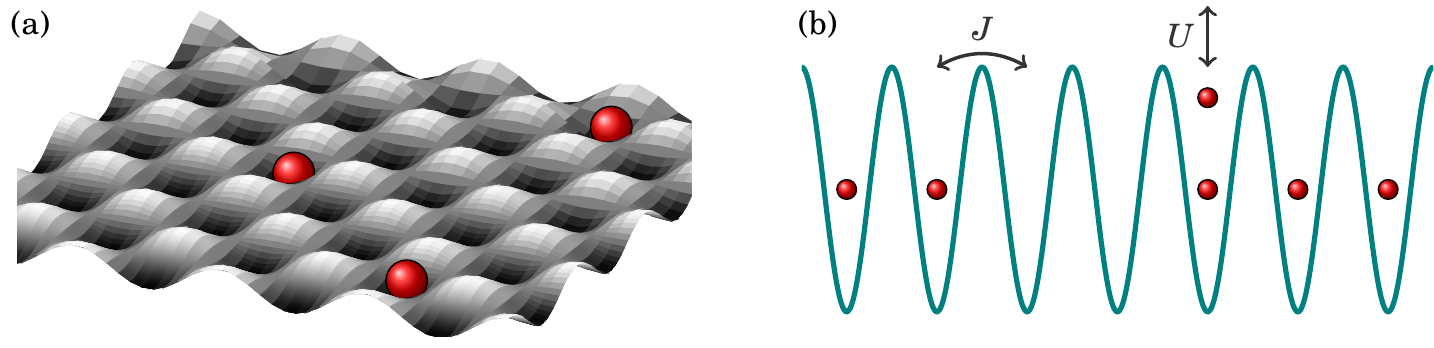}
  \caption{\label{fig:optical lattices}
    (a) Sketch of the potential landscape of cold atoms in a two-dimensional optical lattice potential. 
    (b) The Bose-Hubbard model \eqref{bose hubbard} features a hopping term with energy gain $-J$ and a density-density interaction term giving an energy cost of $U$ to two particles located at the same lattice site. 
   }
\end{figure}

An example of a quantum simulation that has been performed using cold atoms in optical lattices is the study of the phenomenon of many-body localisation (MBL) in different dimensions. 
When left alone for a long time, generically, we expect that local subsystems of any quantum system approaches a thermal equilibrium. 
In contrast, in random potential landscapes non-interacting quantum particles are known to localise \cite{anderson_absence_1958}, providing an example of a quantum system that does not thermalise after being left alone for long times. 
However, in nature, particles interact, and it is an open research question whether or not the localisation phenomenon persists once the particles are allowed to interact between one another, a state of affairs known as \emph{many-body localisation}. 

In fact, in one dimension, numerical calculations are feasible and allow the study of MBL in different conditions: 
it turns out that once the disorder strength supersedes a critical value, certain interacting systems do localise. 
In higher dimensions, however, there are no known methods for performing efficient numerical simulations that would allow us to determine whether or not many-body localisation persists when more degrees of freedom are allowed for. 
This is the terrain in which quantum simulators such as cold atoms in optical lattices, which naturally simulate interacting particles in a potential landscape, may provide new insights. Significantly, flexibility in the dimensions that are accessible in quantum simulators allows for a validation of the simulation in the classically simulable one-dimensional regime. The experimental findings (detailed below) suggest that MBL is present in two dimensions~\cite{bordia_probing_2017,choi_exploring_2016}. 
This claim is supported by several quantum simulations with similar findings that have been carried out in different experimental platforms~\cite{roushan_spectroscopic_2017,smith_many-body_2016}.  
The experiments are validated by comparing the experimental results in one dimension to exact numerical simulations using density matrix renormalisation group (DMRG) methods~\cite{schreiber_observation_2015}. 

Another example of a cold atom quantum simulation studies the Higgs mechanism in two dimensions \cite{Higgs:2012}. 
The Higgs mechanism appears in the study of spontaneous symmetry breaking,
which lies at the heart of our understanding of various natural phenomena. 
Most famously, the Higgs mechanism appears in particle physics where a spontaneously broken symmetry leads to the emergence of massive particles. The Higgs mechanism is also important in condensed matter physics, where the phases of matter can most often be understood in terms of breaking symmetries. It is a subject of theoretical controversy whether in two-dimensional systems a Higgs mode is present \cite{sachdev_universal_1999,altman_oscillating_2002,podolsky_visibility_2011,podolsky_spectral_2012,pollet_higgs_2012,liu_massive_2015}. The standard methods of solving such problems in condensed matter physics -- analytical solution and Quantum Monte Carlo methods -- fail entirely, and approximations such as methods based on low-order perturbation theory do not provide a definitive answer. 

This is precisely the situation where analogue quantum simulation proves a powerful new inferential tool. 
A toy model for the Higgs mechanism is an $O(2)$-symmetric field theory. 
At a critical superfluid--Mott insulator (SF-MI) transition the equations describing cold atoms in an optical lattice are given by an $O(2)$-symmetric field theory \cite{altman_oscillating_2002}. 
It is thus possible to probe the solution space of the $O(2)$-symmetric field theory by inducing a SF-MI critical transition in the cold atom system. 
\citet{Higgs:2012} hoped to answer the theoretical question of whether a Higgs amplitude mode exists in two dimensions, by performing an experiment on cold atoms in optical lattices. 
The experimental findings, detailed below, suggested the signature of spontaneous symmetry breaking with a two-dimensional Higgs mode via indirect evidence, but did not exhibit the `smoking-gun' feature of a Higgs excitation. However, this feature could be detected in similar experiments using a different physical platform that makes use of superconductors \cite{matsunaga_higgs_2013,matsunaga_light-induced_2014}. 

\section{Bose-Hubbard physics in optical lattices} 
\label{sec:cold atoms in optical lattices}

Motivated by the experimental possibility to trap ultra-cold atoms close to absolute zero temperature in an optical-lattice potential \cite{raithel_cooling_1997,muller-seydlitz_atoms_1997,hamann_resolved-sideband_1998}, \citet{Jaksch1998} showed that such systems can in certain limits be effectively described by a Bose-Hubbard Hamiltonian, which is given by 
\begin{align} 
  \label{bose hubbard}
  H_{\text{BH}} = - J \sum_{\langle j,k \rangle} \left(b_j^\dagger b_k + b_k^\dagger b_j \right)  + U
  \sum_j b_j^\dagger b_j^\dagger b_j b_j  + \sum_j \mu_j b_j^\dagger b_j \, . 
\end{align}
Here, $b_j^{\dagger}$ ($b_j$) denotes a bosonic creation (annihilation) operator at site $j$, $U$ denotes the energy cost from having two atoms on the same site, $J$ is the energy gain when hopping from one site to the next, and 
$\mu_j$ describes the energy offset of each lattice site. 
The derivation of \citet{Jaksch1998} starts from the first-principles Hamiltonian for bosonic atoms at site $\mathbf x \in \mathbb R^3$ with field operators $\psi(\mathbf x)$ interacting via a contact (density-density) interaction
\begin{multline}
    H_{\text{CA}} = \int d \mathbf x\,  \psi^\dagger(\mathbf x ) \left( - \frac {\hbar^2}{2m} \nabla^2  + V_0(\mathbf x ) + V_T(\mathbf x )  \right) \psi(\mathbf x ) 
    \\
    + \frac 1 2 \frac { 4 \pi a_s \hbar^2}{m} \int d \mathbf x \,\psi^\dagger(\mathbf x ) \psi^\dagger(\mathbf x ) \psi(\mathbf x ) \psi(\mathbf x ) , 
    \label{eq:cold atoms hamiltonian}
\end{multline}
where $a_s$ denotes the $s$-wave scattering length of the atoms and $m$ their mass. 
The atoms are confined by an external potential $V_T(\mathbf x)$  as well as a spatially dependent periodic trapping potential $V_0(\mathbf{x}) = \sum_{j =1}^3 V_{j0} \sin^2(k x_j)$ with wave vector $k = 2 \pi/\lambda$ determined by the wavelength of the laser light producing the potential. 

In terms of the distinction between system and simulation model, the Bose-Hubbard Hamiltonian is therefore the simulation Hamiltonian, while the first-principles Hamiltonian \eqref{eq:cold atoms hamiltonian} is the basic system Hamiltonian which may be supplemented with additional terms in a concrete experimental situation.
\citeauthor{Jaksch1998} show that $H_{\text{CA}}$ reduces to the Bose-Hubbard Hamiltonian $H_{\text{BH}}$ in the limit of low-energy system dynamics by expanding the field operators in Wannier functions, a complete basis describing localised orbitals of crystalline structures, and keeping only the lowest vibrational states, i.e., a single excitation per site. 
$J$, $U$ and $\mu_j$ are then expressible as integrals of the Wannier functions: 
the expression for $\mu_j$ is approximately given by the trapping potential $V_T(\mathbf x_i)$, the expression for $U$ depends on the $s$-wave scattering length as well as  the `self-overlap' of a Wannier function and the expression for $J$ is given by the overlap of those functions between distinct sites depending on the lattice potential $V_0$. 
Idealising the details of the concrete system, it is assumed that this overlap is negligible for next-nearest neighbours and beyond, and  that it does not depend on the external trapping potential $V_T$. 
The validity of this idealisation can be quantified in terms of energy separation to the first excited band and the relation between the on-site interaction $U$ and the excitation energy to the next band.

It is no exaggeration that by today, quantum simulation of bosonic Hubbard physics based on the relation just described is a significant part of contemporary scientific practice. 
The formal properties of the target (simulation) model that may be studied in the context of Hubbard physics include the non-equilibrium quantum dynamics leading to an equilibrated or thermalised state \cite{BlochEisertRelaxation,schreiber_observation_2015,choi_exploring_2016}, 
quantum phase transitions \cite{greiner_quantum_2002,Braun-pnas-2015,
esslinger2016phases}, 
magnetism \cite{Struck2011magnetism,murmann2015heisenberg}, metal-insulator
transitions, and high-temperature superconductivity \cite{ColdFermions}. 

A key feature of cold atom systems that enables the quantum simulation of such diverse systems is the possibility to manipulate and probe the cold-atom system in a variety of ways. 
Using the response of the atoms to an external magnetic field via a so-called Feshbach resonance, it is possible to tune the $s$-wave scattering length $a_s$ of the atoms and thereby the coupling parameter $U$ by orders of magnitude. 
By varying the amplitude and phase of the generating laser beams, one can tune the potential-dependent hopping parameter $J$ and local potentials $\mu_j$. 
Thus, all model parameters can be varied independently. 
By superimposing several lattice potentials with different wavelengths one can
even realise next-nearest-neighbour interactions and lattices with higher
periodicity \cite{Foelling-Nature-2007}. 
Likewise, it is possible to implement optical lattices \cite{schreiber_observation_2015} with a quasi-random site-dependent energy offset $\mu_j$ in this way as well as fully random energy offsets using an additional potential generated by electro-optical light modulators \cite{choi_exploring_2016}.
It is even possible to realise low (one and two) dimensional systems by increasing the potential barriers in the orthogonal directions to suppress tunnelling.
In such a way it is possible to access a large parameter regime and probe a variety of phenomena that occur in the many-body system. 

There are also a large variety of methods available to probe the cold atom simulator system. 
This stands in stark contrast to real solid state systems. 
Using highly
focused microscopes one can probe individual atoms in the lattice with single-site resolution~\cite{Bakr-Science-2010, Sherson-Nature-2010}. 
Using so-called time-of-flight imaging, and
variants thereof, one can also measure the free-space and quasi-momentum
distribution of the atoms \cite{Foelling-Nature-2007,Bloch-RMP-2008}. 
In experiments with cold atoms in optical lattices one is thus able to learn about the \textit{formal properties} of a target model, by exploiting that this system constitutes a well-controllable and accessible many-body system. 
Significantly, knowledge of these formal properties often cannot be gained by other means, since the models are intractable via other types of computation.
Indeed, neither the bosonic nor the fermionic variant of the Hubbard Hamiltonian
admits an analytical solution or is amenable to numerical simulations outside
of certain parameter regimes.

\section{Many-body localisation}

\subsection{Many-body localisation in one and two dimensions}

One of the core postulates of classical thermodynamics is that, independent from the initial conditions, any system eventually thermalises to an equilibrium state that is uniquely characterised by the internal energy of the system. 
This process is irreversible since the microscopic details of the initial state are lost in the process of thermalisation, giving rise to a statistical ensemble. In contrast, in quantum theory, it is expected that any isolated system will evolve unitarily in a reversible fashion, such that the state remains pure after arbitrarily long times.

At first sight, these two representations of time  evolution of a system appear incompatible. Given the overwhelming evidence for the empirical adequacy of both thermodynamics and quantum theory in their respective domains, there is then an obvious requirement to explain how the relevant `novel behaviour' of irreversibility `emerges' in the cross-over regime.\footnote{There is a wealth of philosophical discussion of questions relating to the emergence of novel behaviours in the thermodynamic limit, usually with a focus on classical statistical mechanics rather than quantum mechanics; see, for example, \cite{batterman:2002,Mainwood2006,butterfield:2011,saatsi:2018,Palacios:2018,palacios:2019}. Oddly, philosophers seem, as of yet, not to have focused much on the question of the emergence of time asymmetry in thermodynamics from quantum statistical mechanics. For philosophical discussion of the status of the arrow of time in thermodynamics see \cite{brown2001origins,sep-time-thermo}. For discussion of the time reversal invariance of quantum theory see \cite{roberts2017three,Roberts:2019}. For fascinating recent work on time's arrow and initial quantum states see \cite{chen:2020}.} The obvious strategy to answer this challenge, based upon the  assumption of quantum fundamentality, is to attempt to derive the irreversibility of time evolution in thermodynamics from the reversible unitary quantum evolution of the microscopic constituents of matter in the appropriate limit~\cite{neumann_beweis_1929,gogolin_equilibration_2016}. Solving the long-standing problem of thermalisation in the quantum regime has proven extremely challenging. In particular, an exact solution of the problem hinges upon knowledge of the full quantum state, which is inaccessible in all practical scenarios. More reasonably one can consider local reduced states, which may in particular become thermal mixtures during time evolution.

An insightful and more tractable approach to the problem is to study situations in which quantum systems \emph{do not} come to equilibrium, even after very long times and the conditions that foster this state of affairs. In this case, information encoded in the initial state of the system remains present even after waiting for a long, in fact infinite, time. 
A prominent example of such behaviour was found by \citet{anderson_absence_1958}: 
in a regular lattice structure, as found in solid states, the eigenstates of non-interacting particles are described by so-called Bloch states, which are extended in real space. 
In contrast, if an arbitrarily small disordered potential is added, their wave functions become localised in a small region of space and remain there for infinite time, giving rise to the phenomenon of \emph{Anderson localisation}. 

Intuitively speaking, this localisation phenomenon can be understood through the phenomenon of destructive interference of long-range hopping processes~\cite{abanin_colloquium:_2019}. 
However, the situation changes dramatically as soon as the particles are allowed to interact since scattering processes increase the mobility of particles even in the presence of strong disorder. 
One may therefore ask the question whether localisation phenomena persist even in the presence of interactions between particles -- a state of affairs dubbed \emph{many-body localisation (MBL)}.
In fact, early on interactions were expected to destroy the interference effect giving rise to Anderson localisation.

It is remarkable that localisation phenomena seem to persist even in the presence of interactions. This has been suggested by several heuristic, analytical and numerical studies. Phenomenologically, MBL is manifested in both the \emph{static} and \emph{dynamical} features of a disordered quantum many-body systems. 
Dynamically, that is in the situation in which an initial product state is evolved under a disordered Hamiltonian, it is characterised by the localisation of local observables such as the absence of particle transport \cite{schreiber_observation_2015,choi_exploring_2016} and a logarithmic growth of the entanglement entropy~\cite{znidaric_many_2008,bardarson_unbounded_2012}. 
Statically, MBL can be captured in terms of spectral properties of the disordered Hamiltonian. 
More precisely, the distribution of the consecutive gaps between its energy levels is expected to be Poissonian (as opposed to Wigner-Dyson distributed)~\cite{luitz_many-body_2015,oganesyan_localization_2007} . 
What is more, its eigenstates obey a so-called area law for the entanglement entropy \cite{bauer_area_2013,friesdorf_many-body_2015}. 
Intuitively speaking, this puts bounds on the amount of entanglement present in the eigenstates of such a Hamiltonian, indicating that those states are a superposition of few orbitals which are localised in space. 

Establishing reliable and robust results, or even obtaining a full answer to the question of MBL, is an extremely challenging task, as it requires solving the full interacting quantum many-body problem in its exponentially large Hilbert space. 
Localisation phenomena do permit certain simulations, even for long times, due to the fact that entanglement grows only very slowly in time. However, this is only in one dimension, where tensor-network based algorithms such as DMRG~\cite{SchollwoeckDMRG} and exact diagonalisation methods are feasible. 
These simulations provide strong evidence that a localised phase exists in one dimension if the strength of the disorder potential is above a certain threshold. 
More recently, however, the debate has taken another turn with arguments that localisation might eventually break down in the infinite-time limit even in one dimension~\cite{de_roeck_stability_2017,suntajs_quantum_2019}: 
it might be the case that in the presence of interactions the approach to thermal equilibrium is \emph{very slow} but not stopped altogether. 
This highlights that dynamical features of a system can in principle never confirm the existence of a stable MBL phase, since one can neither distinguish extremely slow relaxation from localisation, nor rule out that the system localises merely in an intermediate regime. 

However, for higher dimensions, that is, two and three dimensions all known simulation algorithms, in particular, tensor-network methods fail and only heuristic arguments are available for the absence or presence of MBL.
It is therefore an open question, whether or not disordered, interacting two- or three-dimensional systems localise.

\subsection{MBL in optical lattices}

\begin{figure}[t]
  \centering
  \includegraphics[width = .8\textwidth]{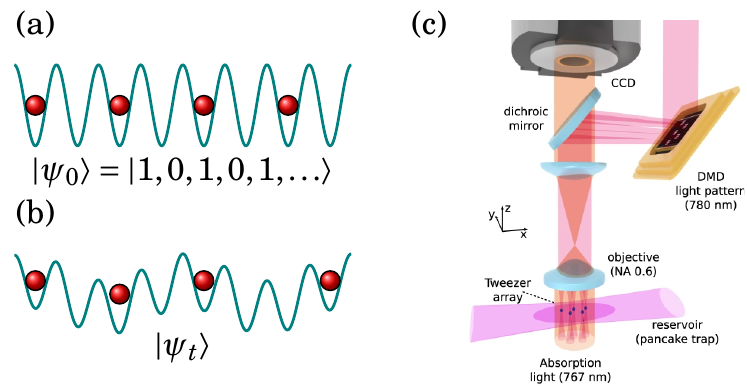}
  \caption{
  \label{fig:mbl setup}
  In optical lattice MBL experiments, an initially perfecly imbalanced state \emph{(a)} is evolved under a disordered Hamiltonian \emph{(b)}, and the final value of the imbalance tracked over time. 
  \emph{(c)} While in the experiments of \citet{schreiber_observation_2015,bordia_probing_2017} a quasi-random lattice is used, \citet{wang2020preparation} apply a digital mirror device (DMD) that is mapped to the plane of the optical lattice via a microscope objective to create an arbitrary disorder pattern (Figure (c) reproduced from \cite[Fig.~1d]{wang2020preparation}).
  }
\end{figure}

Optical lattice quantum simulators are perfectly suited to tackling this challenging theoretical terrain. 
A prototypical model for MBL is in fact given by the interacting Hubbard Hamiltonian \eqref{bose hubbard} with a random energy offset  $\mu_j$, also called  local disorder, that is drawn uniformly from a range $[- \Delta, \Delta] $.
Above, we already mentioned two ways to implement a disordered potential, which is crucial for MBL, in optical lattices: quasi-random optical lattices and electro-optical light modulators. 
Quasi-random optical lattices are created by superimposing two optical lattices with incommensurate lattice spacing \cite{schreiber_observation_2015,luschen_observation_2017}. 
This gives rise to a site-dependent potential $\Delta \cos(2 \pi \beta j + \phi)$ at site $j$, where $\beta$ denotes the ratio of the two distinct lattice periodicities, the disorder strength $\Delta$ can be tuned using the laser amplitude, and $\phi$ is a constant phase offset between the two lasers. 
More recently, a truly random disorder potential with local disorder has been implemented using a random light potential created by a so-called digital mirror device. 
The specific device used in the experiment of \citet{choi_exploring_2016} consists of an array of 1024x768 micromirrors each of which can be rotated by a small angle. 
This device can be used to map an arbitrary grayscale intensity pattern onto the plane of the optical lattice using an additional objective lens (which leads to slight correlations in the final intensity pattern)~\cite{wang2020preparation}, see Fig.~\ref{fig:mbl setup}(c).

The simplest way to study dynamical thermalisation phenomena in optical lattices is to prepare the system in an imbalanced state, that is, one in which only every odd lattice site of the optical lattice is occupied by an atom. 
One then lets the system evolve in a so-called \emph{quantum quench} under the disordered Hamiltonian.
Under normal conditions, the particles are expected to quickly diffuse over the lattice so that the probability of finding a particle at any given site is independent of the lattice site and given by the overall fraction of occupied sites in the entire lattice~\cite{BlochEisertRelaxation}. 
In the state preparation, an optical lattice with twice the intended periodicity $2\lambda$ is fully filled with atoms and the depth of the potential wells is increased to inhibit hopping of the atoms to neighbouring sites. 
In a second step, a super-lattice of double-wells is created by adding a second lattice with periodicity~$\lambda$ but with lower well heights. 
Those double wells are then tilted to let the atoms fall into every odd site of the $\lambda$-lattice and the lattice depth of the $\lambda$-lattice is increased to match that of the $2\lambda$-lattice, creating an imbalanced state in a lattice with periodicity~$\lambda$ \cite{Foelling-Nature-2007}. 

The quench is initiated by suddenly lowering the barrier heights and adding the random potential using additional laser beams, see Fig.~\ref{fig:mbl setup}(a) and (b).
One can then track the \emph{imbalance} $\mathcal I$ between the number of atoms on even ($N_e$) and odd sites ($N_o$) as given by 
\begin{align}
    \mathcal I  = \frac{N_e - N_o}{N_e + N_o} . 
\end{align}
If the system thermalises, $\mathcal I$ will converge to $0$ after very few tunnelling times, corresponding to exactly half a particle per site; 
if it localises, $\mathcal I$ will remain positive, retaining memory of the initial state with imbalance $+1$. 

The imbalance can be measured using a variant of time-of-flight imaging techniques. 
This variant works inversely to the state preparation, wherein two neighbouring potential wells are merged via a double-well lattice in such a way that particles from even sites are in a lower energy level than particles from odd sites. 
Since time-of-flight imaging measures the distribution of the particles in Fourier space or, equivalently, the population of the distinct energy bands of the system, particles from even sites will appear at a different place of the image than particles from the odd sites. 
Alternatively, one can use site-resolved real-space imaging techniques to measure the atom distribution after time evolution. 

Using the dimensionality-reduction methods described above one and two dimensional disordered systems can be implemented straightforwardly in optical lattices.
This allows probing the MBL transition in regimes that are classically intractable (two dimensions) as well as regimes that are classically tractable (one dimension), and therefore allows to validate the experiment in the tractable regime by comparing to theoretical predictions.

\subsection{Experimental findings }

To provide first steps towards resolving the question whether MBL persists in two dimensions, a series of experiments have been conducted in optical lattices. 
In all experiments, a quantum quench was performed to a disordered Hubbard model, starting from a perfectly imbalanced initial state. 
Subsequently, the evolution of the imbalance is tracked over time and its stationary value after long times compared for different interactions $U/J$ and disorder strenghts $\Delta/J$.
One can then read off a transition from a fully thermalising regime to a localising regime from the dependence of the stationary value of the imbalance on the disorder strength. 
Since it witnesses only dynamical features of MBL, this type of measurement data provides only rather limited information about the onset of a ``truly'' many-body localised phase of matter. 
It does, however, provide evidence for the existence or absence of such a phase and allows one to answer questions about the behaviour in an intermediate-time regime.

\begin{figure}[t]
  \centering
  \includegraphics[width = \textwidth]{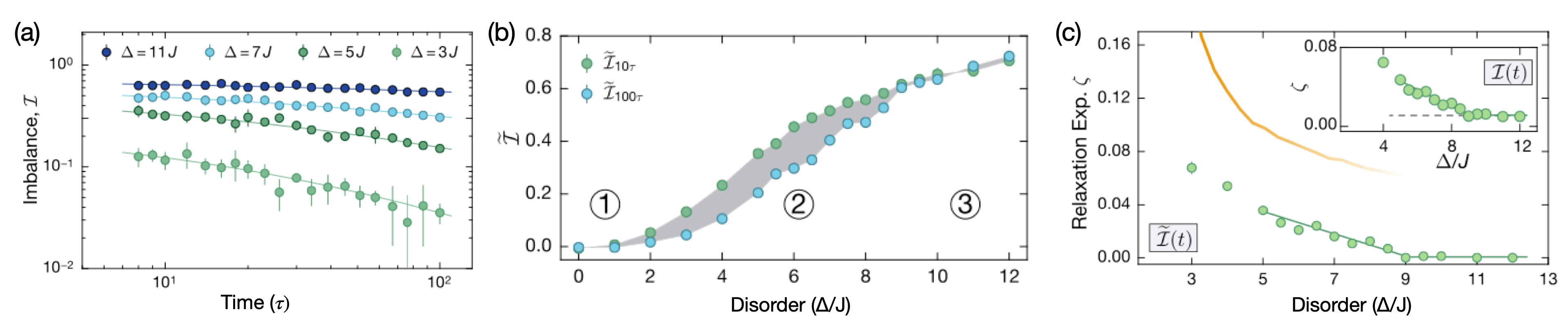}
  \caption{
  \label{fig:mbl imbalances}
  \emph{(a)} Time evolution of the imbalance $\mathcal I$ for different interaction strengths in terms of the tunneling time $\tau = \hbar t/J $ after a quench to a disordered Hamiltonian in two dimensions. 
  The experimental results (circles) are fitted to a phenomenological model of the system \cite[Fig.~2b]{bordia_probing_2017}.
  \emph{(b)} Value of the imbalance at short times $10\tau$ and long times $100\tau$ \cite[Fig.~3c]{bordia_probing_2017}.
  \emph{(c)} In two dimensions, no numerical calculations are possible.  However three regions can be experimentally identified: a rapidly thermalising regime for $\Delta/J < 2 $, a localised regime for $\Delta/J > 9$, and an intermediate regime of slow relaxation, where $2 \leq \Delta/J \leq 9$ \cite[Fig.~3b]{bordia_probing_2017}.
  This is witnessed by the relaxation exponent $\zeta$ of the imbalance, which is found to behave as $\mathcal I (t) \sim t ^{- \zeta \log(t)}$. 
  }
\end{figure}

The first series of results~\cite{schreiber_observation_2015,luschen_observation_2017,bordia_probing_2017} used quasi-periodic randomness of the type described above. 
The first experiment for one-dimensional systems~\cite{schreiber_observation_2015} can be viewed as a benchmark of the experimental source system, since numerical DMRG simulations are feasible in the localised phase. 
Ironically, this is because entanglement grows only very slowly in this phase enabling the use of DMRG for longer times than in thermalising systems. 
This allows for a direct comparison of the measured data with model predictions and hence for a \textit{certification} of the source system in the classically simulable regime, building trust in its correct overall functioning outside of that regime.

In both one and two dimensions, a transition from a fully thermalising regime to a localised regime has been observed, see Fig.~\ref{fig:mbl imbalances}. 
However, even in one dimension the system does not localise perfectly. The authors attribute this to the fact that the system is not perfectly isolated from its environment. 
In two dimensions, there are three distinct regions corresponding to different disorder strengths. 
At low disorder strengths there is a rapidly thermalising regime. At high disorder strengths there is a many-body localised regime. 
In between, there is a region in which the imbalance decays very slowly, namely as a power law with an exponent that decreases with increasing disorder. 
At a critical value of $\Delta/J$ a sharp transition occurs from a nonzero decay exponent to zero, see Fig.~\ref{fig:mbl imbalances}(c).
This transition can be viewed as the phase transition from the thermalising to the dynamically localising regime. 

The second result has been obtained in a different two-dimensional experimental setup, where an arbitrary disorder pattern is mapped onto the setup via a digital mirror device~\cite{choi_exploring_2016}. 
In this experiment, a sharp MBL transition is found at a critical disorder that depends on the interaction strength of the particles. 
Here, the MBL phase is also witnessed in terms of density-density correlations, that is, correlations between the particle number at lattice sites that are a certain distance apart from each other. Those correlations decay with distance and the decay length measures the way in which they decay. 
The MBL phase is now witnessed by a diverging decay length of the density-density correlations in the lattice.

Complementing those results about dynamical features of MBL in optical lattices, a more elaborate analysis of a potential MBL phase has been done using different architectures such as ion traps \cite{smith_many-body_2016} and superconducting qubits \cite{roushan_spectroscopic_2017}. 
While optical lattices feature several hundreds to thousands of coherently interacting particles, the means of probing those particles remain restricted. 
In particular, spectral features that serve as more definitive probes of static features of MBL, that is, a many-body localised \emph{phase of matter} are not accessible in such systems. 
In contrast, ion trap or superconducting qubit architectures are much smaller in size but allow for universal operations and measurements, giving rise to a much more flexible layout. 
Indeed, in a much smaller system of $9$ superconducting qubits on a line, an elaborate analysis of spectral features of MBL witnesses a many-body-localisation transition in the distribution of energy level splittings \cite{roushan_spectroscopic_2017}.  

Altogether, the experimental findings provide strong evidence that a localised phase that persists for long times is implied by the Bose-Hubbard model, even in the presence of strong interactions that act over a long-range~\cite{smith_many-body_2016} and in higher dimensions~\cite{bordia_probing_2017,choi_exploring_2016}. 
In all settings, an elaborate analysis of the behaviour of disordered systems can be performed by tuning the model parameters over a wide range that would be unattainable using only classical simulation techniques. 
In the absence of rigorous certification methods for the individual quantum simulators such findings can act as cross-platform checks of the observed phenomenon: 
an MBL phase is observed independently of the microphysics of the respective simulation platform. 

These impressive results notwithstanding, it is worth noting that the answer to the question of whether dynamical MBL persists \emph{at all} -- even in one dimension -- or whether the decay is merely very slow, remains out of reach using quantum simulators, due to their finite coherence times.

\section{The Higgs mode in two dimensions}
\label{sec:higgs_case_study}

As the second example of an analogue quantum computation performed in an optical-lattice setup, we now elaborate the experiment performed by \citet{Higgs:2012} to explore the existence of a Higgs mode in two dimensions. 
We first explain the theoretical basics of the Higgs mechanism in terms of symmetry breaking, and then move on to explain how the Higgs mode can be probed in a cold-atom setup. 

\subsection{Broken symmetries in $O(2)$-symmetric field theory.} 

Important examples of broken symmetries are the emergence of magnetic or
superconducting states of matter below a certain critical temperature $T_c$.
For example, iron is paramagnetic above $T_c$ and thus only responds to
external magnetic fields. However, below $T_c$ it is ferromagnetic and thus remains magnetised even in the absence of external magnetic fields. 
In this case the rotational symmetry of the elementary magnets present in the metal is broken spontaneously as the
temperature sinks from above $T_c$ to below $T_c$ and the elementary magnets align spontaneously in a fixed direction. 
Above $T_c$ there is no
preferred direction, while below $T_c$ there is a preferred axis, which is spontaneously chosen. 
Thus the rotational symmetry described by the symmetry group $O(3)$ is broken. 
A witness for a broken symmetry is called an order parameter. This is a quantity that turns nonzero only as the respective symmetry is broken. 
For example, in the case of the broken rotational $O(3)$ symmetry of the elementary
magnets in iron, an order parameter is given by the total magnetisation of the system,
a quantity that is zero in the paramagnetic phase and nonzero in the ferromagnetic phase. 
Intuitively, in the paramagnetic phase the elementary magnets point in random directions so that their total magnetisation averages to zero, while they point towards the same direction in the ferromagnetic phase, resulting in a nonzero total magnetisation. 

Suppose we would like to rotate the direction of
magnetisation by an angle. 
The energy cost of a global rotation of the magnetisation is zero as the
relative orientation of the spins does not change at all. 
However, such a global rotation is unlikely to take place as all spins would need to `coordinate'. 
In contrast, infinitesimal rotations of individual spins have an infinitesimally small energy cost. 
As an infinitesimal rotation of a single spin is excited there are two infinitesimal-energy excitations (called Goldstone modes) trying to restore the original direction of magnetisation as dictated by the remaining (aligned) spins.
This is because for an $N$-dimensional order parameter, there are $N-1$ directions in which rotations are possible. 
On the other hand, we can also try and excite an amplitude change of the magnetisation. 
An amplitude excitation, in turn, requires a finite excitation energy due to the fact that the energy levels in a potential well are quantised. 
This results in a single so-called Higgs mode. 
Famously, in particle physics the Higgs mechanism explains the emergence of
particles' masses in terms of the broken symmetry. 
However, it also features in the explanation of superconductivity and superfluidity as a fundamental collective excitation. 

\begin{figure}[t]
\centering
    \includegraphics[width = \textwidth]{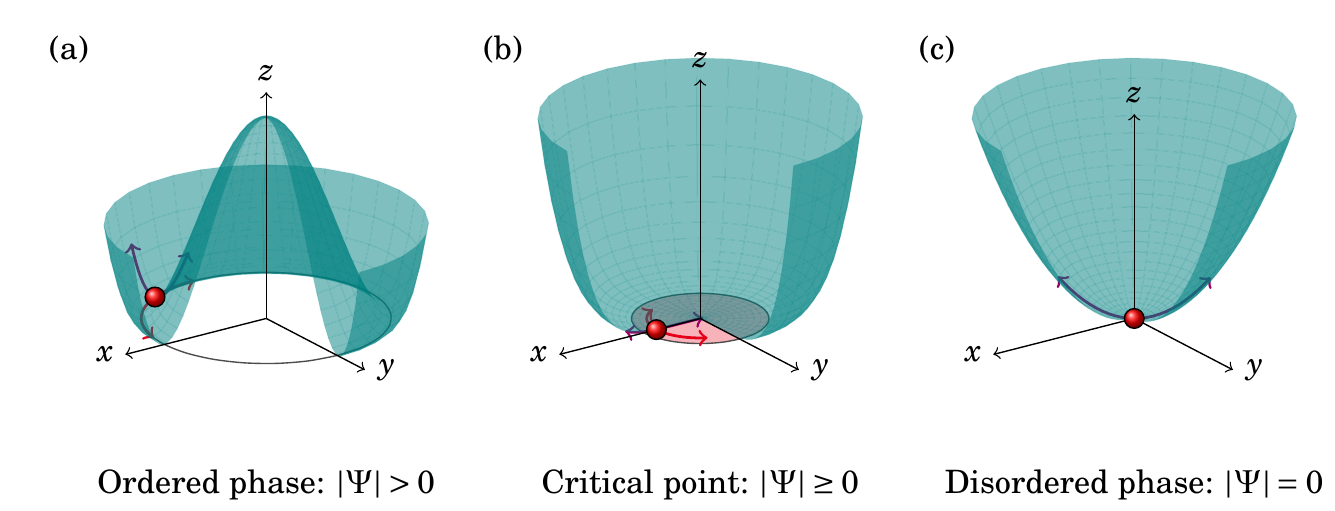}
    \caption{
        \label{fig:higgs potentials}
    (a) Illustration of the Higgs mode (purple arrow) and Nambu-Goldstone modes
    (red arrow) of a system with complex field $\Psi = |\Psi|\mathrm{e}^{\mathrm i \phi}$ in a Mexican Hat potential. 
    The rotational $U(1)$ symmetry is spontaneously broken as the system takes on a particular ground state (red ball) with a specific value of the phase $\phi$. 
    In contrast, the value of the amplitude $|\Psi|$ is fixed by the minimum of the Mexican Hat. 
    (b,c) The Goldstone modes are associated with a spontaneously broken symmetry: The ground state is symmetric with respect to a rotational symmetry around the $z$-axis (c). As the critical point (b) is crossed, where an entire area of states (red) has the same energy, the ground state moves away from the center of the potential to its brim and the rotational symmetry is broken spontaneously, as there are many states with the same energy. 
        }
\end{figure}

Let us make this intuitive picture more precise using a simple model for Higgs and Goldstone modes: an $O(2)$-symmetric complex field $\Psi = |\Psi| \mathrm{e}^{\mathrm{i} \phi}$ that is governed by a potential with a characteristic Mexican-hat shape (Fig.~\ref{fig:higgs potentials}) in the ordered phase \cite{SchollwoeckASP}. The order parameter for this phase $A = |\Psi| $ takes a non-zero value in the minimum of this
potential and the phase acquires a definite value via spontaneous symmetry breaking of the rotational symmetry on the circle. 
We can now expand the field around this symmetry broken ground state and find
two elementary excitations: i) a transversal (Goldstone) mode along the
minimum of the potential; and ii) a longitudinal amplitude (Higgs) mode along the radial direction. 
The Goldstone mode is gapless (and therefore massless) as it requires an infinitesimal amount of energy to excite it. 
In contrast, the Higgs mode requires a finite excitation energy and is therefore gapped (has a mass). 
This is because the excitation energies of a quantum system in a potential well are quantised, giving rise to a finite gap between the ground state at the bottom of the well and the first excited state. 
These correspond precisely to the excitations possible for the case of the elementary magnets discussed above. 
As the phase transition point is approached, the central peak of the hat drops down
to zero which leads to a widening of the potential well and consequently to a  characteristic `softening' of the finite excitation energy (the gap) of the Higgs mode. 
On the other side of the phase transition, where the order parameter $|\Psi|$ is zero, the excitation gap will rise again as the potential deepens around $|\Psi| = 0$, see Figs.~\ref{fig:higgs potentials}(a-c).

The `smoking gun' feature of the Higgs mode is a resonance in the spectral response function as the system is excited~\cite{huber_dynamical_2007,huber_amplitude_2008,podolsky_spectral_2012}. 
This quantity is proportional to the energy absorbed by the system in response 
to an excitation 
with energy $\hbar \nu $
and
can be measured via the temperature change of the system \cite{liu_massive_2015}. 
Consequently, we expect a resonant feature at the excitation energy of the Higgs mode. 
The resonance of the spectral response lies above a background response, whose position and width also depend linearly on the size of the Higgs gap. 

The Higgs mode may be excited by a quantum quench type experiment wherein the shape of the (Mexican Hat) potential is altered quickly so that the state of the system does not have time to adapt and finds itself in an excited state of the new potential. 
Another way to excite the Higgs mode is to directly excite the state, for example via external driving at frequency $\nu$.

It has been a subject of controversy
\cite{sachdev_universal_1999,altman_oscillating_2002,podolsky_visibility_2011,podolsky_spectral_2012,pollet_higgs_2012}
whether in two-dimensional systems such a Higgs mode is present, 
or whether it decays into lower-lying Goldstone modes, resulting in a low-frequency divergence \cite{Higgs:2012}. 
The controversy arises due to the fact that even the simplest relativistic field theory as described
above remains elusive to analytical or numerical treatment. 
Both analytical solution and Quantum Monte Carlo methods
fail. 
Hence, approximations in the form of mean-field theory or perturbative treatment are necessary in the vicinity of the quantum phase transition and only certain parameter regimes are accessible. 

\subsection{The Higgs mode in the superfluid--Mott insulator (SF-MI) transition} 

In order to clarify the question whether a Higgs amplitude mode persists in two dimensions, \citet{Higgs:2012} performed an experiment using ultracold atoms in optical lattices. 
At unit filling and zero temperature this system undergoes a quantum phase transition between a
superfluid (ordered) and a Mott insulating (disordered) phase at a critical value $j_c$ of $j = J/U$. 
While in the Mott phase the atoms are strongly localised to the minima of the
potential wells, in the superfluid phase they are delocalised across the entire
lattice. 
This phase transition is effectively described by a relativistic
$O(2)$-symmetric quantum field theory, as explained above, with order parameter $\Psi(x_i) = \sqrt{\overline{n}} \langle b_i \rangle$ given by the macroscopic condensate wave function with mean atom density $\overline{n}$ at the position $x_i$ of the $i$th lattice site \cite{SchollwoeckASP,altman_oscillating_2002}. 
The expectation value $\langle b_i \rangle$ of the annihilation operator $b_i$ is zero in the Mott phase since
the particles are strongly localised and therefore removing a single
particle from a lattice site creates an orthogonal state. 
Conversely, in the
superfluid phase the system is in a coherent superposition of $n$-particle
states, giving rise to a macroscopic quantum state. 
Therefore removing a single particle at one lattice site
does not significantly affect the overall state. 

This quantum phase transition is experimentally accessible in an optical lattice setup \cite{greiner_quantum_2002} also in two-dimensions via the methods discussed previously.
Moreover, a frequency-dependent external perturbation that is well described using linear-response theory can be achieved via a small modulation of the lattice depth of roughly $ 3\%$
close to the quantum phase transition \cite{Higgs:2012}. 
This modulation of the lattice depth directly  corresponds to external driving of the quantum state of the system. 
Thus, via a measurement of the temperature of the system in response to the external driving, the Higgs mode can be probed in the superfluid-Mott insulator transition. 
This temperature measurement can be achieved at the necessary precision via single-atom imaging~\cite{Sherson-Nature-2010}. 

\begin{figure}
\includegraphics[width=.45\textwidth]{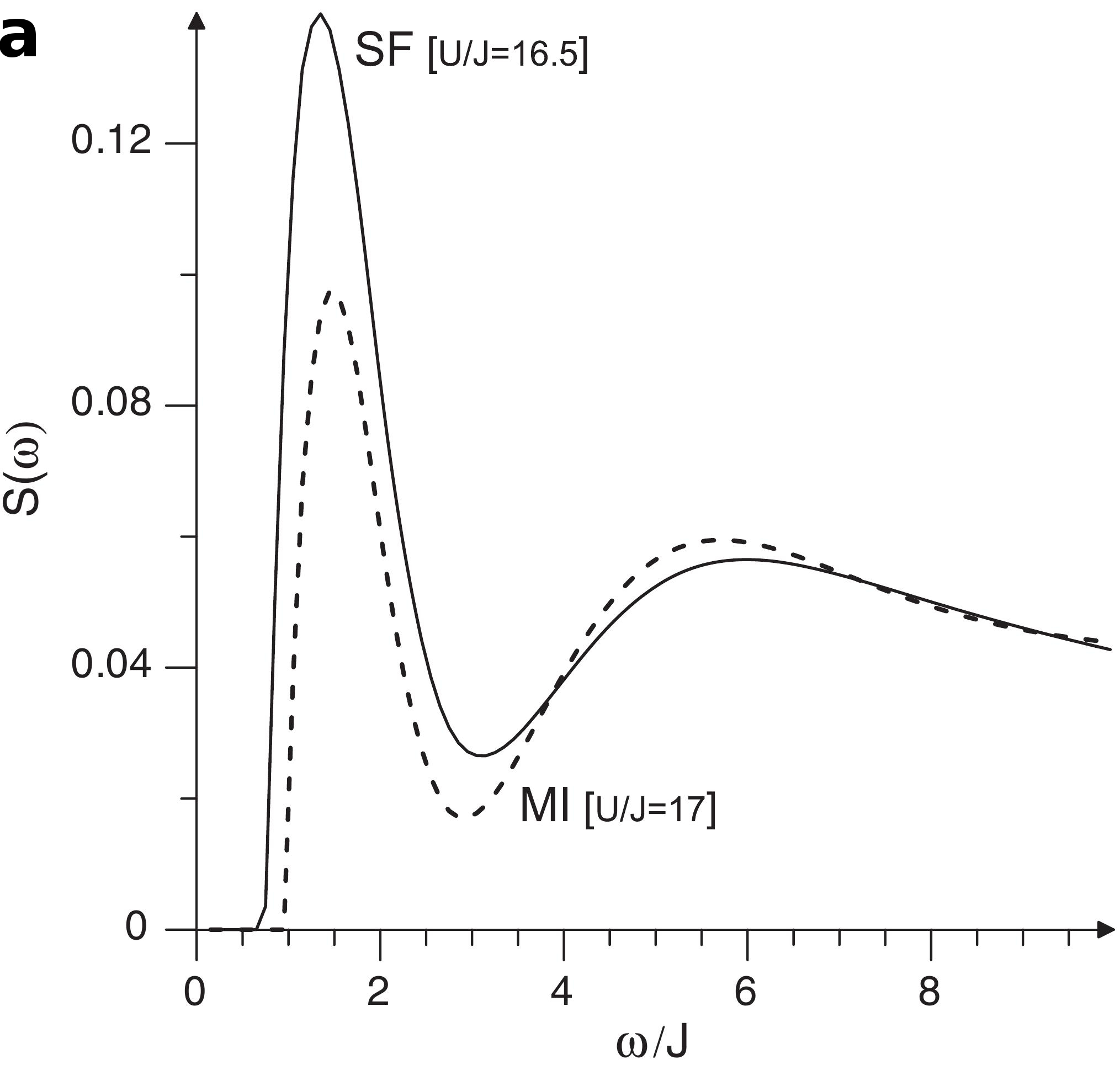}
\hfill
\includegraphics[width=.45\textwidth]{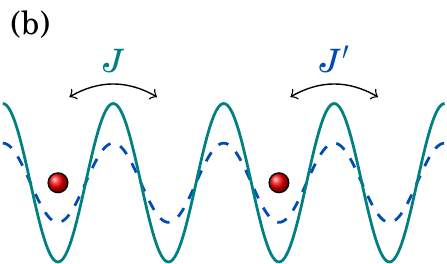}
\caption{
\label{fig:higgs theory} 
(a) Expected resonant feature in the spectral response.  Reprinted figure with permission from \cite{pollet_higgs_2012}. Copyright (2012) by the American Physical Society.
(b) In the experiment of \citet{Higgs:2012} the lattice depth is modulated
by 3\% with modulation frequency $\nu_{\text{mod}}$ whereby the hopping strength $J$ is modulated similarly. }
\end{figure}

For a massive Higgs mode in the superfluid regime of the simulation, one expects the spectral response $S(\omega)$  as a function of the driving frequency $\omega$ to show a response at values of the driving frequency that corresponds to the excitation of the Higgs mode, see Fig.~\ref{fig:higgs theory}. 
Moreover, at the onset of the response, a sharp resonant peak emerges as the driving frequency corresponds exactly to the energy gap of the Higgs mode in a given potential \cite{liu_massive_2015}. 
This sharp spectral response lies on top of a background response whose position and width also depend linearly on the value of the Higgs gap. 

\subsection{Experimental findings}

\begin{figure}
  
\includegraphics[width=\textwidth]{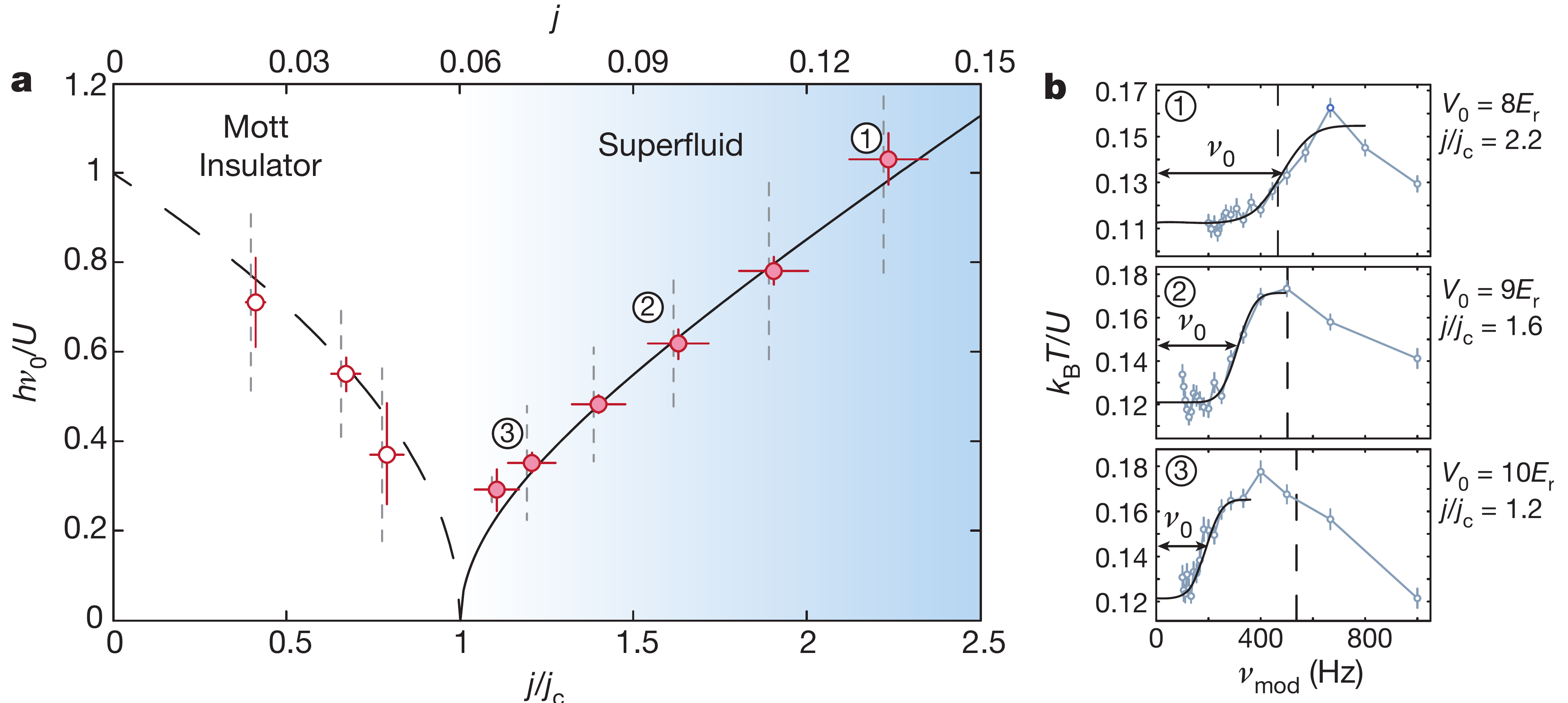}
  \caption{
  \label{fig:higgs data}
  Experimental data for different, fixed values of $j = J/U$ \cite[Fig.~2]{Higgs:2012}. 
  (a) The onset of the spectral response at values $\nu_{\text{mod}} =  \nu_0$ of the modulation frequency $\nu_{\text{mod}}$ decreases (`softens') as the experiment is repeated for varying values of $j$, approaching the critical point $j/j_c = 1$ from above and below. This characteristic softening of the response is in quantitative agreement with theoretical predictions \cite{liu_massive_2015}. 
  (b) Single shots of the spectral response functions $S(\omega)$ for fixed values of~$j/j_c$. 
  It is apparent that the resonant-like feature, which is present in Fig.~\ref{fig:higgs theory}(a) and considered the `smoking gun' feature of a Higgs mode, cannot be observed in the experimental data.  }

\end{figure}

In the experiment reported by \citet{Higgs:2012} the authors measured the temperature response to external driving 
as a function of the lattice modulation frequency~$\nu$. 
The experimental data exhibit a broad spectral response at frequencies~$\nu_0(j)$ that are in quantitative agreement with the analytical predictions for the gap of the Higgs excitations (see. Fig.~\ref{fig:higgs data}(a)).
Moreover, as a function of $j$ they observe the characteristic softening of this frequency
when approaching the critical point $j_c$ both from the superfluid and the
Mott insulating phase. 

However, the experimental data did not feature the expected resonance at the value of the Higgs gap which is the `smoking-gun' signature of the Higgs mode, see Fig.~\ref{fig:higgs data}(b). 
It has been argued that this might be due to effects of the harmonic confining potential \cite{pollet_higgs_2012,liu_massive_2015} and might be overcome by limiting the driving only to a small region in the center of the trap where the trapping potential can be approximated by a flat one \cite{liu_massive_2015}. 

While not fully conclusive yet, the data obtained in the quantum simulation of a relativistic $O(2)$-symmetric field theory is compatible with the existence of a collective Higgs amplitude in such a theory. 
Indeed, given the data, the authors were able to pick out certain theoretical predictions that correctly reproduce this data. 
This is a nontrivial task: many of the available predictions were conflicting since they were based on approximations, perturbation theory or numerical data in an attempt to tackle this computationally intractable problem. 

To summarise, the experiment suggests the existence of a gapped response and thus a Higgs mode. 
This is in particular true given the (later) evidence that the broadening of the resonant peak can be explained by the inhomogenieties in the experimental setup. 
This example of analogue quantum computation unquestionably allows us to \textit{understand} features of the $O(2)$-symmetric field theory in response to external driving by manipulating the ultracold atom source system.  

Let us conclude this case study by noting that similar experiments have been performed in other simulation platforms. 
In $s$-wave superconducting films, which bear qualitatively similar features to the two-dimensional Bose-Einstein condensate in an optical lattice, a resonant peak corresponding to a Higgs mode was in fact detected in the spectral response function \cite{matsunaga_light-induced_2014}. 
Here, the response of the system was measured in a pump-probe experiment, where the system is optically excited at one (THz) frequency and its response measured at a different frequency. 
The results revealed qualitative agreement with predictions for the Higgs gap. However, here, too, the width of the resonant peaks was significantly broadened. 
In this scenario, this broadening of the resonant peaks may be attributed to scattering processes or the finite spectral width of the pump pulse \cite{matsunaga_light-induced_2014}.

In another experiment using superconducting films, dynamical features of the Higgs mode were investigated~\cite{matsunaga_higgs_2013}, and in particular, the oscillation of the order parameter after the system is excited. 
The results are in quantitative agreement with the predictions for the gap of a Higgs mode and therefore serve as another piece of indirect evidence for the existence of such a gapped mode. 

\section{Philosophical Schema}

In Chapter \ref{Distinctions}, we introduced the sub-type of analogue quantum simulation that we defined as analogue quantum computation. The goal of analogue quantum computation is to \textit{understand} formal properties of a mathematical target model. This is particularly relevant and interesting in situations in which the target model is not solvable by either numerical simulation algorithms implemented on a classical computer or analytical calculations. Both case studies considered in this chapter illustrate our notion of analogue quantum computation. This is because in both cases source phenomena (time evolution of the imbalance and the Higgs signature in the ultracold atom system) are being appealed to for the specific purpose of gaining understanding of formal properties of a target simulation model (many-body localisation and the Higgs mode). Moreover, it is not a case of the sub-type of analogue quantum simulation that we defined as analogue quantum emulation above, since the intention of the experimenters is not primarily to gain understanding of physical target phenomena. 

\begin{figure}[t]
\centering
    \includegraphics[height=0.4\textheight]{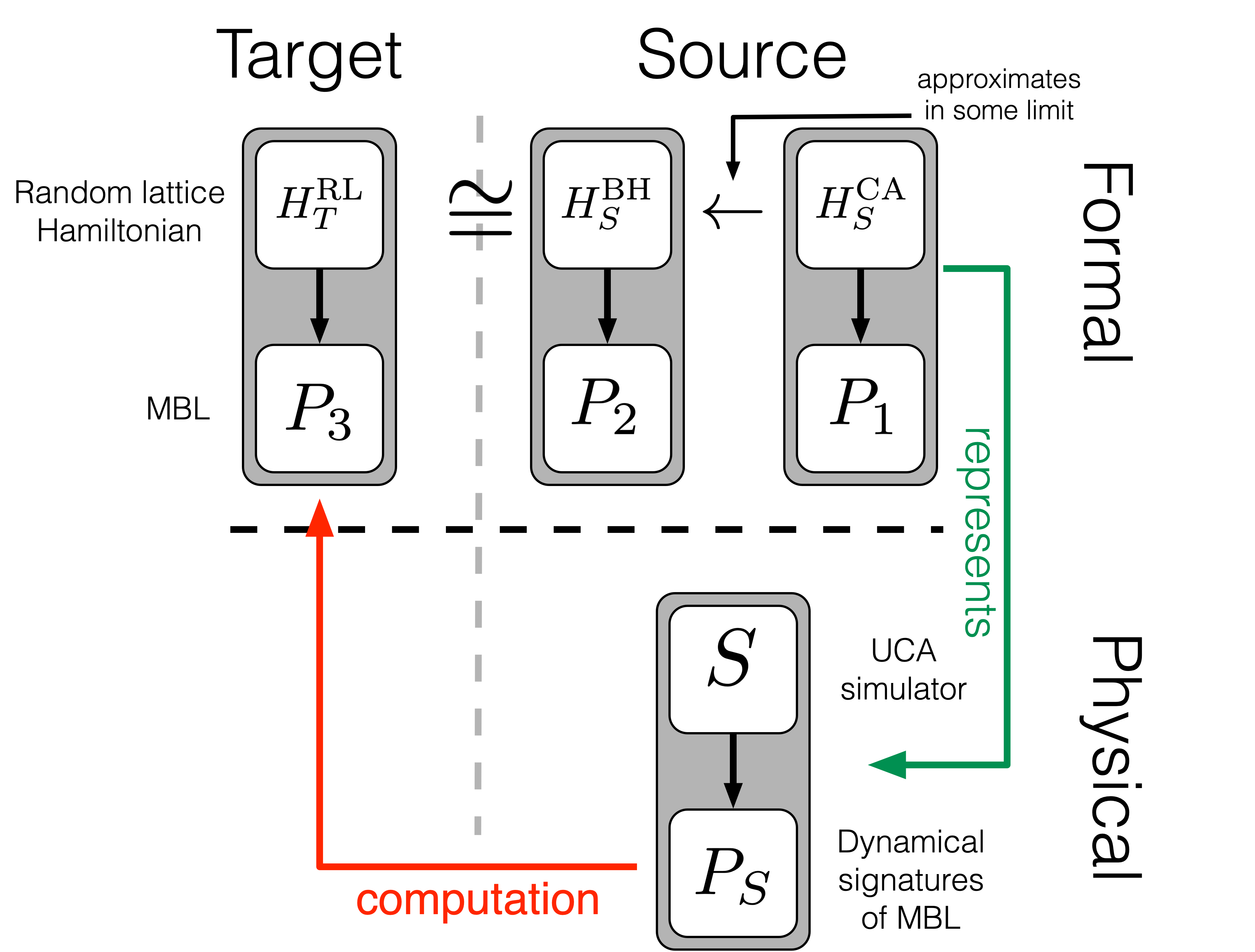}
\caption{
    \label{fig: MBL simulation schema} 
    Schema for analogue computation MBL case study. 
    The source system model of a cold atom system, $H_S^{\text{CA}}$, is approximated in a certain parameter regime by the source simulation model, the Bose-Hubbard Hamiltonian, $H_S^{\text{BH}}$ given in Eq.~\eqref{bose hubbard}. 
    For disordered local fields $\mu_j$ the Bose-Hubbard Hamiltonian $H_S^{\text{BH}}$ is an instance of a generic two-dimensional random lattice Hamiltonian $H_T^{\text{RL}}$ which is expected to exhibit MBL. 
    $H_S^{\text{CA}}$ stands in a representation relation with the ultracold atom source system, $S$. 
    In an analogue quantum computation the target system $S$ is manipulated to learn about the target (simulation) model $H_T^{\text{RL}}$. Specifically, observing dynamical signatures of MBL in $S$ would imply (within the approximation relations) the existence of dynamical MBL in random lattice Hamiltonians.
}
\end{figure}

\begin{figure}[t]
\centering
    \includegraphics[height=0.4\textheight]{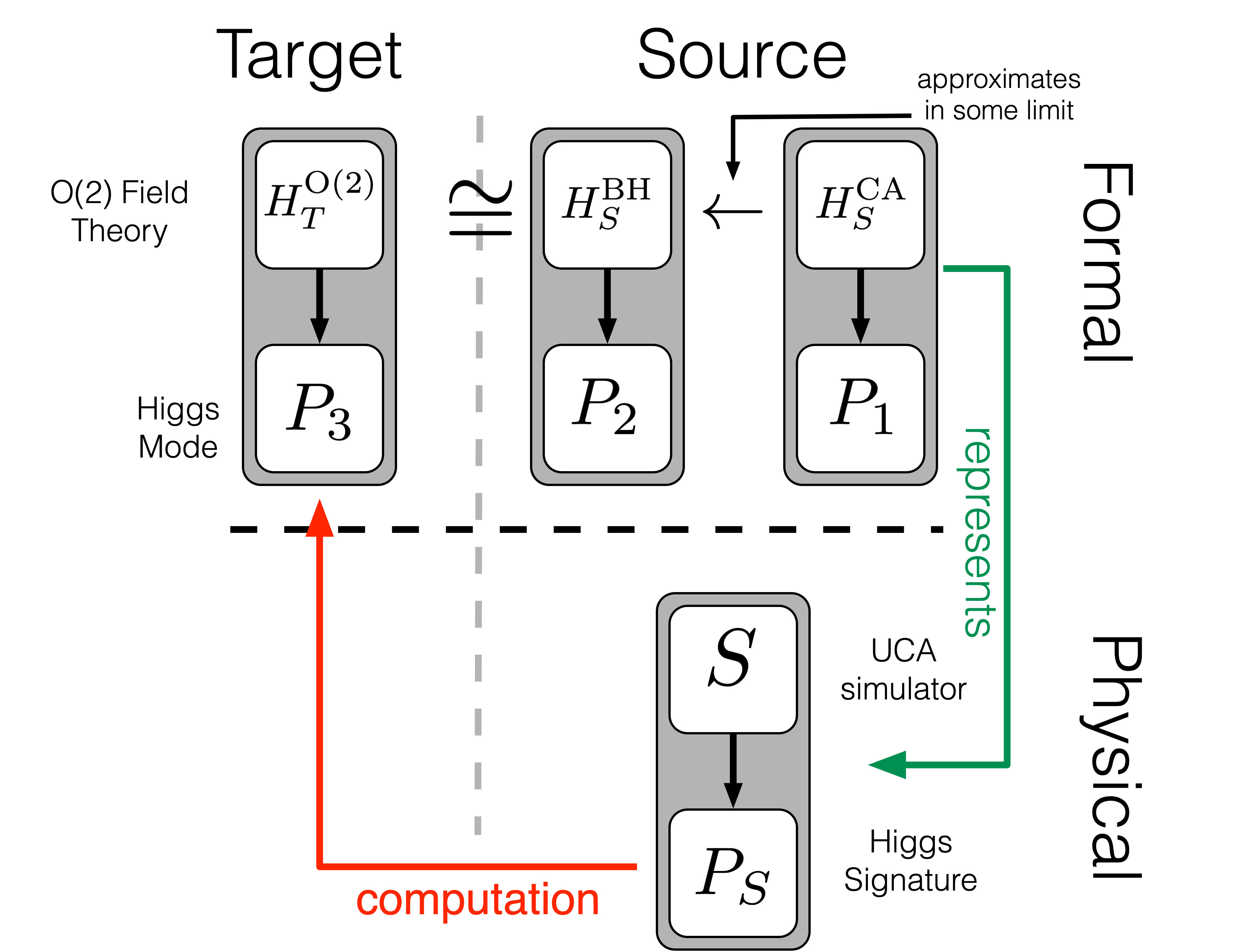}
\caption{
    \label{fig: higgs simulation schema} 
Schema for analogue computation Higgs case study. 
The source system model of a cold atom system, $H_S^{\text{CA}}$, is approximated in a certain parameter regime by the source simulation model, the Bose-Hubbard Hamiltonian, $H_S^{\text{BH}}$. 
In the vicinity of the critical point the long-wavelength, low-energy dynamics given by $H_S^{\text{BH}}$, then corresponds to the target simulation model, that is, the model of an $O(2)$-symmetric field theory with Hamiltonian $H_T^{O(2)}$. $H_S^{\text{CA}}$ stands in a representation relation with the ultracold atom source system, $S$. Analogue quantum computation is then a relation between $S$ and $H_T^{O(2)}$.
    }
\end{figure}

Recall also from Chapter  \ref{Distinctions} that the Figure \ref{fig: ac_model} provided a  schematic representation of the structure of an analogue quantum computation. Three models are involved: the source system model,  (the complex experimental model of the source system), the source simulation model (the simple idealised model of the source system and the simulation target model (the simple idealised model of the target system). The source system and source simulation models were taken to be related by limiting relations and the two simulation models are typically related by a partial isomorphism. Figures~\ref{fig: MBL simulation schema} and~\ref{fig: higgs simulation schema} each provide an application of our schematic representation of the general structure of inferences in analogue quantum computation to our cases studies. In general terms analogue quantum computation is a relation between a concrete source system and an abstract target (simulation) model. Crucially, there is no concrete target system and so the bottom left hand side is empty.

In Chapter \ref{ch:assessing case studies} we will return to the two case studies presented in this chapter in order to frame their epistemic purpose within scientific practice. We will argue that in such cases analogue quantum computation is being employed by scientists with the aim of obtaining \textit{how-actually understanding} of formal properties of the target model. In the case studies these formal properties are the 2D Higgs mode in $O(2)$-symmetric field theories and the existence of many-body localisation. The conditions under which this aim should be understood to be have been achieved together with the wider methodological significance of our analogue quantum computation will then be considered in detail in the remainder of the book. 
\newpage


\chapter{Photonic Emulation and Quantum Biology}
\label{ch:enaqt}



Environment-assisted quantum transport is a proposed mechanism in which energy transfer in certain photosynthetic processes, such as those observed in green sulphur bacteria, is enhanced by environmental noise and disorder. Environment-assisted quantum transport has been the subject of analogue quantum simulation via the platform of quantum photonic emulators which encode information onto individual photons and process that information using complex photonic circuits. Detailed analysis of this case study shows it to exemplify the notion of analogue quantum emulation. That is, the type of analogue quantum simulation in which the source system that is being manipulated is being appealed to for the specific purpose of gaining understanding of features of a physical target system.

\section{What is Quantum Biology?}

Biology, the science of living organisms, on the face of it appears to be antithetic with quantum mechanics.
Biological systems (e.g.\ cells or bacteria) are warm, wet and disordered, and rely on complex interactions of many thousands of atoms.
In contrast, quantum mechanical effects are typically seen at the single or few particle level and observing these effects in the lab requires significant efforts to isolate these systems from their environment (see Chapter \ref{ch:cold atoms}).
Notwithstanding, biology relies on chemical reactions (e.g. the production of enzymes and proteins), which are themselves inherently quantum mechanical processes, so it is not unreasonable to ponder the role, if any, quantum mechanics may play in biology.

Indeed, this question has a long history dating back to the early days of quantum mechanics where
Schr\"{o}dinger, in his book `What is Life', sought to examine the relationship between quantum mechanics and the recently developed theory of genetics.  
He remarked: ``the mechanism of heredity is closely related to, nay, founded on, the very basis of quantum theory'' \citep[pg. 47]{schrodinger1992life}.
His goal was to explain how the molecules responsible for hereditary traits could survive thermodynamic decay as they are passed between generations in the warm, wet biological environment.
As our understanding of genetics progressed in the 1940s and 50s, spurred on by the discovery of DNA, it emerged that some of the specific mechanisms Schr\"{o}dinger posited weren't quite correct.
However his work set the stage for a rich and fruitful research direction, at the intersection of biology, quantum mechanics and computer science: the field of quantum biology.\footnote{See \cite{mcfadden:2018} for a fascinating discussion of the historical roots of quantum biology.}

In recent years, a vast body of experimental and theoretical work has emerged examining the role that quantum mechanical effects may play in \textit{biological function} \citep{marais2018future}.  That is, are there processes in biological systems that not only fundamentally rely on quantum mechanics (e.g. chemical reactions), but actually leverage uniquely quantum mechanical effects to perform a function in a more efficient manner?
A number of such quantum biological effects have been proposed, including olfaction \cite{turin1996spectroscopic}, avian navigation \cite{schulten1978biomagnetic} and even cognition \cite{fisher2015quantum}.
Many of these effects are hotly debated, largely because in biological systems we rarely have a complete description of all the physical and chemical processes at play.
However, here a quantum simulator could offer a unique opportunity:
given our best model of a biological system of interest, the quantum simulator could efficiently estimate whether a particular phenomenon pertains or not.  This in turn serves to verify a known model.  
In addition, the simulator can also test new models to see under what conditions the phenomenon of interest will occur.  This prediction can guide scientists searching for this phenomenon in nature. 
Following the nomenclature of Chapter \ref{Distinctions}, as we are manipulating a source system to gain understanding of a concrete physical system,  we refer to this process as \textit{quantum emulation}.

\begin{figure}[t!]
\begin{center}
\includegraphics[width=1.0\textwidth]{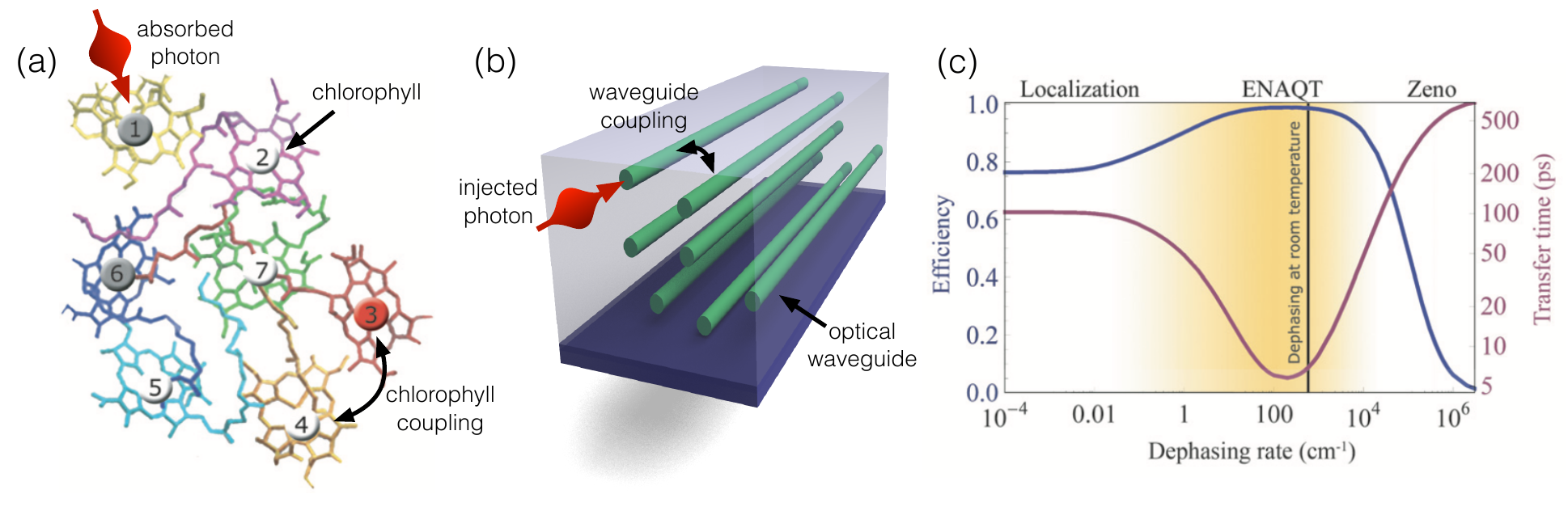}
\caption{Emulation of environment-enhanced quantum (ENAQT) transport in photosynthetic complexes.  (a) One of three FMO complex sub-units consisting of eight chlorophyl molecules (only seven shown) which enable exciton hopping between neighbouring sites. (b) A photonic waveguide emulator. (c) An optimal dephasing rate gives rise to an ENAQT.  Images (a,c) from \cite{Rebentrost:2009hu}.}
\label{fig:FMO}
\end{center}
\end{figure}

In the following chapter, we examine the quantum emulation of a biological phenomenon known as environment-assisted quantum transport (ENAQT). ENAQT describes how the coherent transport of energy can be enhanced within certain regimes of environmental noise, and is proposed as an explanation for the exceptional efficiency of certain photosynthetic complexes [Fig. \ref{fig:FMO}(a)].
Notably, this model is particularly amenable to simulation by a class of simulators known as `quantum photonic emulators'.  These emulators encode information onto individual photons and process that information using complex photonic circuits [Fig. \ref{fig:FMO}(b)].
In the framework of our schema as shown in Fig.~\ref{fig:enaqt_schema}, the quantum photonic emulator is the source system and the target system is given by the biological system exhibiting ENAQT, namely the photosynthetic complex.  In the following pages we explore these concepts in more detail.

\section{Science Summary} 
%
%
In the past decade a vast body of experimental and theoretical work has emerged, suggesting that quantum mechanical coherences (viz.\ the coherent addition of probability amplitudes) play a critical role in many biological and chemical processes \cite{Scholes:2017gw}. 
One particular phenomenon that has received significant interest is ENAQT, which shows that energy transfer, under certain conditions, is not only robust to, but enhanced by environmental noise and disorder; running counter to the maxim that quantum coherences are easily destroyed. 
In the context of biology, ENAQT is proposed as an explanation for the remarkable efficiency of certain photosynthetic processes, such as those observed in green sulphur bacteria, an organism which can live in exceptionally low light environments. 

At the heart of the green sulphur bacteria is the Fenna-Matthews-Olson (FMO) protein complex.
%
This complex is a trimer, with each sub unit consisting of eight chlorophyll molecules separated by a few nanometers (see Fig.~\ref{fig:FMO}(a)).
If a photon is absorbed by the light harvesting antenna (site 1), an exciton (i.e. an electron-hole pair) is transported via the neighbouring chlorophyll molecules towards the reaction centre (site 3) where a charge separation occurs and a biochemical reaction takes place. 
Whilst exceptionally long-lived coherences have been experimentally observed in photosynthetic complexes \cite{Engel:2007hb, Yang:2007kk}, understanding the role that coherences play in functionality is an outstanding challenge, and the subject of on-going experimental \cite{Panitchayangkoon:2010fw, Duan:2017cy, thyrhaug2018identification} and theoretical \cite{Mohseni:2008gp, Rebentrost:2009hu, Wilkins:2015hv} research.

Photonic quantum technologies are systems which precisely generate, manipulate and detect individual photons \cite{Obrien:2009eu}.
Photons are appealing as a carrier of quantum information due to their inherent noise tolerance, light-speed propagation and ability to be manipulated by a mature integrated photonics platform \cite{Silverstone:2016gha}.
However, no quantum technology platform is without its drawbacks, and the inherent noise tolerance of photonics complicates the deterministic generation of entanglement which requires measurement and fast active feedforward \cite{Knill:2001vi}, or atom mediated interactions \cite{duan2004scalable}.
Given this dichotomy, and owing to the ease of the high-fidelity manipulation of individual photon states, photonic quantum emulators are exceptionally well suited to exploring complex single particle dynamics, and have therefore recently emerged as a promising platform with which to explore ENAQT \cite{AspuruGuzik:2012ho}.
In the following sections we describe experiments which `program' in an approximation of the Hamiltonian for the FMO complex into a photonic quantum emulator, and measure photonic transport efficiencies under certain models of artificially applied noise. 
Studying ENAQT in a noisy, physical setting allows experimenters to identify real circumstances in which ENAQT is promoted.
Quantum emulation of ENAQT may thus guide theoretical research as well as experiments aiming to provide evidence for ENAQT in real biological systems.

\section{Environment Assisted Quantum Transport}  

Let us consider a specific example of biological ENAQT.  Following the analysis of \citet{Mohseni:2008gp} the FMO complex may be approximately described by a tight binding Hamiltonian with $N=7$ sites, the system Hamiltonian
\begin{equation}\label{eq:fmo}
	H^{\text{FMO}}_T = \sum_{m=1}^N \epsilon_m \ket{m} \bra{m} + \sum_{n<m}^N V_{m,n} (\ket{m}\bra{n}+ \ket{n}\bra{m}),
\end{equation}
where $\ket{m}$ represents an exciton at site $m$, $\epsilon_m$ the energy at site $m$ and $V_{m,n}$ the hopping potential between sites $m$ and $n$ (due to Coulomb interaction or electron exchange).
In this simplified model it is sufficient to consider a conserved single exciton tunnelling between sites as the recombination lifetime of the exciton (i.e. time until the exciton is lost) is significantly longer than the relaxation time of the chlorophyll (i.e. time taken to go from a high energy to low energy state).

In multichromophoric arrays, coupling to a fluctuating protein and solvent environment via the electron-phonon interaction induces time dependent variations in on-site energies and an irreversible dephasing of coherences; effectively describing the state by classical probabilities (real numbers) rather than quantum probability amplitudes (complex numbers).
Under certain assumptions\footnote{Specifically, that the phonon correlation times are short compared to the relaxation lifetimes, and that fluctuations at different sites are uncorrelated.} this site dependent dephasing can be described by a site independent pure dephasing rate, which can be used with the Lindblad master equation to describe the evolution of the system in the presence of environmental noise. 
Changes in dephasing rate are typically caused by variations in temperature.
In the study of environment-assisted quantum transport (ENAQT), scientists analyse the effect of the dephasing rate in transitions from a given input site [e.g.\ $m=1$ Fig~\ref{fig:FMO}(a)] to a given output site ($m=3$).
\citet{Rebentrost:2009hu} show that in the limit of zero-dephasing, i.e.\ purely coherent exciton hopping, variations in on-site energy restrict exciton transfer due to coherent interference between paths, which as discussed in Chapter \ref{ch:cold atoms} is a phenomenon known as Anderson localisation \cite{anderson_absence_1958}. 
This same effect causes sugar water to appear opaque, even though microscopically it is transparent to light.
Note that in contrast to the many particle localisation effects described in Chapter \ref{ch:cold atoms}, the model considered here is a single exciton, thus coherent particle-particle interactions can be neglected.
As the temperature increases, dephasing disrupts this coherent interference; effectively unsticking the exciton and enabling transfer. 
However, if the temperature rises further, dephasing destroys all coherences within the system which, analogous to a quantum Zeno effect, suppresses transport.
\citet{Rebentrost:2009hu} show that at room temperature an optimal dephasing rate exists whereby transport efficiency is maximised towards unity, the so called `quantum Goldilocks effect' \cite{Lloyd:2011wg} [see Fig.~\ref{fig:FMO}(c)].

\section{Photonic quantum emulators} 

Photonic quantum emulators typically comprise arrays of single mode waveguides, consisting of a high refractive index core surrounded by a lower refractive index cladding. This set-up effectively confines the light and allows the construction of on-chip optical wires.\footnote{Precisely the same operating principle which enables fibre optical communication and long distance communication between Baleen and Whales \cite{Payne:1971kr}.}
Connections between waveguides is achieved via evanescent coupling, whereby waveguides are brought close to one another such that the light leaking out of the wave guides, called `evanescent fields', overlap, enabling photon tunnelling between neighbouring waveguides \cite{Politi:2008tl}.
Given a particular configuration of $N$ coupled waveguides, the system is described by the simulation Hamiltonian
\begin{equation}\label{eq:wg}
	H^{\text{WG}}_S = \sum_{m=1}^N \beta_m \ket{m}\bra{m} + \sum_{n<m}^N C_{m,n}(\ket{m}\bra{n}+\ket{n}\bra{m}),
\end{equation}
where $\beta_m$ is the propagation constant for waveguide, $m$, determined by the refractive index of the mode; and $C_{m,n}$ is the coupling between waveguides, $m,n$, determined by the geometry and separation of the waveguides.
A single photon injected into mode $m$ evolves via $\ket{\psi(t)} = \exp(-i H^{\text{WG}}_S t)\ket{m}$, where time $t$ is related to the length $z$ of the coupling region via $z=ct/n$, where $c$ and $n$ is speed of light in a vacuum and the refractive index of the material respectively.
A further advantage is afforded by the fact the Schr\"{o}dinger equation is a wave equation, and therefore single particle dynamics can be simulated by injecting bright laser light.
Systems which exhibit coherent hopping between connected sites are known as quantum walks, and capture a very general class of phenomena \cite{Kempe:2003df}.

Critically, the isomorphism between equations \eqref{eq:fmo} and \eqref{eq:wg} means the task of building an ENAQT photonic emulator is two-fold: (T1) engineering the appropriate couplings between waveguide modes and (T2) engineering on-site dephasing. 
Recently, two complementary experiments were performed which addressed these tasks in turn.
\citet{Biggerstaff:2016bj} addressed (T1) by leveraging femtosecond-laser direct writing technology, which directly draws three-dimensional waveguides into glass, therefore enabling arbitrary couplings $C_{m,n}$ [see Fig.~\ref{fig:FMO}(b)].
The second approach by \citet{Harris:2017hi} lithographically patterns thermo-optic phase shifters on top of the waveguide circuit allowing them to control on-site dephasing (T2).
By engineering the magnitude of this dephasing, they observe both the peak and fall off in transport efficiency: the full quantum Goldilocks effect.

Let us make some remarks on these results. 
Each ENAQT quantum emulator has its respective advantages and drawbacks: the 3D waveguides of \citet{Biggerstaff:2016bj} enables arbitrary coupling between sites, but lacks the active control necessary for arbitrary dephasing; the lithographically fabricated waveguides of \citet{Harris:2017hi} enables high levels of control, but is inherently limited to one-dimensional connectivity.
Future ENAQT quantum emulators may use some combination of the two technologies to more closely mimic biological structures \cite{Smith:2009hm}, or, by engineering coupling to on-chip phonon modes \cite{Merklein:2017df}, to emulate a realistic protein environment. 
In terms of scaling, each system emulates single particle dynamics and can therefore be modelled by classical wave dynamics (such as water waves).  
Therefore, these analogue quantum emulators provide no more than polynomial computational speedup over emulation on a classical machine.
Notwithstanding, multi-photon quantum walks have become an interesting and active research line \cite{Peruzzo:2010tq,carolan_experimental_2014, Carolan:2015vga}, and are closely related to the boson sampling problem which proves that mimicking the dynamics of a many-photon state is intractable on a classical machine (assuming a few reasonable conjectures).
Mapping this exponential speed-up onto a useful physical system is an outstanding open question, but recent evidence has suggested that a modification to the many-photon input state --- alongside Hamiltonians of the form \eqref{eq:wg} --- enables the calculation of salient properties of molecular systems, such as vibrational spectra, which are critical for quantum chemistry \cite{Huh:2014vk, sparrow2018simulating}.

As we will see in Chapters \ref{ch:understanding} and \ref{ch:assessing case studies}, one of the key goals of quantum emulation is to provide understanding of a target physical phenomenon.
In the case of photonic ENAQT, there are two notions of understanding at play.
The first is understanding the relationship between the target system model and the target phenomena. In particular, is the mechanism by which ultra-efficient photosynthetic transfer occurs truly ENAQT?
While this might be the ultimate goal, it seems unlikely a definitive answer will be provided by a photonic simulator alone.  
Significant experimental effort is already concerned with directly observing signatures of coherent transport in biological systems, and unraveling the relevant time-scales and physical mechanisms (e.g. excitonic, vibrational or vibronic).
However, what the emulator can offer is precise control over critical parameters (such as environmental noise).  
So by varying these parameters and observing under what conditions ENAQT emerges, the emulator may identify particular parameter regimes to look for such features in complex `target system models' that describe real biological systems, thus \textit{guiding} biological experiments.

An example of one such approach can be seen in recent work exploring ENAQT on a trapped ion quantum simulator \cite{maier2019environment}.
Here, the authors examine how different environmental noise models effect quantum transport properties, and observe that non-Markovian noise (i.e. structured noise) can maintain coherences for longer time-scales than Markovian noise (i.e. unstructured, or white noise).
This may mean that to observe ENAQT in nature, one should look in environments with non-Markovian noise structures.
Results like this may therefore offer heuristics that narrow down the search for an explanation of a particular target phenomenon and therefore \emph{guide} theoretical work done towards this goal. 

The second notion of understanding concerns \emph{features of the target system model}.  
Much like in analogue quantum computation, the source simulation  model can glean useful information about the dynamics of a target simulation model. However, unlike in analogue quantum computation, the ultimate goal is to understand the extent to which these features are also relevant to the target system model, and thus the target system itself. 
In the context of ENAQT this is particularly important, as the target simulation model may be relevant for other physical systems, such as semiconductor devices \cite{sho2013quantum}, or be leveraged to design new technologies, such as highly effective photovoltaics \cite{falke2014coherent}. Thus there is considerable scope for multiple distinct target system models to be connected to the same target simulation model via differing de-idealisation procedures. %

\section{Philosophical Schema}

In Chapter \ref{Distinctions} we introduced the sub-type of analogue quantum simulation that we defined as analogue quantum emulation. In analogue quantum emulation a scientist is interested in gaining understanding of physical target phenomena, that is, quantitative or qualitative properties of a target system. That the case study considered in this chapter plausibly illustrate the form of inference can be seen as follows. The intentions of a scientist undertaking photonic emulation are two-fold: to understand (1) whether quantum coherences enhance functionality in real biological systems, and (2) whether understanding these effects can lead to technological breakthroughs in the development of new materials such as ultra-efficient photovoltaics \cite{Bredas:2017cq}.  The status of quantum functionality in biological systems is hotly debated\footnote{See \cite{Lambert:2012fj} for a balanced discussion.}, and in this context analogue quantum emulation has the potential to play a powerful inferential role. This case study illustrates our notion of analogue quantum emulation since features of the source system that are being manipulated (ENAQT into a photonic platform) are being appealed to for the specific purpose of gaining understanding pertaining to a physical target phenomenon (ENAQT in a biological FMO complex). This is not a case of analogue quantum computation since the intention of the experimenters is not to gain understanding directly pertaining to features of an abstract theoretical model.

\begin{figure}[t!]
\begin{center}
\includegraphics[height=0.4\textheight]{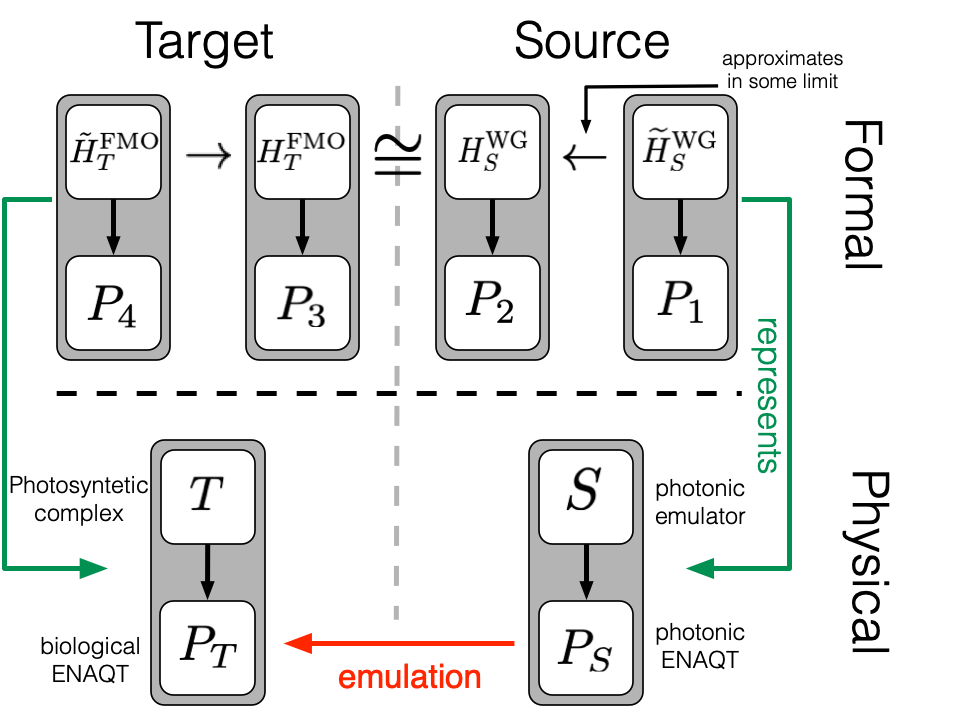}
\caption{Schema for quantum biology case study (see main text for figure explanation).}
\label{fig:enaqt_schema}
\end{center}
\end{figure}

Recall also from Chapter  \ref{Distinctions} that Figure \ref{fig: ae_model} provides a  schematic representation of the structure of an analogue quantum emulation. Four models are involved: the source system model (the complex experimental model of the source system), the source simulation model (the simple idealised model of the source system), the target system model (the complex experimental model of the target system), the target simulation model (the simple idealised model of the target system). Each of the system and simulation models are related by limiting relations and the two simulation models are typically related by a partial isomorphism. Figure~\ref{fig:enaqt_schema}  provides an application of our schematic representation of the general structure of inferences in analogue quantum emulation to our cases study. 

The right hand side of  Fig.~\ref{fig:enaqt_schema} relates to the photonic waveguide source system. First we have some abstract Hamiltonian $\widetilde{H}^{\text{WG}}_S$ that provides the system model of the waveguide system, $S$, that is the model which includes experimental imperfections such as fabrication error, waveguide loss, dispersion and detector noise. This Hamiltonian can be approximated, in the appropriate limit, by the idealised source simulation model, provided by the waveguide Hamiltonian $H^{\text{WG}}_S$. 

The left hand side of the diagram then relates to the photosynthetic complex target system. The concrete target system, $T$, is the FMO complex and the phenomena $P_T$ is ENAQT.  In general, this system will be described by some Hamiltonian $\widetilde{H}_T^{\text{FMO}}$, which includes a non-Markovian phonon bath, relaxation effects and spatial correlations, this is the target system model. 
It is conjectured that the simulation model provided by the tight binding Hamiltonian $H_T^{\text{FMO}}$ approximates the target system model $\widetilde{H}_T^{\text{FMO}}$ within some parameter regime such that the salient transport phenomena $P_T$ are sufficiently reproduced. Establishing the veracity of this conjecture is precisely the role of experimental and theoretical quantum chemistry.
The target simulation model, $H_T^{\text{FMO}}$, is partially isomorphic to the source simulation model, $H^{\text{WG}}_S$. The source system model provided by  $\widetilde{H}^{\text{WG}}_S$ then stands in a representation relation with the photonic emulator source system, $S$, and the target system model provided by  $\widetilde{H}_T^{\text{FMO}}$ stands in a representation relation with the photosynthetic complex target system, $T$. 

Analogue quantum emulation is a relationship between $P_S$ and $P_T$. 
The goal of analogue quantum emulation is to gain understanding of actual phenomena in a concrete physical system. 
This is the key distinguishing feature between computation and emulation that shall be the major focus of our analysis in the context of philosophical treatments of understanding in science. In Chapter \ref{ch:assessing case studies} we will return to the case study presented in this chapter in order to frame its epistemic purpose within scientific practice. We will argue that in such cases analogue quantum emulation is being employed by scientists with the aim of obtaining \textit{how-actually understanding} of target physical phenomena. The conditions under which this aim should be understood to be have been achieved together with the wider methodological significance of analogue quantum emulation will then be considered in detail in the remainder of the book.


\chapter[Emulation of Hawking Radiation]{Emulation of Hawking Radiation in Dispersive Optical Media}
\label{ch:hawking}



Hawking radiation is a thermal phenomenon which there are good theoretical reasons to associate with black holes. The temperature of the radiation is so low, however, as to make experimental detection of Hawking radiation from astrophysical black holes next to impossible. Various platforms for analogue quantum simulation of Hawking radiation have been proposed, and, in some cases, implemented. The focus of our case study is quantum simulation of Hawking radiation in dispersive optical media.  Detailed analysis of this case study shows it to exemplify the notion of analogue quantum emulation. That is, the type of analogue quantum simulation in which the source system that is being manipulated is being appealed to for the specific purpose of gaining understanding of features of a physical target system.

\section{Science Summary}
\label{sec:5.1}

The field of analogue gravity \cite{barcelo:2005} includes a growing number of experimental groups seeking to emulate gravitational phenomena, both classical and semi-classical, via table top experiments on (effectively) non-gravitational systems.\footnote{The semi-classical modelling framework applied to analogue and astrophysical black holes make use of a distinction between a quasi-stationary background and small perturbations on that background. An outstanding and extremely important theoretical challenge (particularly important for black hole evaporation) is to understand backreaction phenomena whereby the perturbations alter the background structure. It is worth noting that recent work indicates that analogue classical (and potentially quantum) simulations may provide a powerful new tool to understand such phenomena \cite{goodhew:2019,liberati2020back}.} In particular, there has been a recent a proliferation of experiments designed to probe the phenomenon of gravitational Hawking radiation via analogue acoustic black hole (or ``dumb hole'') systems. Just as gravitational Hawking radiation corresponds to a thermal photonic flux associated with an event horizon, acoustic Hawking radiation corresponds to a thermal phononic flux associated with a sonic horizon. Of particular note are experiments leading to the observation of classical aspects of acoustic Hawking radiation in an analogue white hole created using surface water waves \cite{rousseaux:2008,weinfurtner:2011} and experiments leading to the observation of the quantum effect via the correlation spectrum of entanglement across an acoustic horizon in a Bose-Einstein Condensate (BEC) \cite{steinhauer:2016,Nova:2018,kolobov:2021}.\footnote{A fuller set of references to the various existent experimental and theoretical treatments of analogue Hawking radiation is given below.}

A further platform for quantum emulation of gravitational Hawking radiation is provided by dispersive optical media such as optical fibres \cite{philbin:2008,jacquet:2018,drori:2019}. 
The physical basis for the fibre optical platform for emulating black hole phenomena is very different to that used in the case of an acoustic horizon in fluids. In the fluid type platform, the physical basis is the correspondence between an acoustic horizon and an event horizon. This can be pictured intuitively as a `waterfall' set-up where the horizon is created by the surface at which the flow of the fluid exceeds the speed of sound in the fluid. Beyond this sonic horizon sound waves are `swept away' and thus sonic contact upstream of the horizon is forbidden. 
It is not possible to manipulate light in optical media such that it is brought to a standstill in the laboratory reference frame, as is the case in an astrophysical event horizon.
Rather, such an approach involves creating an analogue optical horizon based upon the interaction between light waves and matter. 
In particular, experimental work on analogue optical horizons is based upon exploiting the properties of a \textit{moving} boundary between two regimes with different refractive index. 
 \begin{figure}
\label{hawking}
\centering
    \includegraphics[height=0.2\textheight]{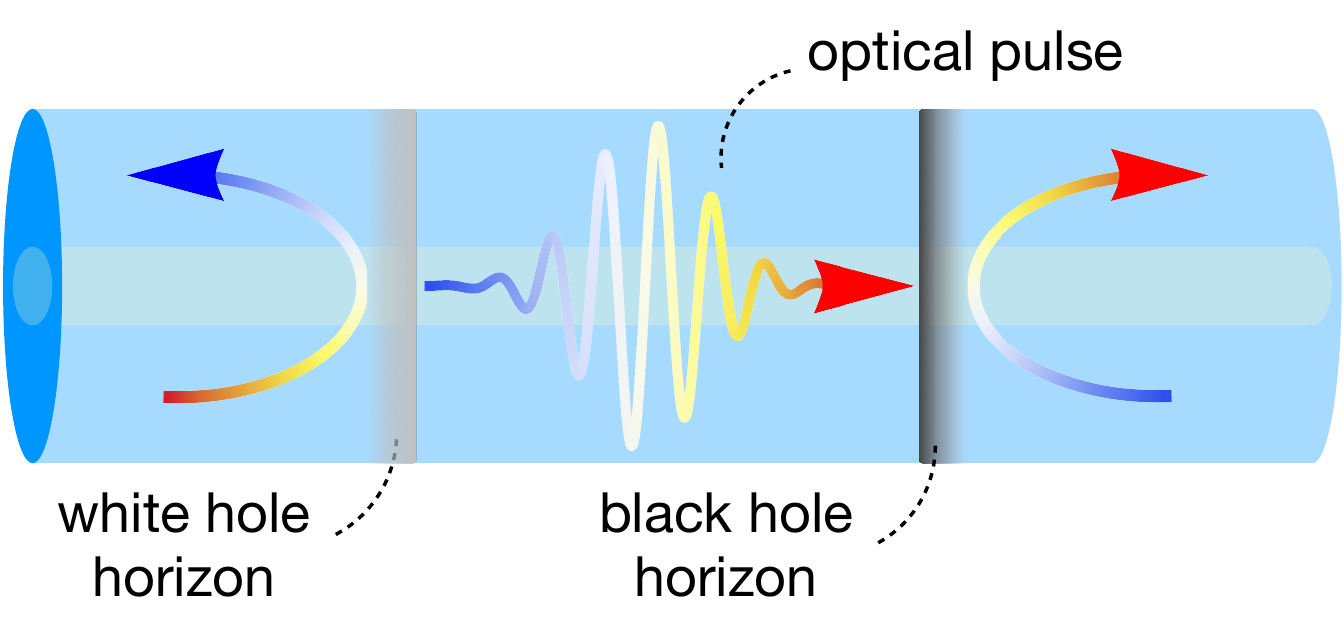}
\caption{Laser pulse in fibre optics represented in lab frame. 
The back of the pulse blueshifts incoming light forming a white hole optical horizon. 
The front of the pulse redshifts incoming light forming a black hole optical horizon. 
(Figure after original due to M. Jacquet)}

 \label{fig: fibre} 
\end{figure}

The moving boundary is created via the Kerr effect, whereby an intense laser pulse increases the refractive index of the medium through which it travels. Light is thus exploited twice: in terms of an intense `pulse' that creates the moving horizon and a weak `probe' that is incident on each side of the horizon. In front of the pulse there is the analogue of a black hole horizon, a surface beyond which light in the medium is trapped. Behind the pulse, there is the analogue of a white hole horizon, a surface beyond which light in the medium cannot pass (see \ref{fig: fibre}). The latter is easiest to conceptualise since it corresponds to the incident light (the probe) travelling \textit{behind} the propagated modification (the pulse) and being unable to catch up and overtake, since the modification is itself changing the properties of the medium such that light travels slower. The point of significance is that the closer to the white hole horizon the incident light gets the greater is the relative slow down, such that eventually the incident light is reflected by the horizon. The story is then the same in time reverse for the front black hole horizon. Light that falls behind the horizon at the front of the pulse is trapped behind a surface beyond which it cannot pass. 

To date, the optical analogue Hawing radiation experiments have not focused upon emulation of spontaneous Hawking radiation itself. Rather, they have focused on the emulation of the event horizon and of stimulated Hawking radiation.
A breakthrough result of the observation of stationary spontaneous Hawking radiation in atomic Bose-Einstein condensates was published in the final stages of completion of this book \cite{kolobov:2021}. It remains to be seen if the spontaneous effect can also be simulated via optical platforms.

\section{Hawking Radiation in Continuum Hydrodynamics}

In this section, following the discussion of \citet{barcelo:2005} and \citet{jacobson:2005}, we will briefly review both the gravitational effect due to \citet{hawking:1975}, and the analogue fluid effect due to \citet{unruh:1981}. This will provide the reader with the relevant physical intuitions necessary to conceptualise the optical case.  

\subsection{What is astrophysical Hawking radiation?}

In a semi-classical approach to gravity we consider a quantum field within a fixed spacetime background. Significantly, this is not the modelling framework of quantum gravity. Rather, we consider quanta of wavelengths much larger than the Planck length and energy densities such that back reaction of the quantum field against the spacetime can be neglected. 

In the simplest semi-classical model we consider a massless scalar field operator $\hat{\phi}$ that obeys a wave equation of the form:
\begin{equation}
\label{waveequation}
g^{\mu \nu}\nabla_\mu \nabla_\nu \hat{\phi} = 0
\end{equation}
where $g^{\mu \nu}$ is a (possibly curved)  spacetime metric. We can expand the scalar field in a basis of orthonormal plane wave solutions
\begin{equation}
\hat{\phi} = \int d\omega   (\hat{a}_\omega f_\omega + \hat{a}^\dagger_\omega f^*_\omega),
\end{equation}
where $ f_\omega = 2^{-1/2} e^{-i(\omega t- \pmb{k}\pmb{x})}$ are plain waves with frequency $\omega$ and wave vector $\pmb{k}$ expressed in spacetime coordinates $(\pmb{x},t)$. 
$\hat{a}^\dagger_\omega$, $\hat{a}_\omega$ are the standard field theoretic creation and annihilation operators of the corresponding modes, and we have assumed a linear dispersion relation of the form $|\pmb{k}|\propto \omega$. The creation and annihilation operators allows us to define both a vacuum state, $\hat{a}_\omega |0\rangle=0$, and a number operator, $\hat{N}_\omega =  \hat{a}_\omega^\dagger \hat{a}_\omega$, in this particular basis. 

\begin{figure}[h]
	\begin{center}
		{ \includegraphics[width=0.4\textwidth]{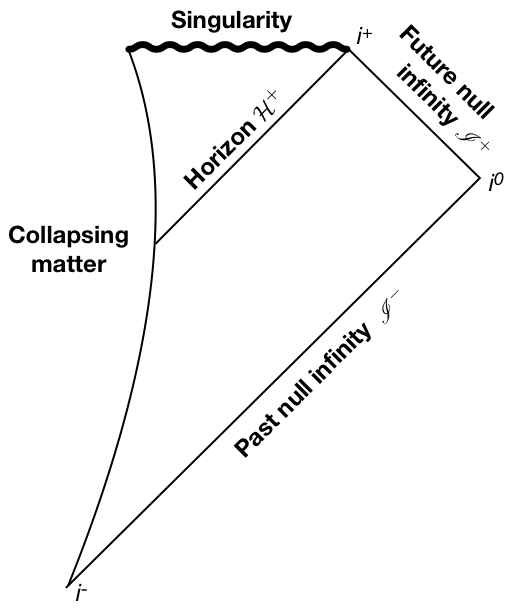}
  }
	\end{center}
	\caption{\label{fig:bh diagram} The conformal diagram of a spacetime external to a spherically symmetric distribution of collapsing matter (after figure due to Sean Gryb).}
\end{figure}

The definition of vacuum states in spacetimes of non-trivial curvature is in general mathematically ambiguous. For this reason the semi-classical approach typically proceeds by defining vacuum states in asymptotically flat regions. For a black hole spacetime, which can be thought of as corresponding to a matter shell collapsing down to a singularity, the two most natural such regions are the distant past long before the black hole formed, corresponding to the earliest possible surface where light can originate, and the distant future long after the black hole formed, corresponding to the latest possible surface that light can reach. These asymptotic regions are assumed to be flat and thus suitable for us to define unambiguous vacuum states. Formally, the distant past region is designated \textit{past null infinity}, $\mathcal{J}^{-}$ and we designate the relevant vacuum there as the `in' state. Correspondingly, the distant future region is called \textit{future null infinity}, $\mathcal{J}^{+}$, and we designate the relevant vacuum there as the `out' state. See Figure \ref{fig:bh diagram} for a representation of $\mathcal{J}^{\pm}$ on the \textit{conformal} diagram of a spacetime external to a spherically symmetric distribution of collapsing matter. 

Hawking's derivation relies on the crucial general observation that the `in' vacuum state need not appear as a vacuum state to observers at positive null infinity: it may contain a flux of `out-particles'. One can calculate this flux by determining the Bogoliubov coefficients, $\beta_{ \omega \omega'}$, which characterise the canonical transformation between the solutions expressed in the `in' basis and `out' basis. The expectation value of the flux, in terms of the number operator of out particles, takes the form
 \begin{equation} 
 \label{Hspec}
_{in}\langle 0| \hat{N}^{out}_\omega  |0\rangle_{in}   
 =  \int  d\omega' |\beta_{ \omega \omega'}|^2.
\end{equation}

What Hawking's 1975 calculation shows is that, for a spacetime which features the establishment of an event horizon via gravitational collapse leading to a black hole (Figure \ref{fig:bh diagram}), one can derive the asymptotic form of the Bogoliubov coefficients and show that it depends only upon the \textit{surface gravity} of the black hole denoted by $\kappa_{G}$. Intuitively speaking, surface gravity is the force per unit mass that must be applied at
infinity in order to hold a zero angular momentum particle just outside the horizon. Hawking's calculation thus implies that a black hole horizon has intrinsic properties that are connected to a non-zero particle flux at `late times' -- i.e as measured by observers in the distant future. 
This flux is known as \emph{Hawking radiation} and its spectrum obeys the relation:
\begin{equation}
\label{bbs}
\langle \hat{N}_{\omega} \rangle = \frac{1}{e^{2\pi\omega/\kappa_G}-1} 
\end{equation}
Crucially, the functional form of this spectrum is \textit{thermal} in the sense that it takes a characteristic Planckian black body energy form for the temperature $T_{BH}=\hbar \kappa_{G}/ 2\pi$. 

Following \citet{Gryb:2018}, two important (and connected) features of the semi-classical model of black holes used to derive Hawking radiation are worth highlighting at this stage. First, the details of the \textit{non-stationary} collapse process that leads to the formation of the black hole are assumed not to matter. That is, the effect is assumed to be identically realised in an idealised model of an eternal black hole as described by the Schwarzschild metric, which corresponds to a stationary spacetime, i.e., a spacetime the geometry of which does not change in time.\footnote{Formally, a spacetime is stationary when there exists a one-parameter group of isometries whose orbits are timelike curves \cite{Wald:2010}. Significantly Kerr spacetimes which describe eternal rotating black holes are also stationary spacetimes in the relevant sense even though they feature rotation, and thus are not static.} Second, it is assumed that there is no back-reaction between the classical spacetime geometry and the quantum field. That is, there is assumed to be no coupling between the gravitational degrees of freedom and the quantum field. Each of these idealisations can be justified (to a limited degree) by reference to a set of robustness or universality arguments which will prove of great significance to our discussion and will be discussed in Section~\ref{HRheuristics} below. Given that these idealisations can be justified, we can then establish an appropriate limiting relation between the simplified target `simulation model' of a black hole (where both back-reaction and the details of the collapse are not included in the model) and the target system model, which we can take to correspond adequately to real astrophysical objects.

\subsection{What is acoustic Hawking radiation?}

Let us now consider a classical fluid as a continuous, compressible, inviscid medium and sound as an alternate compression and rarefaction at each \textit{point} in the fluid. The points are \textit{volume elements} and are taken to be \textit{very small} with respect to the overall fluid volume, and \textit{very large} with respect to the inter-molecular distances. The modelling framework of continuum hydrodynamics is thus only valid provided fluid density fluctuations of the order of molecular lengths can be ignored. 

The two fundamental equations of continuum hydrodynamics are the \textit{continuity equation}, which expresses  the conservation of matter and the \textit{Euler equation}, which is essentially Newton's second law. These standard equations relate the mass density of the fluid at a particular point, $\rho$, the velocity of a fluid volume element, $\vec{v}$, and the pressure, $p$. If the fluid is \textit{barotropic} and \textit{locally irrotational} Euler's equation reduces to a form of the Bernoulli equation. Introducing a velocity potential $\vec{v}=\nabla \psi$, we can then consider the linearisation of the solutions to the equation of motion for the entire fluid about a background, $(\rho_0, p_0, \psi_0)$. We identify the sound waves in the fluid with the first order fluctuations $(\rho_1, p_1, \psi_1)$ about the background, $(\rho_0, p_0, \psi_0)$, which is interpreted as bulk fluid motion. The linearised version of the continuity equation then allows us to write the equation of motion for the fluctuations as
\begin{equation}
\frac{\partial}{\partial t}  \left( \frac{ \rho_0}{c^{2}_{\text{sound}}} \left(\frac{\partial \psi_1}{\partial t}   +  \vec{v_0}   \cdot \nabla \psi_1\right)  \right)
=
\nabla \cdot  \left( \rho_0 \nabla \psi_1  -  \frac{ \rho_0  \vec{v_0} }{c^{2}_{\text{sound}}}\left(\frac{\partial \psi_1}{\partial t}   +  \vec{v_0}   \cdot \nabla \psi_1\right)\right)
\end{equation}

where $c_{sound}$ is the speed of sound in the fluid. 

Most significantly, this equation for the fluctuations can be re-written in a form where the bulk flow plays the role of an `effective spacetime' by introducing the acoustic metric
\begin{equation}
g_{\mu \nu}^{\text{acoustic}} = \frac{\rho_0}{c_{\text{sound}}}  \left( \begin{array}{ccc}
-(c_{\text{sound}}^2-v_0^2) & \vdots & -(v_0)_j    \\ 
\ldots & \cdot & \ldots \\ 
-(v_0)_i  & \vdots & \delta_{ij}
\end{array} \right).
\end{equation}
where the Greek indexes $\mu$ and $\nu$ run over space and time (i.e, $0,1,2,3$ or $t, x, y, z$) and the Latin indices run over space (i.e., $1,2,3$ or $x, y, z$) 

Just as light in a relativistic spacetime is guided by spacetime geometry, sound in an effective spacetime is guided by the `acoustic geometry' given by this metric. Re-writing the equation of motion in these terms gives us:
\begin{equation}
\label{eomfluc}
\frac{1}{\sqrt{-g}}\frac{\partial}{\partial x_\mu}    ( \sqrt{-g}  g^{\mu \nu }    \frac{\partial}{\partial x^\nu}  \psi_1 )  =0, 
\end{equation}
where the metric tensor $g_{\mu\nu}$ is the acoustic metric $g_{\mu \nu}^{\text{acoustic}}$ and $g$ is the determinant of the metric tensor. The close similarity between the acoustic case and gravity can be seen immediately if we consider the \textit{Schwarzschild metric}, which describes an eternal, non-rotating, changeless black hole -- the simplest example of a stationary spacetime. This metric can be written (in so-calledd `Painleve-Gullstrand' coordinates) as:
\begin{equation}
g_{\mu \nu}^{\text{black hole}} =   \left( \begin{array}{ccc}
-(c_0^2-\frac{2 GM}{r}) & \vdots & -\sqrt{\frac{2 GM}{r}}{r_j}    \\ 
\ldots & \cdot & \ldots \\ 
-\sqrt{\frac{2 GM}{r}}{r_i}  & \vdots & \delta_{ij}
\end{array} \right).
\end{equation}
where $c_0$ is the speed of light, $G$ is Newton's constant, $M$ is the mass of the black hole, and $R$ is the raidal distance. 

This similarity between the two metrics can be made rigorous in terms of an isomorphism given certain conditions on the speed of sound in the fluid and the fluid density and velocity profiles. The role of the black hole event horizon is now played by the effective acoustic horizon where the inward flowing magnitude of the radial velocity of the fluid exceeds the speed of sound.  The black hole is replaced by a \textit{dumb hole}.

Unruh's crucial insight in his 1981 paper was that once the relevant fluid--spacetime geometric identification has been made, there is nothing stopping one from repeating Hawking's 1975 semi-classical argument, only replacing light with sound. The result is that, while in the gravitational Hawking effect a black hole event horizon is associated with a `late time' \textit{thermal photonic flux} as measured in the distant future, in the hydrodynamic Hawking effect a dumb hole sonic horizon can be associated with a \textit{thermal phononic flux} at `late times' -- that is, measured `far' away from the sonic horizon (approximated by a few meters) and `long' after it forms (approximated by a few factions of a seconds).    

\section{Optical Black Holes} 

\subsection{Moving Horizons}

Fluids continue to be used as platforms to emulate classical properties of black holes.\footnote{See in particular, \cite{rousseaux:2008,rousseaux:2010,weinfurtner:2011,michel:2014,weinfurtner:2013,unruh:2014,euve:2016,torres:2017,Euve:2018}.} Work specifically targeted at \textit{quantum} emulation of black hole phenomena utilises inherently quantum platforms.\footnote{We should note here that the extent to which Hawking radiation is a inherently quantum phenomena is, in fact, disputed. Thus, whether or not it need be subject to a distinctly quantum emulation remains to be seen.}  Of particular significance are platforms based upon the combination of continuum hydrodynamics and phononic Hawking radiation that can be reached in a particular limit of the physics of  Bose-Einstein condensates  \cite{garay:2000} and superfluid Helium-3 \cite{jacobson:1998}.\footnote{See \cite{steinhauer:2014,steinhauer:2015,Finke:2016,steinhauer:2016,steinhauer:2016a,Nova:2018,Leonhardt:2018,kolobov:2021}.} An alternative platform for quantum emulation of Hawking radiation is provided by various optical systems.\footnote{See \cite{leonhardt:2002,philbin:2008,Faccio:2010,belgiorno:2010,choudhary:2012,unruh:2012,liberati:2012,nguyen:2015,jacquet:2015,jacquet:2017,jacquet:2018,jacquet:2019,drori:2019,rosenberg:2020}.} Optical systems are particularly attractive platforms since light exhibits quantum properties at any temperature and is well  studied and relatively easy to manipulate. Here we will focus our analysis on the platform provided by intense laser pulses in fibre optics \cite{philbin:2008,jacquet:2018}, not least since this example serves to strengthen the parallels with our other case studies. Our treatment will mainly follow that of \citet{jacquet:2018}.  

Recall that the horizon in a fibre optical spacetime is created via the Kerr effect, whereby an intense laser pulse modifies the refractive index of the medium through which it travels. The Kerr effect is based upon the general physical phenomenon whereby intense light fields induce an anharmonic motion of bound electrons of the material through which they propagate. This results in the polarisation of the medium becoming nonlinear and light is being radiated at harmonic frequencies of the fundamental driving and wave mixing. The effective refractive index of the fibre thus gains an additional contribution that is proportional to the instantaneous pulse intensity. This means that, so long as the pulse's energy is high enough to modify the refractive index, then its propagation in the fibre will act as the `moving horizon' set-up that was described above and illustrated in Figure \ref{fig: fibre}.

The key issue is then whether we can set up the formal analogy between the kinematics `seen' by a weak probe travelling in front of (behind) the pulse and that of a non-back-reacting field outside the horizon of a Schwarzschild black (white) hole. Here things are a little more subtle than in the hydrodynamic case. The next sub-section provides the interested reader with some detail regarding the series of approximations needed to construct the relevant isomorphism. 
In Section~\ref{HRheuristics} we will then discuss the philosophically important question, what kind of knowledge we can obtain from the theoretical analysis of Hawking radiation in analogue systems.

\subsection{Fibre Optic Spacetimes}

An adequate  model for the propagated modification of the refractive index of the fibre induced by the pulse can be derived from a microscopic theory describing interactions of light with an inhomogeneous and transparent dielectric based upon the Hopfield model \cite{Finazzi:2013,jacquet:2015,jacquet:2017,jacquet:2018,jacquet:2019}. 

We represent light as a one dimensional scalar electromagnetic field $A(x,t)$  via $E=-\partial_T A$ in the temporal gauge where $x$ and $t$ are space and time in the lab frame. The medium is represented as a collection of oscillators with eigenfrequencies $\Omega_i$ and elastic constants $\kappa_i^{-1}$ with $i=1,2,3$. It is assumed that the frequency of the light is sufficiently far away from the resonances of the medium that we can ignore absorption. Moving to the frame of the pulse, $(X,T)$, which we take to be moving at speed $u$ in the positive $X$ direction, we can then derive an expression for the Lagrangian density of the interaction of the electromagnetic field with the three polarisation fields of the medium, see \cite[\S3.2.2]{jacquet:2018}. This in turn leads to the generic (non-linear) \textit{Sellmeier dispersion relation} of bulk transparent dielectrics. 

Now, whilst this source system model (or `lab model') with non-linear dispersion is an adequate model of the real optical system with a propagated modification in the refractive index, it does not admit a representation in terms of an effective horizon set up like in the hydrodynamics case. Non-linear dispersion in fact prevents us from writing the equation for a weak optical probe travelling in front (behind) the pulse as an effective metric equation like equation (\ref{eomfluc}) above. Rather, setting up the isomorphism to the metric equation, and thus defining the source simulation model, requires us to move to the linear dispersion regime (i.e. that corresponding to a wave propagating in a dispersionless medium), and then also perform some formal manipulation. It will prove instructive to follow these steps more explicitly -- see \cite[\S2.3]{jacquet:2018} for full details.

The move to the regime of linear dispersion on the basis of the approximation that the dispersion due to the medium is weak in the relevant frequency regime. In this regime, we can assume the medium to have only one resonant frequency. The Lagrangian density in the lab frame is then given by 
\begin{equation}
\label{linearoptics}
\mathcal{L}^{LO}=\frac{1}{2}\Big{(}\Big{(}1+\frac{4\pi\kappa}{\Omega^2 }\Big{)}(\partial_t A(x,t))^2+c^{2}(\partial_x A(x,t))^2\Big{)}
\end{equation}
where $\kappa$ and $\Omega$ are constants depending on $x$ and $t$. Significantly, this Lagrangian is of the simple generic form such that the Euler-Lagrange equation will take the form of a wave equation that is analogous to Eq.~\eqref{waveequation} describing the scalar field propagating in the black hole spacetime. Thus, an analogy is set up between the electromagnetic pulse in the medium and the scalar field in the black hole spacetime.  However, the relevant wave equation still does not admit a representation in terms of an effective metric -- the formal analogy between the two wave equations does not provide an isomorphism. To derive such an expression we must next introduce one of the transverse coordinates $y$ which was hitherto neglected and boost to an inertial `observer's frame' moving with velocity $u$. 

We finally arrive an expression for an effective optical metric that has the $g_{00}$ component characteristic of horizon of the Schwarzschild metric which as we noted above describes an eternal, non-rotating changeless black hole. In particular, when written in the special Painleve-Gullstrand form the zero-zero (i.e. time-time) component of the optical metric takes the form:
\begin{equation}
g_{00}^{\text{optical}} \propto c^{2}-u^{2}\Big{(}1+\frac{4\pi\kappa}{\Omega^2 }\Big{)}=1-\frac{u^{2}}{v_p^{2}}
\end{equation}
where $v_p=c \Big{(}1+\frac{4\pi\kappa}{\Omega^2 }\Big{)}^{-1/2}$ is the phase velocity of the waves in the observer's frame. There is a singularity when $v_p=u$ in correspondence with the conditions for the singularity in the acoustic and black hole cases given by $g_{00}=0$ in the metrics (7) and (8). 
This suffices to set up the formal basis for a emulation of a black hole horizon in fibre optics. 
The corresponding inference is represented in our schema in Figure \ref{fig: hawking emulation schema}. 

What we are really interested in is Hawking radiation, and thus the quantum, thermal effects associated with the horizon. In the next section we will consider the theoretical analysis of Hawking radiation in the light of the \textit{heuristics} provided by the analogue black holes -- in particular, theoretical heuristics relating to the origin of Hawking radiation via positive-negative mode mixing and the role of non-linear dispersion in the arguments for the universality of the effect. 
This will provide us with the conceptual basis to finally consider experiments towards the emulation of Hawking radiation in fibre optical analogues in the final section of this chapter. 

\section{Heuristics for the Theoretical Analysis of Hawking Radiation from Analogue Systems} 
\label{HRheuristics}

In the next section we will describe an experiment designed to probe important aspects of this phenomenology. 
However, before we continue our discussion of quantum emulation in optical spacetimes, it will prove well worthwhile to briefly consider the subtle question of what Hawking radiation is and how the study of analogue systems has provided crucial heuristics for better understanding of the phenomenon.

 Hawking's original approach to the derivation of the effect relies upon very general features of the vacuum state in a spacetime that features the establishment of an event horizon. Since that work, various proposals have been made to provide a local physical mechanism for the production of Hawking radiation. The different proposals vary significantly in terms of where and how the thermal radiation is produced and are largely mutually inconsistent. Following \citet{Gryb:2018}, the most significant possible mechanisms include:  splitting of entangled modes as the horizon forms \cite{unruh:1977,gibbons:1977}; tidal forces pulling apart virtual particle-anti-particle pairs \cite{hawking:1979,adler:2001,dey:2017}; entangled radiation quantum tunnelling through the horizon \cite{parikh:2000}; the effects of non-stationarity of the background metric field \cite{fredenhagen:1990,jacobson:2005} and anomaly cancellation \cite{PhysRevD.77.024018}. The formal rigour of these proposals varies greatly, and none is entirely satisfactory from a physical perspective.
 
 What is most significant for the purposes of the analogue effect is that although the precise local physical mechanism for black hole Hawking radiation is as of yet not fully understood, analysis of our models of the analogue systems can help us isolate the generic formal features that are sufficient for the effect to exist. Furthermore, analogue experiments can allow us to gain understanding of these features by exemplifying them in terms of their concrete and manipulable realisations. What we have in mind here is well illustrated by the case of negative frequency modes and modified dispersion relations the importance of each of which we can explain as follows.  
 
  There are good formal arguments that the thermality of the radiative spectrum of the `out' vacuum state defined in the distant future region, i.e. $\mathcal{J}^{+}$ in Fig. \ref{fig:bh diagram}, depends upon a frequency shift between the components of the outgoing and incoming state. In particular, the effect occurs precisely when the radiative modes in the distant future region $\mathcal{J}^{+}$ which have positive \textit{Killing frequency}\footnote{\label{killing} Following \citet{jacobson:2005}, Killing frequency is defined in terms of the time dependence with respect to the time-translation symmetry of the background black hole spacetime (i.e. the timelike isometries of the metric). Killing frequency is constant along geodesics of a spacetime. Significantly, in the asymptotically flat regions far away from the black hole, such as $\mathcal{J}^{+}$ and $\mathcal{J}^{-}$, the Killing frequency agrees with the usual frequency defined by the Minkowski observers at rest with respect to the black hole. However, close to the black hole, where there is non-trivial curvature, the Killing frequency is very different to that defined by freely falling observers. This difference is key to the existence of the effect.}, depend upon an \textit{exponentially} weighted combination of negative and positive \textit{Killing frequency} components of radiation in the distant past region, $\mathcal{J}^{-}$. The exponential dependence is crucial for the thermality of the state, i.e. the black body spectrum of the form given in Eq.~\eqref{bbs} and thus the insensitivity of the effect to the details of the black hole collapse -- and thus the viability of the stationarity and no-back reaction idealisations mentioned above. Furthermore, the exponential relationship between the frequency of Hawking modes near-horizon and far away from the horizon, means that the flux identified as Hawking radiation by observers in the distant future region will correspond to absurdly small frequencies near the horizon -- that is, frequencies corresponding to wavelengths beneath the Planck scale, and thus far beyond the realm of validity of the semi-classical approximation. This is the so-called trans-Planckian problem \cite{Gryb:2018} that we will return to shortly. 
 
 The important point, for the moment, is that the mixing of the positive and negative norm modes is crucial for the radiation to exist at all, since it implies that the Fock spaces based upon the incoming and outgoing radiation, respectively, are different and thus that the `in' vacuum state is not annihilated by the out annihilation operator -- see \cite[\S3.1.2]{jacquet:2018} for discussion. Thus, to understand Hawking radiation better, it is crucial to gain intuitions regarding the physics of negative norm modes -- and this is one area where fibre optic spacetimes provide a particularly powerful heuristic tool. Conceptualising the difference between positive and negative \textit{Killing frequencies} requires an advanced understanding of Riemannian geometry -- rare outside specialists in general relativity, see footnote \ref{killing} for details. In contrast, for an analogue black hole system the difference between the negative and positive frequency modes is a simple and extremely intuitive one. 
 
 In particular, following \cite[\S1]{jacquet:2018}, we can consider the generic relationships found in the study of classical (non-relativistic) fields between the two families of solutions to a wave equation with positive and negative sign in the time factor term of the exponential. Typically, one ignores the negative frequency components of a classical field since they are entirely dependent on the positive component. Thus, the extra set of solutions are usually ignored as otiose or surplus mathematical structure. However,  there exist experimentally realisable conditions under which the positive and negative
frequency components of a field may be observed independently and can even be made to mix. One case is
in nonlinear optics and another is Hawking radiation. The example of the former is thus a powerful heuristic for understanding the latter since it allows us to understand the generic basis of the Hawking effect -- negative and positive frequency mode mixing -- independently from the subtleties of understanding the concept of Killing frequency.   

The foregoing analysis illustrates the fact that analogue models for Hawking radiation, and semi-classical black hole phenomena, have provided, and continue to provide, powerful \textit{heuristic tools} for theoretical analysis. In particular, physical intuition from analogue systems such as fibre optics and fluid dynamics allow conceptual insight into features such as negative norm modes, dispersion relations, and cross-horizon entanglement in the context of Hawking radiation in Schwarzschild black holes, and superradiance in the context of rotating black-holes \cite{basak:2003,richartz:2015,cardoso:2016,torres:2017}.

The specific history of the dispersion relation case, which will prove relevant later in our analysis, is particularly telling. In the early 1990s a suggestion regarding modelling of the breakdown of the model for the fluid mechanical case was made by  \citet{Jacobson:1991,Jacobson:1993}. In particular, it was suggested that one can focus upon the altered dispersion relation that is relevant to an atomic fluid rather than continuous fluid, and consider whether, in such models, an exponential relationship holds between the outgoing wave at some time  after the formation of the horizon, and the wavenumber of the wave packet \cite{unruh:2008}. 

Approximate answers to such questions can be determined in practice via numerical methods, and it was shown by \citet{unruh:1995} that the altered dispersion relation in atomic fluids does imply that the near horizon quantum fluctuations that can be connected to the Hawking radiation are not in fact exponentially large. Later work then proceeded to generalise from the specific fluid dynamical alterations to the dispersion relation, to a model with a \textit {generically altered} relation, independent of the particular cause of the trans-Planckian breakdown. Of particular relevance are calculations to this end by \citet{unruh:2005}. Their results represent a generalisation of earlier work by  \citet{corley:1998} and provide a basis for a universality claim with regard to the Hawking effect.\footnote{For further work on these issues, using a range of different methodologies, see for example \cite{himemoto:2000,barcelo:2009,coutant:2012}.}

We will return to the relevance of these universality arguments in the context of the \textit{understanding} one can gain via analogue quantum emulations of black holes in Sec.~\ref{sec:understanding analogue emulation}. For the meantime, the key idea that we want to impress upon the reader is that it is plausible to think of Hawking radiation as a universal geometric effect that emerges due to conversion of negative to positive norm modes at a horizon, regardless of the microscopic physics that drive the non-stationarity of the  background space-time geometry \cite{visser:2003,rosenberg:2020}. 
As such, Hawking radiation can be considered akin to something like interference phenomena that are generic to wave systems irrespective of whether their realisation is in terms of water waves, gravitational waves or light waves. This is the key justification for considering the idealised semi-classical model of a black hole in a stationary spacetime with no back reaction as having an appropriate limiting relation with the complex `system model' of a black hole formed via a complicated collapse process. And this theoretical insight into the phenomena has only been possible via exploration of analogue systems.  

\section{Experimental Realisation of Optical Hawking Type Phenomena}

The consideration of mode mixing as the basis for Hawking radiation mentioned in the last section leads naturally to a discussions of the existent experiments that have been performed to emulate Hawking phenomena on optical platforms. The ultimate goal in such experiments would be to observe a thermal spectrum of \textit{spontaneous} emission of light from the vacuum associated with the horizons in front or behind the laser pulse. That is, to observe the moving optical horizon (as induced by the Kerr effect) spontaneously emitting quanta with a thermal spectrum matching that predicted for the astrophysical case via equation (\ref{Hspec}). To date no such experiments have been performed. 

\begin{figure}
\centering
    \includegraphics[height=0.4\textheight]{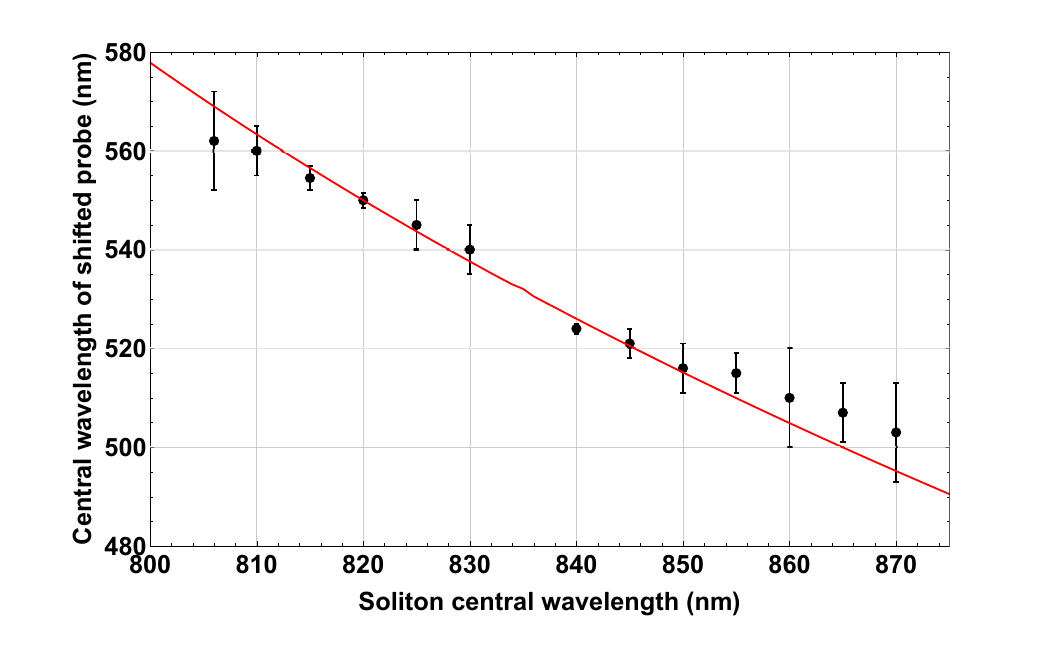}
\caption{
\label{fig:opticaldata}
\cite[Fig.~5.9]{jacquet:2018}. 
Location of shifted probe spectra for different pulse wave-lengths. The red curve is the prediction from the dispersion curves for photonic crystal fibre based upon Sellmeier model and the data points show measure wavelength of the shifted probe spectra.
 }
\end{figure}

What has been achieved, to different degrees in various experiments, is the \textit{stimulated} effect whereby a probe is sent to the white hole horizon behind the pulse and the resulting reflection and frequency shift are observed \cite{philbin:2008,choudhary:2012,jacquet:2018,drori:2019,rosenberg:2020}. In such approaches, the input state is populated with a  finite number of photons, which correspond formally to replacing the vacuum state on the left hand side of \eqref{Hspec} with a coherent state. Such experiments then aim at measuring the relevant Bogoliubov coefficients (i.e. $\beta$ on the right hand side of \ref{Hspec}) for various input frequencies of $\omega'$. The central wavelength of the shifted probe, i.e. $\omega$ in \eqref{Hspec}, then depends upon $\omega'$ via \eqref{linearoptics}.

Focusing on the photonic crystal fibre experiments described in \cite[\S5]{jacquet:2018}, we can consider the data in Figure \ref{fig:opticaldata}. What is shown is the measured centre wavelength of the shifted probe wave as a function of pulse wavelength. These results give  experimental confirmation of frequency shifting as a signature feature of the horizon.  This corresponds to observation of positive-norm to positive-norm frequency shift and thus a close relation to the generic effect taken to be behind Hawking radiation. Crucially, such  correspondence between theory and data can \textit{validate} the relationship between the abstract source system model, non-linear optics obeying the Sellmeier dispersion relation, and the concrete source system of photonic crystal fibre. 

Let us now turn to the more recent experiments of \citet{drori:2019} following the account of  \citet{rosenberg:2020}.  
These experiments followed the approach of \citet{philbin:2008} and also use photonic crystal fibre. The key achievement of the experiment was observing for the first time \textit{both} stimulated Hawking radiation \textit{and} the relevant `partner modes'. 
The partner modes are the (stimulated) fibre optical analogue to the modes that are cut off behind the event horizon in the astrophysical black hole Hawking effect.\footnote{The partner modes are described in \cite{drori:2019,rosenberg:2020} as `negative' Hawking radiation based upon the time reversed `white hole' picture. In particular, in the astrophysical white hole scenario, the partner modes can be understood as occurring outside a white hole when the outgoing radiative modes with \textit{negative} (Killing) frequency, depend upon an exponentially \textit{blue} shifted combination of negative and positive (Killing) frequency components of the incoming radiation. However, just as the positive frequency Hawking radiation outside a black hole is accompanied by partner modes of negative frequency Hawking radiation inside the inescapable causal horizon, negative frequency Hawking radiation outside a white hole is accompanied by partner modes of positive frequency Hawking radiation on the other side of the impenetrable causal horizon. It is probably thus simpler to use the term `Hawking radiation' to apply to both the positive and negative norm partners in both black hole and white hole scenarios. In any case, since we are dealing with bosons in the fibre optics case, the two types of modes are intrinsically identical.} Both the stimulated modes and the partner modes were generated by the mixing of negative-norm and positive-norm modes at the white (black) hole horizon behind (in front of) the pulse.

The key elements of the experimental set up were as follows. First, self-phase modulation due to the nonlinear refractive index is used to counteract anomalous dispersion and form stable solitons, i.e., discrete `pulse' packets travelling along the fibre. The fibre structure is then engineered to change its dispersion relation to have two points with matching group velocities: 
one in a normal dispersion region and another in an anomalous dispersion region, which includes the probe spectra. 
The solitons then create a horizon. Stimulated Hawking radiation and the partner modes were observed. In particular, probe frequency shifts at the horizon were found to be accompanied by Hawking radiation and the relevant  partner modes. Significantly, varying the probe co-moving frequency and its conjugate was found to shift the relevant signals in line with the theory. 
This serves to validate the relationship between the mathematical source system model and the fibre optical system, and thus support the interpretation of the signal as Hawking radiation. 
It is also noteworthy that, according to \citet{rosenberg:2020}, these experiments demonstrate how robust the Hawking effect is: since the effect appears despite the existences of extreme nonlinear dynamics related to the collapse of the soliton.

The challenge remains to observe \textit{spontaneous} quantum Hawking radiation in fibre optical systems. However, this is principally a technical challenge, due to the extremely low power of the effect  relative to spurious radiative noise. As such, it is a plausible near term expectation that the full spontaneous quantum Hawking effect will be observed in fibre optical systems. Even without this full demonstration, we take it that the consideration of this case study has provided strongly suggestive evidence that emulations of black holes using fibre optical platforms can and do provide genuine \textit{understanding} of the Hawking effect in general, in particular the fundamentality of the mixing of positive and negative frequency modes to generation of Hawing radiation. We will return to the status of this understanding claim in Section~\ref{sec:understanding analogue emulation}.

\section{Philosophical Schema}

In Chapter \ref{Distinctions} we introduced the sub-type of analogue quantum simulation that we defined as analogue quantum emulation. In analogue quantum emulation a scientist is interested in gaining understanding of physical target phenomena, that is, quantitative or qualitative properties of a target system. That the case study considered in this chapter plausibly illustrates the form of inference is indicated by the aim of scientists to gain understanding of Hawking radiation in general, including within its astrophysical realisations, through the fibre optical experiments. This is not a case of analogue quantum computation since the intention of the experimenters is not to gain understanding directly pertaining to features of an abstract theoretical model. 
Rather, the goal is to understand features of a physical target phenomenon instantiated in a concrete physical system, namely, Hawking radiation in astrophysical black holes. 

\begin{figure}
\label{hawking}
\centering
    \includegraphics[height=0.4\textheight]{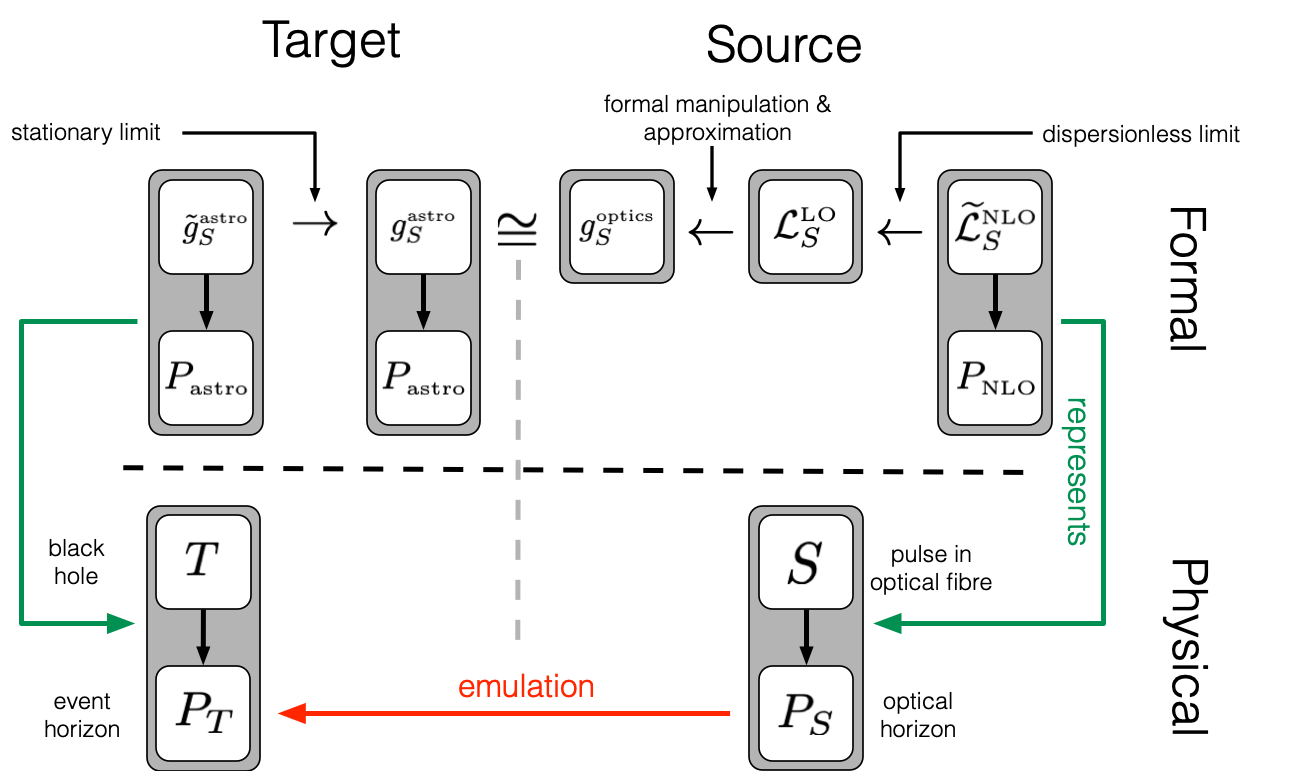}
\caption{Schema for analogue emulation of Hawking radiation case study (see main text for figure explanation).}
    \label{fig: hawking emulation schema} 
\end{figure}

Calling upon our diagrammatic language we can explain the structure of the inferences involved in this example of analogue quantum emulation as follows. The right hand side of Figure \ref{fig: hawking emulation schema} relates to the fibre optical system. First we have some abstract model based upon a Lagrangian $\widetilde{\mathcal{L}}_S^{\text{NLO}}$ for the non-linear optics regime. This complex source system model corresponds adequately to a concrete fibre optic system manipulated in the lab and is thus the source system model or `lab model'. 

In the dispersionless regime this system model can be approximated by a simulation model based upon a Lagrangian $\mathcal{L}_S^{\text{LO}}$ for the linear optics regime. This in turn stands in a formal relation with an effective `optical metric' $g^\text{optics}_S$ -- the analogue to the spacetime geometry `seen' by the light due to the modification of the refractive index. This leads to a partial isomorphism to the target simulation model provided by the Schwarzschild spacetime metric that describes a stationary (or eternal) black hole, $g_T^{\text{astro}}$. 
This metric is a crucial part of the semi-classical model of a black hole and this idealised  semi-classical model then has a limiting relation to a the `target system' model of a real astrophysical black hole, namely, the spacetime metric $\tilde{g}_T^{\text{astro}}$, which includes features such as non-stationarity of the spacetime and back-reaction between matter and spacetime. 
The metric $\tilde{g}_T^{\text{astro}}$, in turn, corresponds adequately to the black hole target system. 

The phenomena in question are as follows: $P_S$ is an optical horizon in the concrete source system; $P_\text{NLO}$ is the representation of that horizon in the full non-linear optical model with dispersion; $P_{\text{astro}}$ is the representation of an event horizon in a semi-classical astrophysical model; and finally $P_T$ is the real event horizon that defines the boundary of a black hole. 

Analogue quantum emulation is a relationship between $P_S$ and $P_T$. 
The goal of analogue quantum emulation is to gain understanding of actual phenomena in a concrete physical system. This is the key distinguishing feature between computation and emulation that shall be the major focus of our analysis in the context of philosophical treatments of understanding in science. In Chapter \ref{ch:assessing case studies} we will return to the case study presented in this chapter in order to frame its epistemic purpose within scientific practice. We will argue that in such cases analogue quantum emulation is being employed by scientists with the aim of obtaining \textit{how-actually understanding} of target physical phenomena. The conditions under which this aim should be understood to be have been achieved together with the wider methodological significance of analogue quantum emulation will then be considered in detail in the remainder of the book.

\chapter[Understanding via Analogue Q. Simulation]{Understanding via Analogue Quantum Simulation}
\label{ch:understanding}


In this chapter we will analyse the sense in which analogue quantum simulations can provide a scientist with understanding. In particular, we will provide a framework for assessing claims of scientific understanding in the context of analogue quantum  computations and emulations that extends the modified simple view of model based understanding, due to \citet{Strevens2008} and \citet{ToyModels}. The scientific details from the case studies of the last three chapters will be drawn upon in this chapter to provide a \textit{preliminary illustration} of the key concepts of our extended framework for model based understanding. The following three chapters will then draw on the details of the case studies to argue that analogue quantum computation and emulation afford different types of understanding.

\section{Understanding Understanding}

Already Goethe's Faust hoped 
\begin{quote}
    \emph{
        [...] zu verstehen was die Welt\\
        im Innersten zusammenh\"alt (...)
    }

    \flushright \emph{That I may understand [know] what the world contains\\
    In its innermost heart and finer vein, (...)}
	\flushright	Johann Wolfgang von Goethe: \emph{Faust; a Tragedy} (Ch.~1). 
\end{quote}
And still today, understanding is one of the primary goals of the scientific endeavour.
No doubt the goals of science are manifold and prominently include predictions and explanations of phenomena. 
Nonetheless, it seems no overstatement to say that the quest for understanding is the driving force behind science as an activity of individuals and groups of scientists. We might therefore place the analysis of the ways in which scientific methods and tools allow individuals to understand at the heart of philosophical enquiry about science.\footnote{For a recent debate on the epistemic value of understanding and knowledge see, for example, \citet{kvanvig_value_2003} and \citet{pritchard_knowledge_2014}. For discussion specifically focusing on understanding as the goal of science and the relation to the broader epistemology of understanding see \citet{grimm2006,grimm2012,grimm2016,rowbottom:2015,bangu:2015,dellsen:2016,stuart:2016,park:2017,khalifa2017,verreault:2019,Dellsen:2021}. For earlier work on understanding in the framework of the unificationist account of explanation see \cite{Friedman-JP-1974,Kitcher-PS-1981}.}  However, when we say that a particular theory, model or experiment provides a scientist with understanding of a physical phenomenon, what do we mean? The standard philosophical strategy for addressing such questions is to attempt to formulate plausible necessary and sufficient conditions that need to be fulfilled for an individual scientist to have understanding of a phenomenon.
Before we get to actually formulating such conditions, however, let us elaborate some of the questions that one might ask when trying to formulate a theory of understanding. 

Clearly, when saying that a scientific tool such as a theory, model, or analogue simulation provides understanding of a phenomenon what we mean to say is that it allows \emph{an individual scientist to understand} that phenomenon. 
Understanding is something that takes place in an individual's mind. In contrast to, for instance, the closely related notion of scientific explanation, we therefore cannot hope to come up with a characterisation of understanding that is independent from the individual scientist.
While referring to the force of gravity and a harmonic oscillator might help a trained physicist to understand the motion of a pendulum, surely it need not do so for a school child without the necessary background to understand those terms.
But the same person might acquire such training and at some later point in time in fact understand the very same concepts that might enable them to understand it. 
For this very reason, understanding might appear a \textit{merely subjective} concept about which no definitive statements can be made.

Now, this idea of the understanding, including scientific understanding, being subjective, does have a clear basis in the fact that understanding is intrinsically \emph{contextual} in nature \cite{Regt-Synthese-2005}. We cannot talk about understanding without also talking about the particular scientist who wishes to obtain understanding of a particular phenomenon in a particular context. However, granting this, and starting from its contextual nature, can we maybe still identify \textit{objective} features of understanding? That is, can we identify features not tied to the particular proclivities of a particular agent? Such features need not universally characterise understanding of every individual, but they might still provide a useful guide as to which circumstances foster scientific understanding in a circumscribed class of relevant cases.  

The importance of understanding scientific understanding for the analysis of motivations behind contemporary scientific practice is particularly relevant in the case of analogue simulation. 
As Immanuel Bloch, a leading scientist in the field of quantum simulation puts it, ``the gained insight (...) [in some analogue simulation experiments is] to understand the phenomenon better.''
His colleague Ulrich Schneider elaborates: 
\begin{quote}
This is where we claim our relevance from: it's about understanding
principles that will be relevant for technology in 10, 20 years from now.
One can compare to the classical wind tunnel. There, the application nowadays is that one scales down a concrete complex object and looks at its behaviour in the tunnel to optimise one concrete application. 
This is not what we do. Rather, we are like the pionieers and take the tunnel 
to test several things in order to find the fundamental laws. 
Building on that we try and create some understanding which the more applied research can then use.
\flushright \emph{Ulrich Schneider, Interview in December 2016}
\end{quote}
Understanding understanding is therefore the first step we need to take to be able to better understand analogue simulation and its potential use in understanding the world around us.

\section{The Simple View of Understanding}
\label{sec:simple view}

Already in the few thoughts above, we have encountered a subtlety about our use of the word `understanding'. 
Take the example of understanding via a model. 
We can both `understand a model' and we can `understand a phenomenon' \emph{using} a model. 
The two senses seem to be quite different and yet interrelated. 
When saying that someone `understands a concept or model' what we mean is that this person is able to comprehend the terms in which the model is phrased and able to derive logical consequences from the model, to apply it to real-world situations, and maybe also to explain it and the underlying mindset to someone else.
The object of understanding, a concept or model, is an abstract entity. 
In contrast, when saying that a person `understands a phenomenon' in the world, the object of understanding are real facts about the world. 
And we might \emph{make use of} abstract theories or models to understand this phenomenon -- having previously understood those. 
To assert that an individual understands a real phenomenon must therefore live up to a much higher standard as it potentially implies much theoretical knowledge. 
What is more, in understanding that phenomenon one needs to make reference to the \emph{actual} mechanisms or reasons behind this phenomenon rather than merely abstract entities. 

The particular view of understanding that will prove most relevant to our analysis follows a vein of thought developed in \cite{trout_scientific_2002,Regt-Synthese-2005,Strevens2008,Strevens2013understanding,regt_scientific_2013,de_regt_understanding_2017}. In particular, drawing upon the work of \citet{ToyModels}, our approach will be to to focus on understanding via models and consider focus on the three necessary conditions:\footnote{There are, of course, exceptions to those conditions, in particular, with regards to the question whether or not understanding a phenomenon requires an explanation of that phenomenon in the first place \cite{lipton2009understanding,khalifa_role_2013,wilkenfeld_understanding_2013,kelp_understanding_2015,dellsen_beyond_2018}. } 
\begin{enumerate}[(i)]
	\item the model must yield an explanation of the target phenomenon (explanation condition); 
	\item  that explanation must be true (veridicality condition); and 
	\item the scientist who claims to understand the phenomenon via the model must have epistemic access to that explanation (epistemic accessibility condition). 
\end{enumerate}

An account of understanding that captures those criteria in a particularly clean way is Strevens' simple view of understanding~\cite[p.~3]{Strevens2008} and \cite[p.~510]{Strevens2013understanding} according to which `An individual has scientific understanding of a phenomenon just in case they grasp a correct scientific explanation of that phenomenon.'
In other words: 
\begin{quote} The simple view: An individual scientist $S$ understands phenomenon $P$ via model $M$ iff.\ $M$ explains $P$ and $S$ grasps $M$.
    \cite[p.~17]{ToyModels}. 
\end{quote}

The simple view of understanding can be viewed as a meta-account of understanding. 
In order to determine whether or not a particular scientist $S $ understands a given phenomenon $P$, it guides us which conditions we need to check: 
Does the model explain? What is the modal strength of the explanation? Does the scientist grasp the model? In order to do so, one needs to resort to specific accounts of explanation, truth, and grasping.
 Explanation and truth are particularly thorny and contested topics in the philosophy of science, with long and largely inconclusive literatures \cite{sep-scientific-explanation,glanzberg_truth_2018}. The simple view of understanding is particularly appealing since it does not depend on specific accounts for each of the three conditions. 
Rather, it remains indifferent to how exactly those conditions are spelled out and thus serves well to analyse understanding using simulations or models. 
While people may not agree on how exactly models or simulations explain, whether they are realists or constructivists, it seems fair to assume that there is a consensus that one can explain phenomena using models \emph{on some account} of explanation and truth. 

The simple view of understanding allows us to go into as much detail on each of the three conditions as needed in our analysis of analogue quantum simulations without committing ourselves -- or the matter of how and when analogue simulations yield understanding -- to a particular account of explanation, truth, or grasping. 
In our analysis of analogue quantum simulations, we will therefore start from the simple view as a basic account of understanding. 

The simple view has the virtue of making explicit the reason why we consider it instructive to focus our analysis on the epistemic goal of obtaining scientific understanding via analogue quantum simulation, as opposed to, say, explanation or prediction simpliciter. As all of the three complementary conditions -- explanation, veridicality, and epistemic accessibility -- are important for scientific understanding, it sheds light on those different aspects of inferences via analogue quantum simulation. 
As we will see, the subjective component of understanding is where the analog nature of analogue quantum simulations plays a crucial role. We will also see that the kinds of explanation provided by analogue quantum computation and emulation are relevantly different, and that different modalities and types of truth play a role for the two types of analogue simulation. 

The important first step in our account is to argue that all analogue quantum simulations can relevantly satisfy two of the three conditions, namely, the explanation condition, and, importantly, the epistemic accessibility condition. We will proceed towards that goal in the next two sections (Sections \ref{sec:explanation condition} and \ref{sec:grasping condition}).

\section{The Explanation Condition}
\label{sec:explanation condition}

\subsection{Explanation via Analogue Quantum Computation}

Recall that the target of an analogue quantum computation is certain formal properties of a target simulation model. 
In an analogue quantum computation, those formal properties are computed in a way that is empirically observable. Insofar as the target simulation model (together with the respective modelling framework) can provide an explanation of its very own consequences analogue quantum computations can do so  too. These properties are abstract \textit{mathematical} facts that are necessary consequences of the model and its modelling framework. 

Most accounts of how mathematical facts are explained are accounts of explanations in mathematical proofs \cite{rav_why_1999}. When applied to formal properties of a mathematical model, how or whether a model explains a property of itself, strongly depends on how this formal property is derived. Some authors have focused on the contrast between inductive proofs and other forms of proof, where there is a contested claim that inductive proofs can never be explanatory \cite{Kitcher:1975,Brown:1997,lange:2009,baker:2010}. Another interesting part of this debate is the connection between scientific explanation and explanation by proof in mathematics. Some authors argue these are distinct kinds of explanation that should not be conflated \cite{Zelcer:2013}. Others have offered accounts that appeal to be applicable to both. In particular, on the unificationist account of Kitcher, a proof counts as explanatory when it instantiates an `argument pattern' from the `explanatory store', which is made of the set of argument patterns that most efficiently systematises our knowledge in a given domain \cite{Kitcher:1989}. Most plausibly, it has been argued that we should expect scientific and mathematical explanations to be distinct species falling under the same genus and thus expect some similarities and some differences \cite{DAlessandro:2019}. A further debate on the nature of explanation via mathematical proof focuses on the contrast between \emph{constructive proofs} as opposed to proofs by contradiction \cite{hanna_proof_2000}. Here it is argued that constructive proofs are able to explain the fact they are a proof, much more so than proofs by contradiction. The idea is that in a constructive proof, one gets a glimpse of \emph{how the property comes about} given the axioms and more fundamental facts. It is the case of constructive proofs and their connection to the counterfactual account of explanation that will prove most relevant for our analysis of explanation via the derivation of a formal property of a model in a physical context.\footnote{For more on explanation in mathematics see the excellent overview provided by \cite{DAlessandro:2019} and references therein.} 

Within a counterfactual account of explanation \cite{Woodward:2003,sep-scientific-explanation,frans_mechanistic_2014}, constructive proofs allow us to answer certain `What-if-things-had-been-different questions' of the type: 
How would the derivatum change if the derivanda were altered in this particular way? 
A good example of how distinct proofs can be extremely different in terms of their explanatory power is the derivation of the phase transition in the two-dimensional Ising model. 
Already in 1936, Rudolf \citeauthor{peierls_ising_1936} provided a rigorous argument for the existence of a phase transition in the model. This argument is based on intuitive yet rigorous reasoning about the energy cost of domain walls at different temperatures. Only eight years later did Lars \citet{Onsager-PR-1944} provide a full solution of the 2D Ising model wherein he also explicitly computed the specific point of the phase transition. 
In his solution, \citeauthor{Onsager-PR-1944} used the so-called transfer-matrix method, developed by \citet{Kramers-PR-1941} some years earlier. 
While this method allowed him to exactly compute an analytical solution, it is also a method that, arguably, does not provide insight into how the phase transition comes about when approaching the critical point. 

\citeauthor{peierls_ising_1936}'s argument provides exactly this: 
it shows \emph{how and why} domains become energetically favourable at low temperatures in two dimensions but not in one based by reasoning via the  free energy of the system. 
To compute the entropy (required for the computation of the free energy) one can make a combinatorial argument, counting the number of possible configurations with the same energy which depends on the size of the domain wall. In two dimensions, this depends on the system size and thus favours domain formation, in one dimension it does not and thus favours a disordered state all the way to zero temperature. This shows that different derivations of the same formal property -- in this case the existence or not of a phase transition in the 2D Ising model -- can be very distinct in terms of their explanatory value and power. 
Peierls's argument is highly intuitive and shows how a phase transition comes about in terms of domain formation. 
And indeed, the formation of those domains can be seen in experiments. 
But it also shows only that: the existence of a phase transition. 
It cannot make a precise prediction for where this transition is going to occur. 
In contrast, Onsager's solution does provide for such a precise prediction but no explanation why the phase transition comes about. 

Precisely how a formal property of a model is derived becomes especially prominent when considering the question whether and, if so, how analogue quantum computations can explain. 
We argue that deriving a formal property in an analogue computation has many of the features that a constructive proof such as Peierls's argument has, too. 
More specifically, in an analogue computation one can directly witness how the formal property dynamically comes about. 
This is specific to the \textit{analog} (i.e. not digital) nature of an analogue quantum computation. 
One can directly observe how the state of the analogue simulator dynamically and \emph{continuously} evolves to the solution of the problem.

For example, when preparing a Hamiltonian ground state via cooling one can observe the properties of the state as it evolves towards the ground state. 
Specifically, in adiabatic quantum computing or quantum annealing \cite{kadowaki_quantum_1998,farhi_quantum_2000,boixo_evidence_2014,albash_adiabatic_2018} one encodes a computation in a Hamiltonian and obtains the solution from measurements on its ground state. 
When gradually cooling into the ground state of this Hamiltonian one not only obtains a glimpse of the vicinity of the ground state but one can also observe which properties of the solution become suppressed at lower and lower energies. 
Similarly, in a quenched evolution -- recall the setting of MBL \cite{BlochEisertRelaxation,schreiber_observation_2015} -- one can directly observe the dynamical onset of an equilibrium by measuring the relevant observables. 
Thus one not only obtains knowledge of the equilibrium properties of a system (the solution of the analogue computation) but also how those properties come about dynamically.
Does the system smoothly approach equilibrium? 
Or does it oscillate back and forth? Are there recurrences? 
Provided that the relevant observables can be measured with high enough temporal resolution, in both examples one can observe the dynamical evolution of the system and thus quite literally \emph{how} the respective formal property or solution to the problem comes about. 

Now, of course, to observe how the solution of a given analogue computation comes about can only ever contribute to an explanation of phenomena occurring in \emph{a single} instance of the model as defined by the parameters used in the computation. 
However, often we are interested in the behaviour of phenomena as a function of the parameters of the model themselves. 
To explain how such a phenomenon comes about naturally requires more than a single instance, but rather the ability to analyze its features in a range of counterfactual situations. 
Indeed, this is precisely the idea of James Woodward's interventionist account of explanation~\cite{Woodward:2003}, who argues that to explain why means to be able to give an account of a wide variety of such counterfactual situations.

To make the comparison with Peierls's argument; analyzing which domain walls are energetically favourable can be done for a fixed setting of the interaction parameter of the model and geometry, but the analysis draws its importance from the possibility to vary  the interaction parameters and model geometry -- from one dimension to two dimensions, from anti-ferromagnetic to ferromagnetic, etc.. 
Likewise, this is where analogue quantum computations have particular inferential value. 
The possibility to access a wide range of regimes by tuning the respective model parameters in an experiment is where their ability to explain (or contribute to explanations of) phenomena rests \cite{Woodward:2003}.

Taken together, the possibility to dynamically observe the onset of a phenomenon in a given instance of an analogue computation and to explore wide parameter regimes makes analogue quantum computations helpful tools for explaining the formal properties of a target simulation model, in the same way as constructive proofs explain mathematical facts. It is important to stress that in this respect analogue quantum computations are not at all singular in the methodological spectrum. For instance, certain discrete numerical optimisation algorithms such as gradient descent bear very similar properties: one can observe how the algorithm explores an optimisation landscape when approaching the optimal point.\footnote{We will elaborate on specific classical algorithms that explore optimisation landscapes, namely, simulated annealing algorithms, and how they might yield explanations in Section~\ref{sec:understanding_formal_properties}.} Explanations via analogue quantum computation therefore becomes particularly relevant in either of two cases: 
(i) there is no efficient algorithm able to derive the target formal property from the model specification, or (ii) the existing derivation of a given formal property does not elucidate how this property comes about. 

\subsection{Explanation via Analogue Quantum Emulation}

The explanatory situation with analogue quantum emulations is a bit more complicated than for analogue quantum computation since the relevant explananda are physical phenomena rather than formal properties. 
Recall that in an analogue emulation, one is probing a source simulator system in order to directly learn about a concrete physical target phenomenon. 
In Section~\ref{sec:computtaionempulation}, we already remarked that emulation is essentially theory or model-mediated.  
In our case studies, we then saw how there are  \textit{four} models which are specifically relevant to analogue quantum emulation: 
first we have the source system model (or lab model) that provides a more detailed and experimentally embeddable description of the source system; 
second we have the source simulation model that has a limiting relation to the source system model in a given parameter regime; 
third we have the target simulation model which both provides a representation of the target system and is at least partially isomorphic to the  source simulator model; 
finally we have the target system model that provides a more detailed description of the target system. 

Viewing analogue emulation as \textit{quadruply} mediated by models in this way gives us a clear basis to assert how emulators can explain -- or contribute to explaining -- a target physical phenomenon. That is, the explanation is in virtue of formal features that are shared between the target simulation model and source simulation models, and which can be linked to the physics of the physical source and target systems via the relevant system models. An explanation of different realisations of the generic physical phenomenon is provided by the formal features that are shared between the target simulation model and source simulation model in combination with our knowledge of the de-idealisation conditions that control the relation between these models and the relevant system models. 

In the case of explanations based upon analogue quantum computations we drew the comparison with mathematical explanation and the accompanying philosophical literature. In the case of explanation based upon analogue quantum emulation the relevant comparison is even more direct. The explanation we are considering is clearly a species of \textit{model-based} explanation. Following \citet[p. 104]{Bokulich2017}:
\begin{quote}
Model-based explanations are explanations in which the explanans appeals to certain properties or behaviors observed in an idealised model or computer simulation as part of an explanation for why the (typically real-world) explanandum phenomenon exhibits the features that it does. 
\end{quote}
The category of a model-based explanation is a fairly broad one, and various further specialisation to different contexts can be given. Our purpose here is not to survey this fascinating literature, but rather draw connections between the mode of model-based explanation found in the context of analogue emulation and existing treatments.\footnote{For more on model-based explanation, the role of de-idealisation and the attendant philosophical debates see \cite{cartwright:1983,elgin:2002,craver:2006,morrison2009,kaplan:2011,rice:2015,jebeile:2015}. For a rich and nuanced discussion of idealisation in the context of understanding and the aims of science which focuses on the role of `causal patterns' see \cite{potochnik:2017}.} 
In this vein, one direct connection is to a core aspect of the account of \citet{mcmullin:1985}, in that on our analysis what is key to the explanation provided by an analogue quantum emulation is a controlled  de-idealisation procedure. 
Moreover, our account of the explanatory power is also clearly in the same spirit as the `bottom-up' species of model-explanations identified by  \citet{bokulich:2008,bokulich:2011}. 
Finally, in the Hawking case study in particular, which we will consider shortly, there is a further direct connection to the account of `minimal model' based explanation via universality arguments discussed by \citet{batterman:2002} and \citet{batterman:2014}. 

These similarities duly noted, as a case of model-based explanation analogue quantum emulation is importantly different from the standard cases considered in the literature. This is for four reasons: (i) the de-idealisation procedure we are considering is between models, not a model and a target system; (ii) there are two \textit{different} de-idealisation procedures involved in the source and target; (iii) the explananda are distinct realisations of a generic physical phenomenon; and (iv) there is a crucial role played by an experimental realisation of the source system within the explanation of the target phenomenon. 

This is not to say that we would like to present analogue quantum emulation as a counter-example to existing accounts of model-based explanation. Nor are we asserting that this case proves insightful or decisive to the much discussed topics of how false models explain or the role of causes, mechanism and structures in explanations. Rather, our point is simply that, in the context of analogue quantum emulation, we evidently are dealing with a form of model-based explanation and, in this context, there are attendant novel philosophical and methodological questions in better understanding how the relevant explanations function that are largely tangential to existing debates. In what follows our focus is to lay bear the specific features of model-based explanation in the context of analogue quantum emulation. We will return to the connections to other forms of model-based explanations found in science in Chapter \ref{ch:method mapping}. We will provide a detailed analysis of the explanatory condition in the context of our two emulation case studies in the next chapter. 

Our focus on model-based explanation points to the following question: what additional value does the experimental side of an analogue quantum emulation have to offer as opposed to a pure model-based explanation? Ultimately to answer this question fully we need to consider the details from our case studies, which we will do in the following chapter. In the meantime, we will point to what we take to be a general feature found in many cases of analogue quantum emulation: the role of the emulator system in mediating understanding rather than explanation. From this perspective the key scientific virtue of analogue quantum emulation is not tied to the superior explanatory power over pure model-based explanations, but rather to us being able to directly \textit{observe} how the explanandum follows from the explanans. 
In such emulations this is not a matter of explanatory power but much rather of the cognitive grasp of the explanation under question. 
Analogue emulation can help better \emph{understand} the target phenomenon. In the simple view, this is captured by the notion of `grasping' which is the focus of the next section.

\section{The grasping condition}
\label{sec:grasping condition}

A key element of the simple view of understanding is the idea that the subjective component of understanding can be subsumed under the notion of `grasping'. 
Strevens uses the word `grasping' to articulate the epistemic accessibility condition.
But what does it mean `to grasp an explanation' of a phenomenon?
In Strevens' original account, grasping is (rather unsatisfactorily) posited as primitive: grasping is `a fundamental relation between mind and world, in virtue of which the mind has whatever familiarity it does with the way the world is' \cite[p.\ 511]{Strevens2013understanding}. Strevens thus adopts a primitive notion of epistemic accessibility in terms of a `subjective component' \cite[p.\ 122]{Bailer-Jones-1997} of understanding that cannot be reduced further.\footnote{In a more recent paper, \citet[pp. 19-20]{strevens:2016} suggest, rather cryptically, that the view that `grasp bottoms out in cognitive capacities' entails that `my grasp of certain basic facts is constituted by my facility in making inferences about, or using, those facts'. However, this view `gets the order of dependence precisely wrong: it is the capacities that are grounded in the epistemic state'. He then provides a marginally less gnomic sketch of a positive account in which `to grasp a fact is like knowing the fact, but it involves a more intimate epistemic acquaintance with the state of affairs in question'.} 
According to this view, all that we need to -- and in fact can -- say about this relation between mind and world is that it must be present for an individual to have understanding. 
This is somehwat unsatisfactory from our view point and does not suffice to fully account for the epistemic value of analogue quantum simulations. 
Let us therefore provide two suggestions for how the grasping condition may be spelled out. 

A specific suggestion for how to spell out the grasping condition from a philosophical perspective has been put forward by \citet{Regt-Synthese-2005}, who offer an account of understanding according to which `a phenomenon $P$ can be understood if a theory $T$ of $P$ exists that is intelligible (and meets the usual logical, methodological, and empirical requirements)' \cite[p.~150]{Regt-Synthese-2005}. 
In their account `intelligibility' plays the role of Strevens' grasping condition.\footnote{We follow \citet{ToyModels}.}
They define a theory $T$ as being intelligible for scientists `if they can recognise qualitatively characteristic consequences of $T$ without performing exact calculations' (p.~151). 
The intuition behind their definition of intelligibility is that `in contrast to an oracle (...) we want to be able to grasp how the
predictions are generated, and to develop a feeling for the consequences the
theory has in concrete situations' (p.~143). 
Their example for such a state of affairs is the kinetic theory of gases which they deem intelligible for physicists with a certain background. 
Such physicists, so the argument goes, are able to infer qualitative consequences of the theory such as the following: 
`if one adds heat to a gas in a container of constant volume, the average kinetic energy of
the moving molecules--and thereby the temperature--will increase' (p.~152). 
Notice that De Regt and Dieks posit intelligibility as a property of the theory or model to be grasped rather than as a goal that individual scientists need to achieve. 
Needless to say, like any other philosophical account De Regt and Dieks' notion of intelligibility also comes with its own pros and cons. 
For instance, one could argue that the argument about the kinetic theory of gases remains on the qualitative level, clearly one must \emph{have gone through} exact calculations at some point, or at least have the functional relationship between kinetic energy and gas temperature in mind. 
Nonetheless, suffice it to say here that their account constitutes a prominent example of how the grasping condition might be spelled out. 

But one might also take another, more naturalistic, route to account for the subjective component of understanding following for instance~\citet{Bailer-Jones-1997} and \citet{ToyModels}.  
On this view what physical processes underlie grasping can be -- and should be -- studied via cognitive science. 
For example, that a scientist grasps a model could mean that they construct a
corresponding `mental model' using which they can reason about the target
\cite{Bailer-Jones-1997,bailer:2009,Hangleiter-2014}. 
Mental models
are a model for cognitive processes in cognitive science, in particular, reasoning and knowledge representation which are used to simulate and reason about real-world or abstract objects~\cite{Nersessian:1999}: 
``Broadly construed, (...) a mental model is a structural analog of a real-world or imaginary situation, event, or process that the mind constructs in reasoning.'' \cite[p.~11]{Nersessian:1999}. 
Indeed, there seem to be many parallels between the idea of mental model representations and mathematical models that we use to reason about physical processes and phenomena. 
Those parallels might be understood in accordance with the so-called ``common-coding hypothesis'' or the idea of embodied cognition ~\cite{Chandrasekharan:2009} stating that there is a shared representation between the execution, perception and imagination of movements in the brain~\cite{Hangleiter-2014}. 
On this account, to build a grasp of a model while devising and using it in scientific reasoning is achieved by constructing mental tools using which scientists can reason intuitively about a target system. 

What is important for the philosophical discussion, is that there exists a
notion of grasping on which individual scientists can introspectively reflect. 
Once again, think of one of the most basic and important physical models: simple harmonic motion. 
It seems to us unquestionable that any basic training in physics involves the process of `grasping' the model of simple harmonic motion in the sense that the student acquires some form of mental model corresponding to a pendulum-like process. 
The acquisition of such a model is something we take to be both introspectively available to individuals and (in principle) externally analysable via cognitive science.

On this account, if a scientist wants to obtain understanding of a phenomenon $P$ via an analogue quantum simulation, the quantum simulation has to permit them to grasp the processes bringing about the phenomenon $P$. 
Plausibly, grasping requires at the very least that the dynamics of the simulator be observable in sufficient detail and manipulable to a sufficient degree. 
This is because only by being able to observe and manipulate the processes pertaining to $P$ can the scientist obtain the mental grasp of $P$ that is required for understanding. 
As we will argue below, this mental grasp of a phenomenon constitutes an additional value that analogue simulation can deliver as compared to understanding via a theoretical model alone. 
Leading researchers in the field of quantum simulation in fact express this sentiment. 

\begin{quote}
So there are analogue simulation experiments where I would say:
`The theory is actually known.' 
But the insight gained from the analogue simulation is that we understand the phenomenon better.  
I believe that by seeing a phenomenon with new methods, from a new perspective, one learns to
understand the phenomenon better. 

That is my experience. 

The theoreticians then need to develop new methods. 
And thus you learn again something about a phenomenon you believed understood. 
\flushright \emph{Immanuel Bloch, Interview in December 2016}
\end{quote}

We take it that giving students the chance to grasp theoretical concepts and models is also a major motivation for laboratory classes in undergraduate physics courses: 
when performing an experiment, previously known in theory, students are able to actually observe those key phenomena and manipulate the experimental systems. 
In such a way, they can obtain an intuition about why and how the phenomenon comes about by experiencing which knobs on the experimental system affect the phenomenon in which way. 
Conversely, they might learn what a prediction of a theoretical model precisely means and in which sense that model represents real physical systems. 
Plausibly, this is precisely what it means to obtain an understanding of the physics underlying those phenomena.
Such a process might well be understood in terms of mental models. 

While laboratory courses are a neat example for grasping models in cases in which phenomena are well understood and working models developed \emph{by other scientists} to begin with, the role of analogue simulations becomes particularly interesting in cases in which little or no understanding of the underlying physics is present within the relevant scientific community. 
This aspect is also reflected in the quote above that touches on the interplay between theory and experiment. 
Only by actually observing and manipulating a phenomenon with new methods and from different angles can certain of its features be brought to light.\footnote{This aspect of our account is naturally aligned with the discussion of the difference between explanation and understanding recently provided in \cite{Dellsen:2021}. In particular, remarks that `understanding brings with it cognitive benefits other than explanation...chief among these are manipulation and prediction' (p. 12) are in a very similar spirit to what we have in mind here.}

Importantly, in analogue simulation scientists are able to transfer their intuition about processes in the highly controlled and well understood source systems to the target system.
In this respect, both source systems that are entirely \emph{different in type} (as in the example of hydrodynamic or photonic simulations of Hawking radiation in black holes) and source systems that are \emph{similar in type} (as in the example of cold atoms in optical lattice as simulators of condensed-matter systems) may be advantageous for different reasons. 
In the former case, analogue simulations allow one to transfer knowledge and intuition about a variety of different source systems to the target phenomenon. 
This allows us to both find parallels between the different realisations of the same phenomenon and single out specifics to particular source systems. 
On the other hand, if the source system is of similar or even the same type as the target system, one can exploit common mechanisms. 
For example, in both systems noise processes may be very similar and thus the analogue simulation may be particularly faithful in a way that goes beyond what can reasonably be modelled \cite{Cubitt:2017ti}. 
This may well be the situation that we face in the context of cold-atom experiments as analogue simulations of condensed-matter systems (Chapter~\ref{ch:cold atoms}).

For an example of the former type of heuristic, take the case study on emulating Hawking radiation (Chapter~\ref{ch:hawking}). 
There, we saw how thinking about an analogue source systems can help to isolate generic features of the target phenomenon. In particular, we saw how Hawking radiation can be understood as resulting from mixing between positive and negative norm modes and that   the thermality of the spectrum crucially depends upon an exponential red-shifting between near the horizon region and the distant future region. Furthermore in analogue black holes, in contrast to real black holes, both sides of the horizon are experimentally accessible and thus we can experimentally probe and explore important theoretical concepts such as entanglement between outgoing Hawking modes and in falling partner modes across the horizon. Furthermore, it is also possible for us to emulate time reversed `white hole' systems. Analogue systems can thus be a platform to `play' with the physics of black holes. 
Observing similar effects when experimentally manipulating various analogue black hole systems may not only further bolster our confidence in the robustness of Hawking radiation but also help scientists to better grasp the mechanisms that are presumably at play in real black holes -- emulation provides enhanced cognitive access to Hawking radiation phenomena. 
In this context, the intuition developed when dealing with real, noisy physical systems as opposed to ideal models is crucial. 

This leads us straight on to another example of the same type (Chapter~\ref{ch:enaqt}): in the case study on emulating environment-assisted quantum transport (ENAQT), we saw how experimenting with the analogue system can help scientists grasp the phenomenon at play in realistic environments as opposed to idealised model scenarios. 
As an example, we saw that in an analogue simulator one could tune and manipulate noise processes (Markovian vs.\ non-Markovian noise) in an analogue system and observe the effects thereof.
This allowed the experimenters to identify regimes in which the quantum transport phenomenon can maintain long coherence times as opposed to other regimes in which transport is quickly suppressed. 
Not only do such observations increase our grasp of the processes at hand but they also help guide experiments in the respective target system, in this case, real biological systems, as well as the search for explanations of certain target phenomena. 

We therefore take it that observability and manipulability of the source system in a quantum simulation can plausibly function to permit scientists epistemic access not only to 
an abstract target simulation model, but also how this model represents real target systems. We can thus see how  analogue quantum simulations can contribute to establishing the grasping condition, which is a prerequisite for understanding of target phenomena.\footnote{This way of thinking about grasping via the observability and manipulability of a source system might be thought of in terms of an `active' mode of understanding analogous to the `active' notion of explanation discussed in a different context by \cite{Evans:2020b}.}

\section{The veridicality condition: the modified simple view}
\label{sec:modified simple view}
\label{sec:veridicality condition}

The third and final ingredient in the simple view of understanding is the veridicality condition -- the explanation of the phenomenon has to be true. 
As for the other two conditions, it does not commit to any particular account of truth. 
This permits the flexibility required to analyse analogue computation and emulation with respect to their capacity to yield understanding of their targets. 

Remember from the preliminary analysis of Section~\ref{sec:computtaionempulation} that the difference between analogue quantum computation and analogue quantum emulation was characterised via a distinction between the possible `targets' about which a scientist wishes to gain understanding. While the objects of study in an emulation are real physical phenomena, in an analogue computation the intended goal is to learn about formal properties of a target mathematical model. Clearly, there are very different roles of `truth' at play in the different cases. 

In an analogue emulation we are aiming at truth in the sense of true descriptions of physical phenomena. Significantly this is truth with regard to physical phenomena rather that truth simpliciter. Thus for our purposes if an analogue emulation `saves the phenomena' such that it is \textit{empirically adequate}, then it is veridical in the sense that is relevant for our purposes. In this sense our approach will be broadly empiricist in nature \cite{vanFraassen:1980}. We will talk interchangeably about truth in this sense and empirically adequacy in what follows. 

In contrast, in an analogue computation a much more abstract sense of `truth' plays a role, namely, that the outcomes of the computation should be correct in that they are logically implied by the target model. Clearly such a notion of truth is not tied to empirical adequacy since it does not relate to phenomena. Rather, it will typically relate to quantitative or qualitative formal properties derived from a mathematical model. We will talk interchangeably about truth in this sense and derivation or implication in what follows

In sum, the veridicality condition can be implemented \textit{either} in the context of true explanations of formal properties of a target mathematical model or true explanations of physical target phenomena. In the first case the condition can be understood in terms of the derivability of formal properties in the second case to the empirical adequacy of a model of physical phenomena. A full analysis of this subtle distinction would take us into a rather lengthy digression into the much contested analysis of the nature of truth, without ultimately gaining much towards our core purpose. Our aim in what follows will always be to interpret the veridicality condition in the manner most natural to the scientific context and, as such, we take the rather thin characterisation just provided to be a feature rather than a fault. 

The most significant aspect of the veridicality condition within the simple view of understanding is that it allows us to introduce so-called modalities of explanation:  we can import the distinction between `how-actually' and `how-possibly' explanations into the context of understanding.\footnote{For more on how-actually and how-possibly explanations in general see \cite{dray:1968,hempel:1965,Reiner:1993,Forber:2010,Bokulich2014,Bokulich2017,Zuchowski2019}. For a discussion specifically related to quantum computation see \cite{Cuffaro:2015}. For a discussion in the context of so-called analogue illustrations -- which bear important similarities and differences to analogue simulations -- see \cite{Evans:2020b}.}  
That is, whether or not the explanation via which $S$ understands is true allows us to distinguish between: (i) how-actually explanation that is required to be true; and (ii) how-possibly explanation is not required to be true.\footnote{The how-possibly vs.\ how-actually explanation distinction, and thus the how-possibly vs.\ how-actually understanding distinction, could plausibly be taken to come in degrees. How to make this idea concrete is not entirely clear however. Options include i) putting how-possibly and how-actually explanations on a spectrum with how-actually explanations having a strong evidence for their truth and how-possibly explanations having a weaker evidence for their truth \cite{Resnik:1991,Brandon:2014,Forber:2010,Bokulich2014}; ii) introducing the idea of degrees of truth \cite{smith:1998}; or iii) introducing some notion of probable truth. For our purposes a binary distinction will prove most insightful and we will not explore such ideas further here.} 
Following~\citet{ToyModels}\footnote{For the sake of 
clarity, here, we neglect the contextual nature of understanding.} one can refine the simple view of understanding accordingly. 

\begin{enumerate}
	\item A scientist $S$ has \textit{how-actually understanding} of phenomenon $P$ via model $M$ iff model $M$ provides a how-actually explanation of $P$ and $S$ grasps $M$.

	\item A scientist $S$ has \textit{how-possibly understanding} of phenomenon $P$ via model $M$ iff model $M$ provides a how-possibly explanation of $P$ and $S$ grasps $M$.
\end{enumerate}  

Let us use understanding via toy models \cite{ToyModels} as an example to illustrate the binary distinction between how-possibly and how-actually understanding.

Toy models are highly idealised and simple tractable models.  They can be contrasted with realistic models that involve a large number of modelling assumptions and parameters used to more accurately describe some concrete target system. Compare, for instance, the Ising model in physics with climate-models. One core aim of both forms of modelling is to obtain some kind of understanding of a target phenomenon \cite{ToyModels,Hangleiter-2014,Regt-Synthese-2005}.\footnote{That is not to say this is the only aim. See \cite{Sugden-JEM-2000,Niss-AHES-2011,Hangleiter-2014,sep-models-science}.} Reutlinger, Hangleiter and Hartmann \citeyear{ToyModels} argue that toy models are studied precisely because they: (i) permit an explanation of formal properties of a target model, that might in some cases be relevantly related to a true explanation; and (ii) are simple enough that individual scientists can grasp them. 
In particular, the simplicity of toy models facilitates analytical solutions or simple computer simulations using which we can gain a grasp of salient formal properties and corresponding physical phenomena. This may take the form of an intuition about `What if?' questions, or of appropriate mental models. 

Following the argument of \citet{ToyModels}, toy models can be categorised into embedded toy models, that is, `models of an empirically well-confirmed framework theory' (p.\ 4), and autonomous toy models that are not embedded. 
On this account autonomous toy models are employed with the aim of providing how-possibly understanding of formal properties of a target model. In contrast, embedded toy models are employed with the aim of providing how-actually understanding of formal properties of a target model.

An example of an autonomous toy model is Schelling's model of residential segregation \cite{Schelling-1978}. 
Schelling's model is a very simple model of segregation phenomena and in particular motivated by residential segregation in Chicago it was devised in 1968. 
The model is an `agent-based model' that relies on simple `mildly segregationist' rules according to which black and white stones can move on a checkerboard. 
The key idea behind the model is that the driving force of segregation need not be racism, but maybe mildly segregationist preferences, according to which no-one likes to be a minority in their surrounding, suffice to cause segregation. 
Arguably~\cite{ToyModels}, Schelling's model can yield how-possibly understanding of the actual segregation in Chicago in 1968. 
The model assumptions---that Chicago is a checkerboard, that residents can be reduced to their skin colour, etc.\---are too unrealistic to plausibly yield any understanding of the concrete and actual target phenomenon. 
At the same time, Schelling's model can provide how-possibly understanding of a wide range of phenomena where a group is divisible into two parts and can move spatially, and Schelling already provided numerous examples thereof: Surfers and swimmers, black people and white people, men and women all segregate in some environment.

In contrast, the Heisenberg model in physics that models electronic solid-state systems by quantum-mechanical spins on a lattice might yield how-actually understanding of magnetism based upon quantum physics and the realistic assumption that the magnetic dipole moment can be modelled as a three-component spin. 
The Heisenberg model is embedded in quantum mechanics, a well-confirmed framework theory, and there are clear routes for de-idealising it in order to obtain a more realistic model of magnetism. 
Hence, the kind of understanding we can obtain from the Heisenberg model may well be how-actually understanding about how magnetism arises in actual solid states.  
But because we cannot deduce the correct solution of the Heisenberg model using our limited computing power we do not know precisely the extent to which the Heisenberg model accurately represents magnetic behaviour as an abstract phenomenon or even real magnetic materials. \\

To summarise, our framework allows us to identify four types of understanding distinguished by the two binary distinctions of the possible targets (formal properties vs. physical phenomena) and the modalities of explanations (how-possibly vs. how-actually). Just as one may have how-possibly or how-actually understanding of physical phenomena, one may also have how-possibly or how-actually understanding of formal properties of a target model, applying the relevant notion of truth in each case. 
In Chapter~\ref{ch:assessing case studies} we will use the framework for understanding via analogue simulations developed above to assess the extent to which understanding is obtained in each of the four case studies above. 
In Chapter \ref{ch:norms} we will take a step back and link the simple account of understanding to the actual scientific practice: We will identify concrete norms that must be met in an experiment for understanding via analogue quantum computation and emulation to be achieved. 
Finally, in Chapter~\ref{ch:method mapping} we will use the four cases to define a `methodological map' that situates analogue quantum computation and analogue quantum emulation alongside traditional forms of scientific activity.

\chapter[Understanding via Analogue Q. S. in Practice]{Understanding via Analogue Quantum Simulation in Practice}
\label{ch:assessing case studies}


In this chapter we will apply the framework for understanding via analogue simulations developed in the previous chapter to our four case studies. 
Whereas the role of the case studies in the previous chapter (and the next chapter) was to illustrate the various aspects of understanding, here the case studies will be used in a more direct evaluative mode. We apply our framework to our case studies of analogue quantum computation  and emulation from Chapters~\ref{ch:cold atoms}-\ref{ch:hawking}, and examine its implications in detail.

\section{Understanding via Analogue Quantum Emulation}
\label{sec:understanding analogue emulation}

We start our analysis with analogue emulation since the application of our framework for understanding is more straightforward in this context. Recall that the two case studies of quantum emulation were environment-assisted quantum transport (ENAQT) in photonic architectures and Hawking radiation in dispersive optical media. 
In ENAQT a photonic source system is manipulated with the goal of obtaining understanding of a physical phenomenon in biological systems, namely, photosynthesis via the ENAQT mechanism.
In the case of analogue Hawking radiation the photonic source system is manipulated with the goal of obtaining understanding of a physical phenomenon in astrophysical systems, Hawking radiation in black holes. In each case, therefore, the goal of the scientists is to gain understanding of physical target phenomena rather than formal features of a model. We can apply the modified simple view of understanding and assess whether the three relevant conditions obtain in each case. Let us consider the two cases side by side and proceed by examining each condition in turn. 

\subsection{Explanation and Emulation}

Recall that model-based explanations are explanations in which the explanans appeals to certain properties or behaviours observed in an idealised model as part of an explanation for why the explanandum phenomenon exhibits the features that it does \cite[p. 104]{Bokulich2017}. Are the scientists carrying out quantum emulations in our ENAQT and Hawking radiation case studies providing model-based explanations of the relevant target physical phenomena? If so, then the explanation condition holds. 

We take the analogue simulation of Hawking radiation case study to demonstrate the potential power of model-based explanation based upon an analogue quantum emulation. 
Recall that, partially based upon exploration of analogue black hole platforms, the key theoretical mechanism thought to underlie Hawking radiation can be identified as the mixing between positive and negative frequency modes in the presence of a horizon that exponentially red (or blue) shifts these mixed modes as seen at late times. Furthermore, recall that from this perspective Hawking radiation is assumed to be a `universal' geometric effect that is independent of the particular micro-physical realisation. 
What we take to be the most physically insightful available \textit{explanation} for the generic Hawking phenomena is thus that based upon the \textit{model} of positive-negative mode mixing. 

This explanation is by design general enough to be applicable to both the astrophysical and fibre optical Hawking effects since it is based upon a formal feature that is shared between the relevant target simulation and source simulation models.
Observing the full spontaneous Hawking effect in an analogue fibre optical analogue system would then provide evidence that the explanation provided by the source simulation model is true, and therefore help establish the veridicality condition.
Indeed, very recently, \citet{kolobov:2021} reported the observation of the full spontaneous Hawking effect in a BEC system, providing an example of an experiment that establishes the veridicality of the source simulation model. 
To the extent that we are convinced of the successful achievement of spontaneous Hawking radiation, we can therefore be convinced of the truth of the source simulation model.

Moreover, we have good reasons, based on the universality arguments, to believe that the explanation is robust under the different de-idealisation procedures necessary to link these models to the relevant system models that describe black holes and fibre optical horizons. 
Thus, the analogue quantum emulation can provide an explanation of Hawking radiation as realised in both fibre optical platforms and astrophysical black holes. 
The explananda are concrete, physical phenomena and the explanans is the relevant set of models together with the de-idealisation procedures. 
This is clearly a model-based explanation of the Hawking phenomena in precisely Bokulich's general sense of the term. 

In contrast, for the case of ENAQT, we have a model-based explanation of the relevant target phenomena in the system that is far from proven. 
In particular, we might plausibly believe that we are not probing the same physical mechanisms in the photonic simulator as are at play in the biological target system. Clearly, the photonic simulator does provide exceptional levels of control that enable us to dynamically observe the onset of ENAQT.  For example, we can vary the noise strength and observe transport efficiencies within different regimes, from the low noise Anderson localised regime to the high noise quantum Zeno regime.  We can also observe how quickly these transitions emerge and how sensitive they are to the particular noise model. Moreover, the calculation of these properties (\textit{viz.} transport efficiency between given chlorophyll molecules) may either be inefficient using classical computers or not amenable to analytical methods. 

However, crucially, in the case of ENAQT, although we do have good understanding of the de-idealisation procedure that links the system and simulation models of the source waveguide system, that is precisely what we do not have for the system and simulation models of the target biological system. This means that in this case, while we might consider explanation of physical phenomena wherein the underlying mathematical model (explanans) is that of a tight-binding Hamiltonian subject to some environmental noise and the physical phenomena (explananda) is excitation transport in photosynthetic systems, the modal strength of this explanation is unclear without better basis to justify the belief that the phenomena inferred in the target simulation model are robust under de-idealisation to the target system model. Furthermore, in the case of ENAQT it is not clear that there is an additional explanatory value of the analogue quantum emulation, as opposed to a pure model based explanation.

In general, the model-based explanation provided by emulation occurs in virtue of four models and their interrelation. That is, the target model-source simulation model relation in combination with the story regarding de-idealisation that connects the source simulation model to the source system model and the target simulation model to the putative target system model. In the case of ENAQT, only one side of this story is complete. The source simulation and system models do provide an explanation for ENAQT in Hamiltonians of the form $H^\text{WG}_S$ --- we dynamically observe the onset of efficient quantum transport as a function of the applied noise. 

With regard to the target system, however, the explanation is less clear. This is because it is not yet known under what conditions the idealised target simulation model provided by the nearest-neighbour Hamiltonian $H_T^{\text{FMO}}$ best approximates the target system model describing an actual photosynthetic complex. This question is the subject of vigorous biochemical research directly on the photosynthetic system itself. However, what the emulator can offer is a platform to implement hypothetical noise models and test whether the physical phenomena still persists. This, in turn, may guide the biochemical research by determining particular parameter regimes to look for the phenomena: for example, can ENAQT be observed under a Markovian or non-Markovian noise model?  What is the strength of the noise that is required to observe ENAQT and is this plausible in a biological setting? So while the form of explanation in the photonic emulation is more in line with explanation due to computation (whereby we explain properties of an abstract target model), the explanatory scope of the emulation certainly has scope for expansion, and by directing biochemical experiments, may contribute indirectly to an explanation for efficient photosynthetic transport.

\subsection{Emulation and Veridicality}

As noted in the previous chapter in an analogue emulation we are aiming to provide true explanations of physical target phenomena. Can we support such a truth claim in the case of the emulation of Hawking radiation? That is, what evidence can we provide that the model-based explanation of Hawking radiation is veridical as opposed to merely hypothetical?

The first crucial step towards veridicality in this context is leveraging the full power of the universality arguments. In particular, it has been argued by \citet{dardashti2015confirmation,Dardashti:2019}, that in the context of analogue emulations of Hawking radiation the existence of the universality arguments mean that evidence in favour of the simulation model of the source system can thus be used to make inferences about the target system. The argument has been formalised in terms of Bayesian confirmation theory wherein the role of the universality arguments is understood in terms of support for a background assumption that is common between the source and target simulation models and, crucially, which relates to the robustness of the Hawking phenomena under de-idealisation from these simple models to the source and target system models.\footnote{This means that there is a binary variable that can be assumed to be positively correlated with the empirical adequacy of both the source and target simulation models.}

It is in this context that the role of the experimentally realised emulator system becomes crucial. In particular, following the arguments of \citet{dardashti2015confirmation,Dardashti:2019}, it is evidence from the experiment on the source system that supports the existence of astrophysical Hawking radiation,  given the plausibly of the universality arguments.\footnote{The structure of these arguments closely parallels inferences scientists have recently started to apply in the context of other cases of analogue quantum emulation. In particular, a range of recent experiments have shown the ultracold atomic systems far from equilibrium exhibit universality in which measurable experimental properties become independent of microscopic details \cite{prufer:2018,erne:2018,eigen:2018}. Most vividly one of the experimental teams claims that one may use this universality to learn, from experiments with ultracold gases, about fundamental aspects of dynamics studied in cosmology and quantum chromodynamics \cite{prufer:2018}.} That is, we can establish a confirmation relation between the experimental evidence for the source phenomenon and the existence of the target phenomenon given the universality arguments \cite[Theorem 1]{Dardashti:2019}. 
If such strategies for confirmation via analogue simulation are viable then the veridicality condition can be established using inductive means, as per any other model-based explanation which we believe to be true in virtue of a successful experimental demonstration. 
This suggests a direct connection between establishing the veridicality condition in the context of an analogue quantum emulation and the notion of validation in the context of the epistemology of experimentation. 
We will return to this connection in the next chapter.   

 It is worth pausing briefly here to consider possible sceptical doubts regarding this chain of inferences just outlined.  In particular, if one is sceptical regarding the mode of inference between source and target systems based upon universality arguments, then the account just given would appear to have all the benefits of theft over honest toil. Such a sceptic has been given no further reason to change their mind.\footnote{This feature of the arguments for `confirmation via analogue simulation' \cite{dardashti2015confirmation,Thebault:2016,Dardashti:2019} has lead to a dispute in the literature, with \citet{Crowther2019} arguing for supposed `vicious circularity' in the chain of reasoning. Analysis due to \citet{evans:2019} in turn suggests that the argument of \citet{Crowther2019} can only be consistently sustained at the cost of collapse into a general (and implausible) form of inductive scepticism.} However, in this context doubt may be mitigated, by performing multiple independent emulations on platforms of different material constitution to gain trust in the empirical adequacy of the universality claim \cite{evans:2019}. Such additional experiments provide an independent line of inductive evidence for the inferences in question.\footnote{Formally, such multiple sources of confirmation can also be modelled in Bayesian terms and results in a number of interesting features such as the saturation of the confirmation measure after some finite number of distinct emulations, see \cite[Theorem 2]{Dardashti:2019}, and, rather surprisingly, an effect whereby quantum emulations in which we are more confident about the fundamental physics of the source system provide weaker confirmation of the model of the target system than those about which we are less confident. An important outstanding issue is the whether there are limits on the degree of confirmation that is possible for exotic target systems such as black holes. See \cite{Field:2021}.}

This suggest a general strategy for supporting the veridicality of emulations by building multiple distinct types of source system. Such a strategy is certainly not guaranteed to succeed in that, for example, future analogue emulations of Hawking radiation on different platforms may indeed fail. In such circumstances the evidence for the veridicality condition obtaining would be substantially undermined.  

Let us then turn to the photonic emulation case study where the evidence for the veridicality is already weak. In such circumstances, we are limited in the strength of explanation we can reasonably claim.  Without the veridicality condition, we transition the type of understanding we can hope to glean from \textit{how-actually} understanding to \textit{how-possibly} understanding. That is, while we have a plausible model for a photosynthetic complex and thus the emergence of ENAQT, the devil is in the details: the relevant energy and time scales of the various biochemical processes will be critical in determining whether ENAQT is possible. Once more, in this context, it is not clear that there is an any additional explanatory value of the analogue quantum emulation, as opposed to a pure model based explanation. 
If the experiment cannot support the veridicality condition then what was the point of performing it as opposed to simply appealing to the target simulation model \emph{in abstracto}? 
It is at this point that the grasping condition becomes crucial to appreciate the value of analogue quantum emulation for model-based understanding.

\subsection{Emulation and Grasping}

Finally, we need to consider whether the scientists carrying out quantum emulations in our ENAQT and Hawking radiation case studies are able to  grasp the model explanation employed to understand the target phenomenon. Does the grasping condition hold? 
In both ENAQT and Hawking cases, the photonic emulation platform allowed a large flexibility with regards to the parameter regime in which the respective phenomena could be manipulated and probed. The photonic platform also allowed scientists to transfer intuition about the well-understood processes in the photonic source system to the target phenomena. Such trasnfer is crucial in the theory-experiment dialogue, allowing scientists to construct better models and explanations for phenomena.

The grasping condition is particularly relevant in the case of ENAQT, where the phenomenon itself (transport efficiency) is intrinsically a function of a dynamically tunable parameter (noise strength).
Moreover, ENAQT is highly sensitive to both the strength of the noise and the particular form of the noise model (i.e. Markovian or non-Markovian).
If the noise is too weak then the excitation becomes Anderson localised.  If the noise is too strong then the excitation becomes suppressed due to the quantum Zeno effect.
Directly manipulating this noise in a photosynthetic system (i.e. the target) is incredibly challenging.
In contrast, the emulator provides a direct way to explore the dynamic effect of noise in a way direct manipulation of the target system cannot.
One can observe how quickly ENAQT is onset as a function of noise strength, and how sensitive it is once it occurs.  
This builds the experimenters intuition for ENAQT. 
It enables them to apply such models to photosynthetic systems specifically, and other physical systems, such as electron transport in solid-state systems, more generally.

Similarly, in the case of the emulation of Hawking radiation, the theory-experiment dialogue provides scientists with means to construct better models and explanations for those phenomena. For example, as we have already discussed in Section~\ref{HRheuristics}, the study of analogue black holes has allowed scientists to build up intuitions regarding ideas such as modelling the breakdown of the semi-classical model via a modified dispersion relation, construction of measures of cross horizon entanglement, and studying the role of negative-positive norm mode mixing. All of these theoretical insights into astrophysical Hawking radiation built upon scientists ability to `grasp' aspects of Hawking type phenomena via analogue experiments.\footnote{The following quote from a recent paper further evidences this claim: `Although in analogue gravity the back-reaction of the quantum fields on the acoustic geometry would not follow the Einstein equations, the observation of evaporation in analogue black holes could provide information over the interplay of quantum fields and classical geometry within a semiclassical scheme in a broad class of quantum gravity scenarios. If one sees this analogue gravity system as a toy model for emergent gravity, it could then give a physical intuition of how Hawking radiation can be pictured as a feature emerging from an underlying full quantum theory'. \cite{liberati2020back}.}

To summarise, in the examples of analogue quantum emulation the relation between the observable and manipulable source system and the target physical system is exploited in order to obtain a scientific understanding of the physical target phenomenon. 
Emulation is, in this regard, much like experiment: it is an inferential tool aimed at probing a class of physical phenomena by manipulating a concrete system as a surrogate for that class.
We can therefore say that the analogue quantum emulators are employed with the aim of providing how-actually understanding of physical target phenomena. 

\section{Understanding via Analogue Quantum Computation}
\label{sec:understanding analogue computation}

The case of analogue quantum computation is somewhat more ambiguous. Recall that the direct target of the simulation is formal properties of the target simulation model. A target system model is not typically involved in the relevant chain of scientific reasoning.  Indeed, typically, target models of analogue quantum computations are rather simple models comprising many degrees of freedom (such as the Bose-Hubbard model). 
However, often even these highly idealised models cannot be solved using standard tools of inference -- analytical calculations or numerical computations using classical computers. 
Therefore, novel, non-deductive tools of inference such as analogue quantum computation are required. 
It seems very plausible to say that analogue quantum simulators, like the cold atom simulator considered in our example, are employed with the aim to obtain how-actually understanding of formal properties of the target model, such as the 2D Higgs mode in $O(2)$-symmetric field theories and the existence of many-body localisation.
Evidently the targeted function of analogue quantum computations has much in common with that of classical computer simulations.

As above, let us go through the three conditions in the (modified) simple view -- explanation, grasping and veridicality -- in order to argue that how-actually understanding of formal properties of a target model is indeed the epistemic aim researchers seek to achieve when performing analogue quantum computation. 
As before, we consider the two case studies on analogue quantum compuation side-by-side.

\subsection{Explanation by computation}

First, it is clear that in both of the cold-atom case studies, the goal of the experimenters was to answer a question regarding the formal properties of their respective target. 
The questions asked were of the form: Does a certain property $P$ obtain in the target model in two dimensions? 
In both of those cases the answer to this question was known for another dimension; one dimension for MBL and three dimensions for the Higgs mode respectively. 
The open theoretical question was then whether certain types of coupling to excitations in the system would destroy the relevant phenomenon. 
In the case of MBL the excitations would be due to the interaction between the particles, providing the required energy to overcome the destructive interference phenomenon that constitutes MBL. In the case of the Higgs excitations the issue was whether coupling to longitudinal (Goldstone) modes would broaden the Higgs excitation in a two-dimensional system. 
The experimenters' goals when performing each of these analogue quantum computations was to demonstrate clear answers to these theoretical questions, or else uncover yet another effect that would affect the emergence or absence of MBL or a Higgs mode, respectively. 

In the Higgs experiment, such effects would be witnessed by the spectral response of the system to external driving as could be measured in the experiment at hand. 
The precise shape and size would, it was hoped, then allow for the development of a more accurate theoretical treatment and hence an explanation for why the Higgs mode occurs or does not occur in a two-dimensional system. 
Likewise, by observing both dynamical and static properties of transport as a local disordered potential is varied, the experimenters aimed to obtain more information about the transport properties of the target model. 
Given the target model, this would then  provide us with an explanation for how and why many-body localisation occurs or fails to occur in 2D. In both case studies the goal of the analogue quantum computation was to derive, and therefore explain in the sense in which proofs explain, formal properties of the target models in question.

\subsection{Computation and Veridicality}
Furthermore, clearly the goal of the experimenters in each case is to design their experiments in such a way that the result of the analogue quantum computation is also  veridical. Recall, that in the context of an analogue computation we are interpreting veridicality as tied to the  outcomes of the computation, that is, formal properties,  being logically implied or derivable from the target simulation model. How can we ascertain whether such a condition holds in virtue of an experiment on a physical source system? Answering this subtle question involves examining the connection between physical source phenomena, the source system and simulation models and the target simulation models. We defer detailed discussion of veridicality in the context of the crucial notion of internal and external validation to \S\ref{sec:veridicality norms}.

\subsection{Computation and Grasping}
For the time being let us focus on the third condition. 
That is, let us, consider whether or not the goal of an analogue computation in our case studies was to better grasp the phenomena under question. 
We have argued above that analogue simulations in general do indeed achieve this task in many ways. 
First, the \emph{theoretical investigation} of a potentially very different analogue system in regards to a particular target phenomenon may bring to light generic features of that phenomenon or property of a simulation model that go beyond what can be observed in a \emph{particular} target system or model.
While it may not bring to light entirely novel facts about this phenomenon or formal property, it may provide scientists with a better grasp of how a particular phenomenon manifests itself: 

First, theoretically studying a generic phenomenon in a concrete model or physical system may bring to light aspects or features that had not been thought about before. 
It might also allow scientists to better grasp that phenomenon if they are able to integrate it within their area of expertise, or a system that they already have a good physical intuition about.
It is this type of potential for grasping that seems most relevant to the case studies on analogue quantum computation in the cold-atom quantum simulator. 
In both of those studies rather generic phenomena were studied in a concrete physical setting. 
Indeed, many-body localisation is a phenomenon that is discussed in the very broad context of interacting many-body systems with random potentials. 
Concrete examples such as the Heisenberg model often serve as a playground on which that phenomenon is investigated. 
Ultimately, the question is of course, whether MBL also occurs in real physical systems, and whether it may even be exploited technologically. 
Likewise, the question about the existence of a Higgs mode was specified in the very general toy setting of an $O(2)$-symmetric quantum field theory. 
The cold-atom setup at the superfluid--Mott insulator transition is a concrete instantiation of this very general model.

Second, the \emph{experimental investigation} of the target phenomenon or formal property may bring to light features of it or effects that influence it in a way that theoretical investigation cannot. 
By confronting a theory or model with experimental investigation, one is often bound to discover new aspects of that phenomenon and thereby obtain a better grasp of it. 
This type of grasping seems, in particular, to be potentially  fruitful when making use of analogue systems that are \emph{similar in type} to the target system or model. 
Again, this is the case in the MBL case study, where the cold-atom system is a particular instance of an interacting many-body system. 
Presumably, the physical mechanisms in such a system are similar to those of the target system and hence one might also expect that previously unobserved features may provide good heuristics for the study of the target system or model. 
An example of this is the result by \citet{Cubitt:2017ti} that draws parallels between the types of noise that might occur in target and source system. 
In the words of Immanuel Bloch quoted above: `The theoreticians then need to develop new methods. And thus you learn again something about a phenomenon you believed understood.'

We thus take it that analogue simulations are indeed performed with the goal to obtain a better grasp of the target phenomenon. 
Such grasp may be reflected in heuristics for novel theory construction, a better understanding of the mechanisms underlying the phenomenon or intuition borrowed from a completely different realm of physics that helps scientists grasp the phenomenon. \\

In the following chapter, we will analyse more generally the conditions under which analogue quantum computations and emulations actually allow us to achieve the epistemic goals outlined here. 
We will make use of the general framework from Chapter~\ref{ch:understanding} as well as the concrete examples from the case studies which we discussed in this chapter. 
This will put us in a position to formulate some general epistemic norms for practising scientists who perform analogue quantum simulations in order to obtain understanding. 

\chapter{Norms for Validation and Understanding}
\label{ch:norms}


In the previous two chapters we have argued, in general and with specific reference to our case studies, that the goal pursued by scientists when performing analogue quantum computations is to obtain how-actually understanding of formal properties of the target simulation model and the goal when performing analogue quantum emulations is to obtain how-actually understanding of physical target phenomena. 
In this chapter, we will develop \textit{norms} for understanding via analogue simulations with particular reference for the problem of validating analogue simulators. Using those norms, epistemic and pragmatic claims regarding how-actually understanding via a particular analogue simulation may be evaluated. We thereby hope to provide some guidance to future research on analogue simulation.

\section{Validation and Veridicality}
\label{sec:veridicality norms}

The veridicality condition is the most difficult to establish in both analogue quantum computation and emulation. 
This is because analogue quantum emulations are often intended to be probative of a physical target system about which little is known and, at the same time, analogue quantum computations often yield results regarding formal properties of a target (simulation) model in regimes in which it cannot be definitively established that observations on the source system support the intended conclusions about the properties of the target model. 
After all, this lack of direct access to the target system and formal features of the target model using traditional means of inference is typically precisely why scientists are interested in using analogue emulations and computations in the first place. 

In this section, we will return to the problem of  establishing the veridicality condition in analogue simulations or, in other words, consider the question: 
Under which conditions can conclusions drawn based on the result of an analogue simulation go beyond mere  how-possibly to veridical or how-actually inferences regarding their target?  

The key idea for us to conduct this evaluation of the veridicality condition is \textit{experimental validation} as applied in discussions of conventional experiments and computer simulations \cite{franklin:1989,Winsberg-2010}. 
An experimental result is \emph{internally valid} when the experimenter is
genuinely learning about the exemplar system they are manipulating; an experimental result is \emph{externally valid} when the information learned about the exemplar system is relevantly probative about the class of physical systems that are of interest to the experimenters. Consequently, the preconditions for us to gain
how-actually understanding via an experiment clearly must include both internal and external validation. 

The distinction between internal and external validation will prove useful in our analysis of the conditions under which analogue quantum computation and emulation are able to support true inferences about their respective targets and therefore \emph{how-actually} understanding. 
In what follows we will break the relevant validation procedure down into two steps: 
(i) Internally validating the source simulation model; (ii) Externally validating the target-source relationship. 
Whereas the first step is common to both analogue quantum computation and emulation, the second step comprises aspects that differ for analogue computation and emulation. 
To externally validate an analogue computation one need `merely' establish the source simulation model-target simulation model relation. 
In contrast, to externally validate a (model-based) analogue emulation, in addition to establishing the source simulation model-target simulation model relation, the target simulation model needs to be shown to be an empirically adequate model of the target system. 
This is usually achieved by probing the robustness of relevant phenomena under a de-idealisation relationship with the putative target system model. 

It will prove instructive to consider the more general question of internal validation in the context of classical and quantum computation before we proceed to our main analysis of validation in the context of analogue quantum simulation.

\section{Internal Validation in Classical and Quantum Computation}

Analytical derivations or proofs are sequences of steps that deductively follow from the previous step (and the premises of the derivation). 
As such, they can be checked, at least in principle, by any one who is capable of basic logical manipulations, and, in the case of real proofs, relevant mathematical facts that are assumed in the proof. 
Deductive analytical derivation is not, however, the only form of inference which is able to yield statements regarding the formal properties of a model that are true.

Consider, for instance, analytical derivations that involve approximations. 
To the extent that the approximations made in a derivation do not influence its final outcome can we still speak of that outcome as being true? 
To make this claim in a quantitative fashion, the resulting errors must therefore be controlled. 
Indeed, in many situations rigorous error bounds can be derived and thus we can reasonably talk about the derived value approximating the true value.
Likewise, when running a discretised numerical algorithm in order to derive certain mathematical facts, due to the discretised nature of the algorithm, the outcome will often be only approximating the true value. 
However, in many cases any desired degree of accuracy that the outcome of such an algorithm satisfies can be guaranteed at the cost of an increased size or runtime of the computation. 
And this is really what limits the precision of discretised algorithms. 
Likewise numerical algorithms often make use of randomness. 
In those cases, the outcomes of the algorithm might be guaranteed to be approximating the true value with quantifiable probability. 

An important point in the context of computer simulation, which is also relevant to analogue quantum computation, is the fact that the individual logical steps performed by the computer cannot be checked by hand anymore. 
If we are to claim that the outcome of a computation approximates the true value, then we need to make sure that the computer is also functioning as intended, that is, that it actually performs the logical operations which it was programmed to perform. 
We can designate the process of ascertaining the correct functioning of the computer in this sense as the \textit{validation} of the computational model. In the case of a desktop computer that computational model comprises circuits of logical gates such as AND, OR and NOT (recall Section~\ref{sub:analog_vs_digital}) and it is the correct functioning of those gates and their conjunction which needs to be validated to obtain trust in the outcome of a computer. 
We conceive of such validation of computing hardware with respect to the intended computational model as `internal validation' of the computer in the same way as an experiment is internally valid if it indeed probes the intended features of the source system. 

Validating a quantum computation -- analog or discrete -- in the very regime in which it is classically intractable is an extremely difficult task. 
Ideally such validation would not require any a priori assumptions about the correct functioning of the device or parts of it. 
In fact, this task is possible for universal (discrete) quantum computers via elaborate schemes that involve back-and-forth communication between a user and the device under cryptographic assumptions \cite{mahadev_classical_2018}, making use of multiple quantum computers \cite{reichardt_classical_2013} or small, a priori certified quantum devices \cite{fitzsimons_unconditionally_2017,fitzsimons_post_2015}. 

For analog dynamics it is, however, not at all clear whether or not it is possible to rigorously validate an analogue computing device in a way that is feasible and efficient.
Indeed, fully characterising a quantum device via quantum tomography requires a number of measurements that scales exponentially in the size of the system.
Even then, fully characterised and trusted measurement devices are required. 
While for certain situations such as static quantum simulations in which a specific quantum state is to be prepared certification tools have been devised \cite{flammia_direct_2011,Certification}, those tools still require elaborate computing devices with a large degree of control. 

Without such control, often the best way to validate a quantum simulator is to compute the predicted outcome of an experiment according to a model of the system at hand. 
But due to the exponential size of quantum state space, even to validate very specific tasks such classical re-simulation requires exponential time in general. 
This is why, to date, all examples of analogue quantum computations have only been validated in  classically tractable regimes. 
Alternatively, one can validate the individual components of a device in order to build trust in the correct functioning of the whole.
This point will prove relevant in the below.

\section{Internally validating an analogue quantum simulator}

In order to pave the way for the philosophical analysis of internal validation of analogue quantum simulators, we need to make clear what we mean by internal validation of an analogue quantum simulator. 
To do this, let us again take a step back and clarify which relations need to be validated. 
Consider again our schemata of analogue quantum computation and emulation from the case studies (Figs.~ \ref{fig: MBL simulation schema}, \ref{fig: higgs simulation schema} (computation) and  \ref{fig:enaqt_schema}, \ref{fig: hawking emulation schema} (emulation)). 
In each of the cases, we have system models which involve the detailed and messy features of real physical phenomena and simulation models that are simplified and idealised. 

Recall that the system model vs. simulation model distinction is rooted in the fact a particular experimental system may be described on different levels: on a quantitatively accurate level it is described by a system model, qualitatively it is described by a certain simulation model in a certain parameter regime. 
In particular, given a certain simulation model there will often be several concrete experimental systems, or platforms even, whose system model approximates the simulation Hamiltonian in some limit. 

Crucial to the internal validation of the analogue quantum simulator are therefore the relations between: i) the source simulation model; ii) the source system model; and iii) the physical source system; recall Fig.~\ref{fig: source_model} for an illustration. 
Typically the relation that must be established pertains to a specific formal feature of the source simulation model, the robustness of this feature under de-idealisation to the source system model, and then both the relevant de-idealised feature in the source system model and a specific physical behaviour of the source system.  What experimenters are interested in is the stability of the relevant empirical correspondences in a wide and well controllable range of parameters.

Thus the goal in internal validation is to \textit{certify} within a prescribed error tolerance that tuning the experimentally accessible parameters corresponds to the intended variations of the simulation model's parameters in the source system. This means that the internal validation requires both: i) certification of the de-idealisation process which connects the source simulation model to the source system model; and ii)  certification of the accuracy of the source system model as a empirically adequate model of the phenomena of interest in the physical source system.\footnote{Whereas the first function of certification here broadly corresponds to the notion of `verification' applied in scientific and philosophical discussions of the epistemology of classical computer simulation, the second is closer to the notion of `validation' in that context. Within this literature the idea has developed that one  can then separate the `mathematical' process of verification from the `physical' process of validation in our epistemology of classical computer simulation. \citet[p.155]{winsberg:2010} has convincingly argued that this conceptual division can be misleading in the context of classical computer simulation since in practice the two are often not cleanly separable. We take this conclusion to be fully appropriate in the context of analogue quantum simulation and thus apply a broad notion of validation which includes the residual `verification' aspects. Certification is then the process by which internal validation is established. 
}  

It is worth noting that an alternative methodology for internal validation is that rather than the source simulation model being certified via the system model, the  source simulation model could be directly certified. We have witnessed good examples for both routes to the internal validation of a quantum simulator in our two case studies of analogue quantum computation in cold-atom systems.\footnote{
Recall in our analogue quantum computation case studies we considered two examples where scientists use an ultra-cold atom platform. 
In the first to explore the potential emergence of a Higgs mode in two dimensions and in the second to probe the existence of many-body localisation in two dimensions.} 
In both case studies the claim is that by manipulating the physical source system, i.e. the cold-atom quantum platform, it is possible to probe formal features of the Bose-Hubbard Hamiltonian. 

\subsection{Direct certification of the source simulation model}

Let us begin by analysing the first type of certification which involves directly certifying the source simulation model as an empirically adequate model of the source system. 
As we saw above, the difficulty in validating this correspondence directly heavily depends on the extent to which the source simulation model is analytically or computationally tractable. Typically analogue quantum computation is applied in scenarios when such tractability fails, in at least some parameter regimes of interest, making alternative, weaker forms of certification adequate. 

The cold-atom case studies provide clear examples of this. 
Recall, that in the analogue quantum computation of many-body localisation (MBL), it was not clear whether or not MBL would persist in two dimensions, while in one dimension those calculations could be performed using classical computers. 
In order to validate the source system internally, a first experiment
was performed in a parameter regime in which the Bose-Hubbard Hamiltonian -- the simulation model of the source \textit{and} target in  the quantum computation -- is computationally tractable on a classical computer, namely, in the quasi-one-dimensional regime. 
In this regime, the experimental measurement outcomes could be compared with classically computed predictions using the Bose-Hubbard Hamiltonian and a good match was found.\footnote{In more detail: the experiment probed the dynamic and static MBL of the imbalance parameter. 
The outcomes of the experiment in varying parameter regimes were compared  to exact numerical calculations using DMRG and exact diagonalisation in Figs.~\ref{fig:mbl imbalances}(a) and (b), respectively. 
The predictions using the Bose-Hubbard Hamiltonian were found to match the measurement data very well. 
Small deviations could even be accounted for by incorporating effects of the harmonic trapping potential, which affect the hopping parameter $J$ in a site-dependent way, into the calculation. 
Thereby the source system was validated against the target model, namely the Bose-Hubbard Hamiltonian, in the relevant parameter regime.} Thus, the Bose-Hubbard Hamiltonian was certified as an empirically adequate model of the experimental source system, constituting internal validation in the quasi-one-dimensional regime. 

In a next step, the effective dimensionality of the simulator was increased from one to two by decreasing potential barriers in the transverse direction. 
In the resulting regime -- quasi-two-dimensional -- the Bose-Hubbard predictions cannot be efficiently computed on a classical computer anymore. 
At the same time, the differences in experiments performed in one or two spatial dimensions are extremely small; the only difference is that one less confining laser beam is being used in the two-dimensional experiment. 
Hence, the experimenters argue, the experiment in two dimensions can be considered validated by the corresponding experiment in one dimension. 
At the very least, internal validation of the source \emph{in some parameter regime} is an important step towards validating the computation in the computationally intractable regime and builds trust in the correctness of the outcome.

\subsection{Indirect certification of the source simulation model}

Let us now consider an example of the second route for internal validation. The goal is to certify the source simulation model via the source system model.
For this, a chain of relations needs to be valid: 
\begin{itemize}
\item First, the source system model needs to be an empirically adequate model of the dynamics of the physical source system to the relevant degree of accuracy in the relevant regime
\item Second, the solution space of the system model must approximate the solution space of the simulation model in a well-controlled way in the relevant regime 
\end{itemize}


The crux of internally validating the analogue quantum simulation lies in establishing the first claim. 
In order to establish this claim a system Hamiltonian which provides a highly accurate description of the dynamics of the source system must be identified.
Moreover, the experimental control model must be certified, i.e., it must be shown that changes in the experimental parameters (magnetic fields, intensity and wavelength of the optical lattice) lead to the intended changes in the model parameters. 
As discussed above, doing so via the traditional means of comparing the model's predictions with measurement outcomes in the source system is infeasible, in particular, for many-body interacting models such as the Bose Hubbard Hamiltonian, and even more so for the much more complicated system Hamiltonian that in addition includes terms describing experimental imperfections. 

The second claim, in turn, is established on a theoretical level. 
Specifically, in our case studies of cold-atom analog quantum computation, the second relation generally holds in virtue of the result of \citet{Jaksch1998} that relates the source system model $H_S^{\text{CA}}$ \eqref{eq:cold atoms hamiltonian} and the source simulation model $H_S^{\text{BH}}$ \eqref{bose hubbard} and specifications of that result to the source system Hamiltonian describing the particular experimental source system at hand. 

An example of a strategy for verification in these circumstances is the experiment by \citet{Higgs:2012}, where the goal was to determine whether or not a Higgs mode existed in two dimensions. 
In this case, internal validation of the experiment was not pursued directly but \textit{indirectly} in the sense that the experimental outcomes were compared with various theoretical models, including in particular comparison of experimental data with approximate analytical mean-field \citep{Higgs:2012} and numerical Monte Carlo calculations \citep{liu_massive_2015} using the system Hamiltonian. 
In those comparisons, reasonable agreement of the experimental data with the theoretical predictions was found. 

However, while the abovementioned methods are reliable in certain parameter regimes, they are not in the regime targeted in the experiment. 
Consequently, it is unclear whether the experimental data is adequately described by the simulator Hamiltonian in the correct parameter regime. 
In this regime, the hopping parameter $J$ is homogeneous throughout the system as realised in an ideal, flat potential well. 
In particular, prior theory work had shown that a resonance-like feature in the response to external driving constitutes a `smoking-gun' signature of the Higgs mode. 
But in the experiment, such a feature was not observed.
It has been argued that the harmonic confining potential might have lead to an inhomogeniety of the hopping parameter $J$ that lead to a broadening of the resonant peak, making it indiscernible in the data \cite{pollet_higgs_2012,liu_massive_2015}. 
If such an argument holds, then here we have a case in which the experimental system is accurately described by a known system Hamiltonian as certified by matching experimental outcomes with theoretical predictions using that Hamiltonian, but not by the anticipated simulation Hamiltonian in the intended parameter regime (the Bose Hubbard Hamiltonian with spatially homogeneous hopping $J$). 
%
While the Hamiltonian that describes the experiment presumably was the Bose-Hubbard Hamiltonian in \emph{some} parameter regime and thus the second claim is partially established, it is not the \emph{intended} parameter regime that needed to be reached for the computation to be conclusive.\footnote{In particular, in the system Hamiltonian of the source system, $H_S^{\text{CA}}$, besides there being additional
terms to those in the Bose-Hubbard Hamiltonian (such as next-nearest-neighbour hopping terms and coupling to the environment), the hopping parameter $J$ is not homogeneous throughout the trap but displays a dependence on the position of the lattice sites in the harmonic confining potential. While the first type of deviation could be controlled by putting quantitative bounds on the significance of the additional terms, given the estimates of the relevant interaction parameters (see e.g., \cite{BlochEisertRelaxation}), the position-dependence of the hopping seems to have thwarted a successful computation of formal properties of the $O(2)$-symmetric quantum field theory.} 
Together with the lack of a full certification that the experiment was performed in the intended parameter regime to a degree of approximation to which the outcome would not be influenced by potential errors, we must therefore deem the outcome of the simulation indecisive for the existence or not of a Higgs mode in two dimensions. 
We can therefore conclude that the computation of the Higgs mode in two dimension was not fully internally validated in the cold-atom setup. This not withstanding, this example makes clear that there is in principle a powerful methodology for successful indirect certification of  source simulation models.
Moreover, as for analogue Hawking radiation, experiments were performed in different experimental platforms, inductively bolstering the confidence in the outcome of the individual experiments.

\subsection{Internally validating quantum emulators}

Internal validation is, of course, equally important in the case of analogue quantum emulations. 
In order to draw veridical conclusions about the physical target system, it is necessary (but not sufficient) to validate the relationship between the physical source system and the source simulation model. If, as in our case studies, emulation runs via an approximate structural isomorphism between the simulator source and simulator target models, internal validation of the source system takes much the same structural form as in the case of analogue quantum computation. 
However, since the object of interest in an analogue quantum emulation is phenomena manifested by the physical target system rather than features of the target simulation model, it is not unusual to find situations in which the source and target models are analytically or classically computationally tractable. 
This is the case for Hawking radiation where the models of the analogue and astrophysical black holes can be solved analytically, 
thus making internal validation fairly straightforward.

What form does internal validation take in the examples from our case studies? 
In the case study of quantum emulation of ENAQT, the photonic emulator is characterised using both classical and quantum techniques to determine relevant parameters such as waveguide loss, dispersion, light source parameters (brightness, photon indistinguishability) and detector parameters (dark counts, efficiency, shot noise). 
This information is fed into a system model of the emulator, which can then be validated as an empirically adequate model of the source system as it is computationally tractable. 
Provided the errors lie within some bound (a notion which can be made theoretically rigorous) we can establish the relevant approximation relation between the source system model and the simulation model. 
Similarly, in the case of the quantum emulation of Hawking radiation, internal validation requires certification of the empirical adequacy of the system model provided by the `lab Lagrangian' for the full non-linear optics regime and of the limiting relation between this model and the the `simulation Lagrangian' for the linear regime. 
Since each of these models is analytically or  computationally tractable and controllable in the relevant regime, these relations can be established via conventional experimental and mathematical techniques

\section{Externally validating an analogue quantum simulator}

Analogue quantum computation and emulation differ significantly when it comes to the question of what it means to \emph{externally validate} the quantum simulation. 
Recall that the target of a quantum computation is formal properties of a target mathematical model -- the target simulation model. 
Let us assume that the source simulation and system models are internally validated. 
In this case, we know that we can probe formal features of the source simulation model that provide an empirically adequate description the source system in the relevant parameter regime. 
Internal validation therefore warrants reliable inferences from experiments on the source system to formal properties of the source simulation model. 
To validate the inference to the formal properties of the target simulation model of interest we need to further validate the relation and correspondence between the target simulation model and the source simulation model in the relevant parameter regime. 
Making the relation mathematically rigorous might be considered `formal external validation' as it validates that the source system is indeed relevantly probative of the intended class of targets -- in the case of analogue computation, target mathematical models that need not have any corresponding physical target system. 

Given that working out such relations is often the starting point of an analogue computation, rather than constituting a norm for practising scientists formal external validation should rather be considered a precondition for performing an analogue computation in the first place. 
Nonetheless, given the analog setting it is still important to not only work out the correspondence in a given parameter regime, but also the robustness of the correspondence to variations in the model parameters as well as potential sources of noise in the source system. Such analysis has been made on very general grounds for the simulation of Hamiltonian systems by \citet{Cubitt:2017ti}. 

External validation becomes much more interesting and important for quantum emulators; recall Fig.~\ref{fig: ae_model} for an illustration. 
Again, we can draw the distinction between the target system  model which results from characterising the particular complexities of the target system without a focus on tractability, and the target simulation model, which is an idealisation of the system model involving simplifying assumptions, usually aimed towards making the model more tractable. In order to be able to draw true inferences about a target physical system from observations on the source system, one needs to validate the correspondence between the source simulation model and the target system, as mediated by the target simulation model and the target system model: 
the source simulation model must be relevantly probative of the intended class of target physical systems. 

External validation of an emulation thus requires two steps. 
First, as in analogue computation, the target simulation model must be externally validated, that is, it must be validated that the relevant formal properties of the target simulation model approximate formal properties of the source simulation model that can be probed in an experiment. 
We called this `formal external validation' above. 
Second, it must be validated that the target simulation model is an empirically adequate model of the target system in the relevant regime. 
We will call this latter type of external validation `empirical external validation'.
As in internal validation, the (idealised) target simulation model may be certified directly, or via the target system model. 
Only if both the first and the second type of validation are achieved can we hope to draw reliable inferences about the target system from an experiment on the source system. 

How do our case studies fare in this respect? External validation of the model of light harvesting complexes through the ENAQT mechanism is an outstanding challenge in experimental and theoretical quantum chemistry.  
Experimentally, 2D electronic spectroscopy is used to determine the structure of the molecule (e.g.\ the energy levels), and to observe long lived quantum coherences.  This in itself is only half the picture, as the context in which these experiments are performed is critical.
Specifically, it matters whether the cells are in their natural biological context during the experiment (in vivo) or whether they are isolated from that context (in vitro), and whether the experiment is performed at high or low temperatures.
To validate the model, computations are performed on the Hamiltonian to ascertain whether it reproduces known target phenomena. This in general will require approximations, and further work is required to establish the applicability of those. In such cases the emulator may itself help validate the empirical adequacy of the model.\footnote{In particular, it must be shown that $\tilde{H}^\text{FMO}_T$ 
 reproduces known target phenomena $P_4\cong P_T$. This requires that the approximation between $\tilde{H}^\text{FMO}_T\rightarrow {H}^\text{FMO}_T$ can be established (analytically or numerically). When this holds we have a form of self-validation for the empirical adequacy of the model.} Thus, as we have noted earlier, in the photonic emulation case study, where the target system-simulation model relation is not yet established, we are limited in the strength of inferences that the analogue quantum emulation can licence. 
 In particular, without the external validation, the veridicality condition does not hold and the type of understanding is merely \textit{how possibly understanding.} That is, while we have a \textit{plausible} model for a photosynthetic complex and thus the emergence of ENAQT, this model is not explanatory of the target phenomena in the modally strong how-actually sense.  
 
The story for the Hawking case is significantly more complicated. Experimental evidence for (astrophysical) Hawking radiation in real black holes many many orders of magnitude outside the sensitivity of current equipment and no experiment is within sight that might achieve this goal. Consequently, both the system and simulation target models \emph{cannot be certified} as an empirically adequate model of the target system (real black holes) in a direct experiment and hence external validation seems doomed to fail. 
Indeed, the goal of the (experimental) analogue emulation might be considered precisely to obtain evidence that Hawking radiation is a real phenomenon that also pertains in real black holes, that is, to \emph{confirm} (in the sense of gain inductive evidence for) the existence of astrophysical Hawking radiation \cite{dardashti2015confirmation}. 

Nonetheless, there might in fact be ways to circumvent this daunting obstacle to a certain degree. 
Recall from our earlier discussions that the key link in the chain of inferences from the source analogue black hole system to the target astrophysical black hole system was a perspective under which Hawking radiation is assumed to be a `universal' geometric effect that is independent of the particular micro-physical realisation. 
In such a context, explicit arguments for the universality of Hawking radiation (e.g. those of \citet{unruh:2005}) can be understood in terms of support for a background assumption that is common between the source and target simulation models. 
This is a very weak form of theoretical external validation. Crucial to upgrading these arguments to a more robust mode of external validation is providing empirical evidence of the universality arguments' adequacy in real physical contexts \cite{Thebault:2016} and also providing more `integrated' universality arguments \cite{Gryb:2018}. 
To provide empirical evidence for universality in the first context it is important that multiple independent emulations on platforms of different material constitution are performed. 
Such additional experiments provide independent lines of inductive evidence and thus plausibly external validation, for the inferences in question \cite{evans:2019}. We take the appeal to universality arguments in the context of analogue experiments for Hawking radiation to serve as an example of a more general methodology for the external validation of analogue quantum emulations.  \\

Let us summarise the last two sections. 
In analogue quantum computation, the part of the validation which is particularly challenging is internal validation. This is because analogue quantum computations are typically performed with a view to resolving theoretical questions that could not be resolved with classical computers or analytical arguments. 
By the same token, however, the internal validation of the empirical adequacy of the source system and simulation models of the source system in the computationally interesting regime requires methods that go beyond comparing to predictions from theoretical models.
For example, one can sometimes leverage precisely characterised local components of a quantum system to obtain a rigorous validation of the global system. 
In analogue quantum emulation, the part of the validation which is particularly challening is external validation. 
Emulations are performed with the goal to understand features of a physical target system and this is typically applied in situations in which the target system is experimentally inaccessible. 
Here, the question is typically: is the simulation model, be it classically simulatable or not, an empirically adequate description of the target system. 
Thus, the challenging part is to validate that the target simulation model is a good model of the target system. 
This can only be achieved indirectly in situations in which the target system is experimentally inaccessible, and we have seen a neat example for how such an argument might go in the Hawking case study. 

In the next sections, we will now take our analysis as a basis to formulate norms for the validation of analogue quantum computations and emulations. 
Those norms directly correspond to the conditions for understanding -- veridicality, explanation and grasping.

\section{Norms for validating analogue quantum simulators}

The findings of the preceding sections can be summarised in terms of a norm for achieving the veridicality. 
Our characterisation of this condition can be encapsulated in terms of the following norms for the validation of an analogue simulator, presented as \textit{sufficient conditions}: 

\begin{enumerate}
	\item An analogue quantum computation is  veridicial \textit{if}:
	\begin{enumerate}
	 	\item  the source simulation model is validated as an empirically adequate model of the source system across the relevant parameter regime (\emph{internal validation} of the simulator) 
	 	\item [] \textit{and if}
	 	\item formal properties of the source simulation model approximately correspond to formal properties of the target simulation model in that regime (\emph{formal external validation}).
	 \end{enumerate}

	\item An analogue quantum emulation is veridical \textit{if} 
	\begin{enumerate}
		\item  the source simulation model is validated as an empirically adequate model of the source system across the relevant parameter regime (\emph{internal validation} of the simulator)
		\item [] \textit{and if}
	 	\item formal properties of the source simulation model approximately correspond to formal properties of the target simulation model in that regime (\emph{formal external validation}).
	 	\item [] \textit{and if} 
	 	\item the target simulation model is validated as an empirically adequate model of the source system across the relevant parameter regime, either directly or via the relation with the target system model  (\emph{empirical external validation}). 
	 \end{enumerate}
\end{enumerate}

If any of the validation norms is not satisfied, we can invoke the modified simple view of understanding and still potentially obtain how-possibly understanding of the respective target of the simulation.

\section{Norms for Understanding}
\label{sec:explanation norms}

To obtain \textit{scientific understanding} of the respective target of an analogue quantum simulation the explanation and the grasping condition also need to be satisfied. Here we will review the key requirements for these conditions to hold making reference to our case studies once more. Our particular goal is to highlight the significance of manipulation and observation as gateways towards explanation and grasping respectively. 
The ability to manipulate and observe the source system to sufficient degrees is thus the second fundamental norm for scientific understanding via analogue quantum simulation. 
Finally, from a practical perspective, a further key norm is that the explanation provided by an analogue quantum computation should give us some advantage over that provided by a classical computer -- else the explanation will be scientifically superfluous. We will consider this practical norm of speed-up in the context of explanation in the next section also.   

\subsection{Explanation, Manipulation, and Observation} 
\label{sub:explanation_condition}

We have argued already that experiments on the source system of an analogue simulator can provide a form of explanation of the respective target physical phenomenon or formal property of a model (Secs.~\ref{sec:explanation condition} and \ref{sec:understanding analogue computation}). 
Key to this argument was that the dynamical evolution of the relevant observables can be tracked and thereby, an experimenter can obtain information as to how a particular property of the source and target simulation models comes about. 
Clearly, for this to be possible an experimenter must have adequate means to probe the source system in the laboratory. 

Furthermore, we also argued that it is crucial to explanations of more general properties of the target simulation model that the source system be manipulatable in relevant parameter regimes of the target simulation model. 
By manipulating the source system in different settings, the experimenter is able to obtain answers to `what if' questions, a crucial part of (counterfactual) accounts of explanation. 

An analogue simulator -- be it a computer or an emulator -- should therefore satisfy the following norms for explanation, related to observability and manipulability of the simulation system. 
We will present these together as a \textit{jointly sufficient} condition. 
\begin{enumerate}
\setcounter{enumi}{2}
	\item An analogue simulator satisfies the norms for explanation \textit{if}  
\begin{enumerate}[a.]
 	\item it allows faithful measurements of the observables relevant to the target, 
 		\item [] \textit{and if}
 	\item it allows experiments in broad parameter regimes that are relevant to the target at hand such that the functional dependence on those parameters can be explored,
 		\item [] \textit{and if}
 	\item it allows us to probe the dynamical evolution of the relevant observables to reveal how the observed phenomenon comes about. 
 \end{enumerate} 
\end{enumerate}

As already noted, from a pragmatic perspective additional value with respect to the explanation that is obtained from the simulation over a computation using a classical computer or analytical computation can only be gained if the analogue simulation allows access to properties of the target simulation model that are not accessible otherwise. 
Consequently, the simulator should have an advantage over classical computers solving the same task at hand. 
In most cases such an advantage is manifested in a \emph{speedup over classical devices}. 
Such a speedup may take on varying degrees of rigour and extent. We propose that a disjunction of the four currently available options can be consider a (tentative) necessary condition for the practical norm of speed-up:

\begin{enumerate}
	\setcounter{enumi}{3}
	\item The practical norm of speed-up obtains \textit{if and only if, either}: 
\begin{enumerate}[a.]
    \item The problem solved by the analogue simulation is proven to be strictly harder than any problem that is simulable by a classical computer.\footnote{An example (under plausible complexity theoretic conjectures) is boson sampling \cite{Aaronson2005bs}. }
    	\item [] \textit{or}
    \item The best known classical algorithms are not able to solve the problem efficiently \textit{and} the quantum simulator can scale-up to large problem sizes without sacrificing accuracy.\footnote{An example is Lloyd`s digital quantum simulator \cite{lloyd_universal_1996} and Shor's factoring algorithm \cite{shor:1997}.}
	\item [] \textit{or}
	\item The best known classical algorithms are not able to solve the problem efficiently, but it is unknown if the quantum simulator can scale-up to arbitrary problem sizes without sacrificing accuracy.\footnote{An example is the experiment by \citet{Higgs:2012} (see Sec.~\ref{sec:higgs_case_study}). Classical computational methods (e.g.\ quantum Monte Carlo) can be used only to simulate certain very restricted parameter regimes and for the full simulation a quantum device is required.} 
    \item [] \textit{or}
	\item There are \emph{efficient classical algorithms}, but the \emph{scaling of resources} is more favourable in the quantum setting.\footnote{An example is the setting recently discovered by \citet{bravyi_quantum_2017} and the quantum algorithm for database search by \citet{grover_fast_1996}.} 

\end{enumerate}
\end{enumerate}

That the target model falls into one of these four classes is a clear pragmatic norm for the practice of analogue quantum computation.  
That is, we require some form of quantum computational advantage based upon one of these scenarios else there is already available an explanation via classical computation.  


\subsection{Mental-model grasping by observing and manipulating } 
\label{sub:grasping_condition}

The final condition of the simple view of understanding that needs to be satisfied for understanding via analogue simulation we need to consider is the grasping condition. Recall that the grasping condition requires a scientist who aims to obtain understanding of a particular target phenomenon to obtain epistemic access to (the explanation of) the phenomenon. 
In Section~\ref{sec:grasping condition}, we articulated this condition in terms of so-called mental models. 
Given the present, rather rudimentary, stage at which neuroscientific research into the details of human reasoning is at present, formulating clear norms here would be to reach beyond available science. What is more, mental models only constitute one of many possible ways in which one might articulate the grasping condition in the (modified) simple view of understanding. 

Nevertheless, phenomenologically speaking, it appears self-evident that `grasping' of the target phenomenon must be accompanied with some intuition about the physical mechanisms governing the target phenomenon or system at hand. Such intuition might, in a counterfactual spirit, involve recognising `qualitatively characteristic consequences of [those mechanisms] without performing exact calculations' \cite[p.~151]{Regt-Synthese-2005}
or, in a unificationist spirit, being able to relate the phenomenon at hand within a larger physical context. 
It seems to us that certain qualities of an analogue simulation do, if not guarantee, at least benefit scientists aiming to obtain such intuition. 

When discussing the explanation condition, we already mentioned the importance of being able to manipulate the quantum simulator in a broad parameter regime. 
Indeed, not only seeing the results of such manipulations but actually performing those manipulations `physically' can promote grasping of how and why the phenomenon comes about. 
In the context of mental models, and cognitive science in general, an argument why this might be the case is the so-called `common coding' hypothesis (recall Sec.~\ref{sec:grasping condition}). 
According to this hypothesis, we build a common mental representation -- the mental model -- of performing the action of manipulating the source physical system at hand and observing the consequences of that action. 
We take it that together with a theoretical understanding of the underlying mechanisms, models or explanations, performing an action \emph{and} observing its consequences strengthens this mental representation and thus promotes grasping of the corresponding explanation.

From a unificationist perspective, comparing the behaviour of various analogue systems, both in theory and experiments can help identify common features. 
We saw an example of this in the case study on analogue Hawking radiation (Chapter~\ref{ch:hawking}). 
By analysing the effect of Hawking radiation in many different theoretical contexts it becomes clear that Hawking radiation should really be perceived as a phenomenon that is rooted in general geometric features of sonic, optical and wave horizons. The value of such perspectives are supported by the experience of working scientists: ``I believe that by seeing a phenomenon with new methods, from a new
perspective, one learns to understand the phenomenon better. That is my experience.'' (I. Bloch, 2016) (cf.\ Sec.~\ref{sec:grasping condition}).

Physicality is also important in many contexts -- at the very least from a heuristic perspective. Only when actually performing the experiments are scientists confronted with all aspects of physical reality, and in particular those aspects that were not included in their model. 
Even when the physical mechanisms in the source system do not have equivalents in the target system, experiments can enforce an interplay between experiment and theory on real data. That interplay will in many cases provides new insights, and result in a better grasp, a better understanding of a given target phenomenon: ``The theoreticians then need to develop new methods. And thus you learn again something about a phenomenon you believed understood'' (ibid.). 
A good example of such interplay is the case study on the Higgs mechanism in two dimensions. 
Experimenters and theorists alike did not consider the influence of a harmonic trapping potential prior to the experiment of \citet{Higgs:2012} or deemed such influence to be of less importance. 
Only after the experiment yielded inconclusive results were the effects of inhomogenieties across the system on the emergence of a Higgs mode studied in detail \cite{pollet_higgs_2012,liu_massive_2015}. 
Those studies resulted in a better grasp of the physics of the Higgs mode in realistic scenarios in that the broadening of the resonant peaking in the susceptibility due to such inhomogenieties was revealed and quantitatively studied.

These considerations suggest that one might plausibly consider the form of understanding in which the grasping condition is established via observability and manipulability of the source system to be qualitatively different, and arguably superior, to understanding in which grasping is established by other, non-tactile, means. We might hope to draw upon cognitive science research to differentiate satisfaction of the grasping condition via a `haptic' route or via an `non-haptic' route. The presumption would then be that haptic understanding would have some identifiable heuristic advantages, at least in some contexts. We leave exploration of this idea to future work.

In conclusion, the norms for grasping via analogue simulation  coincide to a large degree with those for explanation. That is, 
observability and manipulability of the experimental source system are key for both conditions to obtain. Further work is needed, however, to unpack the key cognitive role that observability and manipulability play in facilitating scientific understanding. \\

To summarise, in this chapter we have analysed the conditions under which the veridicality condition of understanding can be satisfied in analogue quantum computations and emulations. This is achieved through several steps of validating an analogue quantum simulator, specifically, internal validation and external validation. Based on our characterisation of validation, we formulated norms for the validation of analogue quantum computations and emulations. 
We then formulated norms for the other two conditions of understanding -- explanation and grasping -- in terms of observability and manipulability of the source system, as well as a pragmatic norm of speedup for analogue quantum computations. Having analysed how and under which conditions analogue quantum simulations can provide scientists with understanding, in the following chapter we will  take a step back and compare analogue quantum computation and emulation to various other forms of inference. 
\chapter{Methodological Mapping}
\label{ch:method mapping}

In the last chapter we will compare analogue quantum computation and emulation to other, more traditional, types of scientific inference. The guiding principle of this part of our analysis is that the goals a scientist has in using a particular form of scientific inference can be characterised by the \textit{form} of understanding that they expect to acquire. The different forms of understanding can vary in terms of both their \textit{modality} and their \textit{target}. It is not our assumption here that understanding is the only goal of science but rather that consideration of this goal is sufficient to uncover the principal methodological differences between analogue quantum computation and emulation at hand. This in turn allows us to situate analogue quantum computation and emulation on the `methodological map' of modern science \cite{galison1996}. We will once more refer back to our case studies, this time in a principally \textit{illustrative} mode.

\section{Using computers to understand formal properties of a target model}
\label{sec:understanding_formal_properties}

Let us begin by comparing analogue quantum computation as a tool to derive formal properties of a target model to computations of such properties that are carried out on a \emph{classical computer}. 

In the canonical, and rather outdated, view of science, if a scientist wants to understand the formal properties of a target model they would simply deductively derive these properties from the model. 
Traditionally, a deductive derivation of mathematical facts takes the form of a proof or an analytical calculation\footnote{Arguably, proofs are not even strict deductions in a formal language but often contain gaps, to be filled in by the educated reader \cite{sep-proof-theory-development}.} wherein every reformulation of the premises follows easily from the previous step. 
We hope that by now the limitations of such an account in the context of modern science are apparent. 
In a huge number of cases of real scientific models such a deductive derivation is simply impossible in practice because traditional means of deriving a conclusion from the premises fail. 
Thus, such understanding must be gained, if it can be gained at all, via modern forms of deductive inference in which every step of the `proof' cannot be checked by pure inspection anymore, or via non-deductive forms of inference.

A form of inference that, often, scientists seem to deploy with the aim of gaining understanding of the formal properties of a target model is computer simulation.\footnote{It is worth highlighting that in fact there are a range of good reasons \textit{not} take the general aim of computer simulation in science to relate merely to the provision of how-actually understanding of the formal properties of target models. 
Rather, in many cases, computer simulations play an inferential role more similar to experiments. That is, they are deployed with the aim of gaining understanding of actual features of physical target phenomena. 
Moreover, in real scientific practice, Monte Carlo simulations are typically embedded within larger `phenomena orientated'  simulations, and thus their discussion as autonomous inferential tools in what follows is somewhat of an oversimplification. 
These issues notwithstanding, since our space is rather limited we will not engage further with this fascinating debate. See \cite{morgan2002,morgan2003,beisbart2009can,parker:2009,winsberg:2009,Winsberg-2010,Winsberg-SEP-2013,barberousse2009,morrison2009,morrison2015,massimi2015,boge:2019}.}
In some cases computer simulations provably and efficiently yield the correct solution of the task at hand, in other cases they are merely heuristic tools, and are therefore non-deductive. 
Provably exact computer simulations may be considered as a form of deductive inference and can therefore yield how-actually understanding, provided that the algorithm in question is run correctly. 
In contrast, if there is no guarantee with regard to the behaviour of the computer simulation, we can only conceive of it as a form of non-deductive inference so that the form of understanding obtained without further justification is how-possibly understanding.
We will now review the reasoning behind these conclusions with a view to explicating our analysis of analogue quantum simulation later on.

To do this, let us use be more concrete and use a particular class of computer simulation methods, namely, Monte Carlo simulation, as an example. 
Those simulations are based on (pseudo)-random sampling from some probability distribution in order to estimate some mathematical expression. 
As the number of random samples is increased, the variance of the respective estimator decreases and in this sense `converges to the true result in a probabilistic manner as the number of random points is increased.' \cite[p. 408]{beisbart:2012}. 
The mathematical expressions to be computed  `may have a probabilistic meaning or not.' (ibid.). 

While Monte Carlo simulations are guaranteed to be exact in the limit of infinite runtime, we would only like to speak of an exact method if the simulation returns the result up to an arbitrarily small error at a relatively small increase in runtime. We say that such a method is efficient.\footnote{In theoretical computer science, it is the convention that any increase in the runtime in terms of the achieved error $\epsilon$ that scales polynomially, that is, at most as $(1/\epsilon)^c$ for some fixed number $c$, is considered efficient.}
In Monte Carlo methods, this can fail because it is computationally difficult to generate the required samples or because the variance of the Monte Carlo estimator is very large, in both cases obstructing an efficient solution of the task at hand. 

We will now provide two examples of Monte Carlo methods: 
one for which there is a provable guarantee that the algorithm efficiently converges to the correct solution, and one where Monte Carlo methods are employed as a heuristic to approximately solve problems. 
In the latter type of usage we often know a priori that the problems are computationally difficult to solve exactly in the worst case. 
On our account, those simulations with a provable convergence guarantee are therefore deployed to gain how-actually understanding of formal properties of the target model, while the heuristic use of Monte Carlo simulations without further justification yields how-possibly understanding.

There are a vast number of examples of applications of Monte Carlo methods in modern science, ranging from physical science to computational biology and finance. 
In both of our examples (and in many other cases, too), the problem to be solved can be cast in terms of computing formal properties of an Ising model. 
The Ising model is a paradigmatic model of magnetism, where spin-$1/2$ particles with configurations $s_i = \pm 1$ interact pairwise via an interaction energy $V_{i,j} s_i s_j$, and there may be local magnetic fields $b_i$ giving rise to the energy of a configuration $s$
\begin{align}
\label{eq:ising model}
	H(s) = - \sum_{ i,j} V_{i,j} s_i s_j - \sum_i b_is_i. 
\end{align}
When all interaction energies $V_{i,j} \geq 0 $ are nonnegative the system is called ferromagnetic. 

Our first example is the so-called `simulated annealing' method which is a heuristic Monte Carlo optimisation method.\footnote{For a good discussion of simulated annealing see the overview book chapter of \citet[Sec.~12.6]{jerrum_markov_1996}.} 
For this method we cannot guarantee to achieve a small error efficiently and therefore argue that it can be used to obtain how-possibly understanding of a formal property of a target mathematical model. 

The task of simulated annealing is to minimise a cost function over a discrete set. 
A famous example of such a problem is the so-called \emph{travelling salesman problem}. 
Given a set of nodes, representing, for instance, cities for parcel delivery or houses for food delivery, and distances between those nodes, the task of the travelling salesman problem is to find the shortest path that visits all nodes, that is, to minimise the cost of visiting all houses. 
No efficient algorithm exists for the travelling salesman problem.\footnote{Indeed, the travelling salesman problem is as hard as any problem in the complexity class called \textsf{NP} (nondeterministic polynomial time) and thus an efficient solution for it would provide a positive answer the famous millenium problem whether $\textsf{P} = \textsf{NP}$. 
It is one of the most fundamental conjectures of theoretical computer science  that $\textsf{P} \neq \textsf{NP}$, however.} 
Hence, one needs to resort to heuristic algorithms in order to obtain a good guess for the shortest path in a reasonable time. 
The most extreme instance of such a heuristic algorithm would be guessing a random path. 
And, indeed, random guessing is very often an important basis of modern heuristics for solving such combinatorial optimisation problems \cite{Arora:2009vk}.

The idea behind the simulated annealing algorithm relies on the observation that a cost function such as the length of a path can be cast into the form of a certain Ising model with energy function $H(s)$ \eqref{eq:ising model} that assigns an energy to every possible path through the set of nodes.\footnote{Crucially, this energy function will require both positive and negative interaction terms $V_{i,j}$ so that the algorithm of \citet{jerrum_polynomial-time_1993} is not guaranteed to be efficient anymore.}
Intuitively, the process of finding the state of the system with lowest energy is hence simulatable via a slow thermal cooling process into the ground state of the system. 
The formal property one wants to compute is therefore a configuration (a path) with the lowest energy.

In Monte Carlo algorithms, configurations of the corresponding physical system are proposed at random via some heuristic and accepted with a certain probability. 
The crucial part of this cooling process is the choice of acceptance probability, which is chosen such that it depends on the relative change of the cost function in this step of the algorithm, that is, the difference in energy $\Delta E$ between the system configurations at step $k+1 $ and step $k$. 
A typical choice of acceptance probability of worse solutions  would be precisely proportional to a Boltzmann weight factor of the form $e^{-\beta \Delta E}$, where the parameter $\beta$ can be tuned to alter the probability of accepting worse solutions.\footnote{This is precisely the choice of acceptance probability that is used in the algorithm of \citet{jerrum_polynomial-time_1993} discussed below. } 
We may thus view it as an `inverse temperature' of this process. 
A non-zero temperature, corresponding to a non-zero probability of rejecting better solutions is important because it allows moving out of local minima in a very large and rugged optimisation landscape. 
After all, in the end, one aims to converge to a solution which lies in a global optimum of this landscape. 
This is attempted in the method by gradually decreasing the `temperature' or acceptance probability of solutions that are worse in terms of the cost function than the previous one. 

The outcome of a simulated annealing algorithm is often a good approximation of the true solution but not guaranteed to be correct.
And, indeed, the algorithm is also known to fail in some cases. 
Thus, it is a heuristic method and as such an instance of an algorithm providing how-possibly solutions to a mathematical model (in this case the shortest path visiting each node).

Another example of an application of Monte Carlo methods that is \emph{provably efficient} is the computation of the partition function of the ferromagnetic Ising model~\cite{jerrum_polynomial-time_1993}, given by 
\begin{align}
\label{eq:partition function ising}
	Z(\beta) = \sum_s  \exp(-\beta H(s))
\end{align}
at an inverse temperature $\beta = 1/T$. 
All thermodynamic properties of the system, in particular, its phase transitions between ordered and disordered states are characterised by the partition function. 
In our language, the partition function is a certain formal property of the target model -- the Ising model.
The goal of the Monte Carlo algorithm is to  compute that formal property, namely, the partition function $Z(\beta)$ for some large $\beta $. 
Technically, this amounts to estimating the sum \eqref{eq:partition function ising} over $2^n$ many possible configurations $s \in \{-1,1\}^n$ of $n$ spin-$1/2$ particles in the system.
\citet{jerrum_polynomial-time_1993} devised a specific Monte Carlo algorithm that results in an  estimate of $Z(\beta)$ with an error that can be efficiently made arbitrarily small.

Specifically, they prove that for a ferromagnetic Ising model \eqref{eq:ising model} one can efficiently produce Monte Carlo samples from the probability distribution $e^{-\beta H(s)}/Z(\beta)$ over spin configurations $s$ using the sampling method described above and use those samples to estimate a ratio $Z(\beta)/Z(\gamma)$ for inverse temperatures $\beta, \gamma$.  
The ingenious step of their analysis is to use estimates together with the known partition function $Z(0)= 2^n$ at infinite temperature in order to sequentially compute the partition function $Z(\beta)$ of some large inverse temperature $\beta \gg 0 $ as the product
\begin{align}
	Z(\beta)  = \frac{Z(\beta_r)}{Z(\beta_{r-1})} \cdots \frac{Z(\beta_2)}{Z(\beta_{1})} \cdot \frac{Z(\beta_1)}{Z(\beta_0)} \cdot  Z(\beta_0)
\end{align}
of many ratios of partition functions at increasing inverse temperatures $0 = \beta_0 < \beta_1 < \cdots < \beta_r = \beta$ that follow a certain rule. 
They prove that via Monte Carlo sampling, $Z(\beta_i)/Z(\beta_j)$ can be efficiently estimated with a \emph{relative error} and this error can be efficiently made arbitrarily small. 
Consequently, the product of those ratios will also be correct with a certain, arbitrarily small relative error.

Following the account of \citet{beisbart:2012} -- here Monte Carlo simulations function as `arguments' that `deliver their results by transforming presumptions built into their setups in a way that preserves truth (deductive inference) or a way that preserves its probability (strong inductive inference)' (p. 411). 
According to \citeauthor{beisbart:2012}, Monte Carlo simulations `are distinctive in that their steps are governed by random or pseudo-random processes, whereas ordinary argumentation does not include randomly chosen inferences'. 

We have just seen that the Monte Carlo computation of ferromagnetic Ising partition functions can be supplemented with a suitable analysis of the error incurred during that random process. 
It is crucial that this error can be made arbitrarily small at a small, namely, efficient cost in terms of the overall runtime of the algorithm.
In their epistemological function in disclosing the formal properties of the Ising model, Monte Carlo simulations therefore play a role that is identical to that of a derivation. 
Thus, the Monte Carlo algorithm of \citeauthor{jerrum_polynomial-time_1993} enables us to explain the actual formal properties (the partition function) of a target model (the Ising model) in a manner inferentially identical to how pen and paper calculations allow us to explain the actual formal properties of the solutions to a set of equations.

We have now described two applications of Monte Carlo algorithms that yield how-possibly and how-actually explanations of certain formal properties, namely, the ground state and the partition function, respectively, of a target model (the Ising model). 
The crucial question for our analysis is whether those explanations can be supplemented with a suitable `grasping' component such that we get how-possibly and how-actually understanding, respectively. 

We consider it plausible that simulated annealing allows a scientist to grasp how the solution to the model comes about.
In the simulated cooling process one can directly observe how the configuration space is explored, how configurations with certain properties are discarded while configurations with other properties are favoured. 
On the one hand, this can be done in terms of varying optimisation schedules. 
For example, one can vary the proposal heuristic for how to choose configurations in every step, or the temperature decreasing schedule. 
On the other hand, one can also explore very different problem instances in the guise of optimisation landscapes at ease in order to develop an intuitive understanding about the interdependencies between the parameters of a problem. 
These are crucial features of grasping and explanatory power. 
In this respect, simulated annealing in particular, and Monte Carlo simulations in general, are similar to analogue simulation where one can directly observe the dynamics of a physical system. 
Thus simulated annealing therefore may well be considered to provide how-possibly understanding of the solution to combinatorial optimisation problems. 

The case is less clear in the example of computing the partition function of the ferromagnetic Ising model, as it does not have as intuitive an interpretation as the energy of a spin configuration. 
Still, for the computation of the partition function, too, we can observe how the Monte Carlo estimator converges to its true value as the number of samples is increased. 
This allows the scientist a view into how the solution of the mathematical expression comes about and thus provides for a kind of mathematical explanation that might be considered similar to that of a proof. 
Moreover, in many of its applications Monte Carlo simulation is not used by scientists simply to derive a formal property of a model, but rather to map out an entire phase diagram, that is, to answer `What if?'-questions about the dependence of a quantity on the size of certain parameters. 
In such cases, we have a clear example of an inferential tool used by scientists with the purpose of gaining understanding of actual properties of a target model.

To summarise, Monte Carlo simulations of formal properties of a target model allow a scientist to grasp how this formal property comes about as one can directly observe the convergence of the Monte Carlo estimator to its optimal value.
Depending on whether this convergence can be supplemented with an error analysis, we can obtain how-possibly or how-actually understanding of the formal property of the target model in question.

\section{Understanding physical target phenomena in science}
\label{sec:understanding_physical_phenomena}

Let us now illustrate scientific inferences that are deployed to gain understanding of physical target phenomena, that is, means by which scientists can both provide explanations for such phenomena and satisfy the crucial grasping condition. 
We begin by noting that it was argued by Reutlinger, Hangleiter and Hartmann (2018) that toy models, i.e., highly idealised and simple mathematical models, can be used to obtain understanding of physical phenomena. 
In particular, so-called `embedded toy models' may be used to obtain how-actually understanding of such phenomena, while, generally speaking, `autonomous toy models' merely yield how-possibly understanding thereof.
In their terminology, an embedded toy model is one which is in the semantic sense embedded in a well-confirmed framework theory such as quantum mechanics or Newtonian mechanics, and for which there is a clear path by which the model can be `de-idealised'. An autonomous model is one that is not embedded.

In this section we briefly discuss two further forms of scientific inference each of which involves the combination of a model with a concrete source system. 
The first is a conventional experimental inference built upon the combination of an experimental model with a manipulable source system. 
We think it is plausible to take such experimental inferences to be a means of gaining how-actually understanding of physical target phenomena. 
The second example we will consider is the particular form of analogical inference built upon the combination of an analogue model (featuring a `material analogy') with a concrete source system. 
We think it is plausible to take such analogical inferences to be a means of gaining how-possibly understanding of physical target phenomena. 
As before, nothing in the core argument of this book depends upon accepting our analysis here: these are examples provided for illustrative purpose. 

Analogical inferences have long played an important role in science and the topic has been fairly extensively discussed in the philosophical literature \cite{keynes:1921,hesse:1964,hesse:1966,bailer:2009,bartha:2010,Bartha:2019}. An important distinction, due to \citet{hesse:1964}, is between `material analogies', that are based upon to relevant similarity of properties between two systems, and `formal analogies', that obtain when two systems are both `interpretations of the same formal calculus' \cite{sep-models-science}. Here our focus is upon the material analogy. An interesting question regards whether reasoning by material analogy takes the form of a speculative inference or a stronger inductive inference, i.e. one that can be confirmatory. Authors such as \citet{salmon:1990} and \citet{bartha:2010} in particular take arguments by analogy to establish only the \textit{plausibility} of a conclusion, that is, that there is some reason to believe in that conclusion, and with it grounds for further investigation. 

Adopting this speculative inference view, we might then consider one of the myriad examples of scientists deploying a material analogy between a source system, say water flow in a pipe, and target system, say current in wire. Clearly, in such cases, the speculative form of the inference means that the only plausible form of understanding of physical phenomena in the target system that we might be able to gain is the modally weaker how-possibly form. 
In particular, by observing some phenomena in the source system we can, by appeal to the material analogy, make a speculative inference about the same phenomena occurring in the target system. 
Furthermore, in observing the source phenomena we can `grasp' the analogue phenomena in the target system. 

Much more could be said in fleshing out this account of how-possibly understanding via material analogies and its relation to more general issues of gaining understanding via analogical reasoning. For our purposes, what is important is that we take this specific example of gaining understanding via analogical inference to be importantly different from some cases of analogue quantum emulation. Despite the similarities in that both involve a source system and some general notion of `analogy', often the goals of scientists in deploying these two forms of inference are very different. In some cases, we take analogue quantum emulation to bear a closer resemblance to that of conventional experimental inference, which we will analyse briefly now.       

Inferences built upon experiment are the gold standard of scientific reasoning. If the account of scientific understanding considered in the previous section is to have any general applicability to science then surely it must allow us to characterise the form of understanding that scientists hope to gain in carrying out an experiment. Let us consider arguably the most famous \emph{experimentum crucis} 
of the twentieth century: Eddington's 1919 measurement of the deflection of optical starlight as it passed the sun during a solar eclipse \cite{kennefick:2009,will:2014}. A simplified reconstruction of the reasoning that this experiment entailed runs as follows. Eddington was interested in explaining the observable target phenomenon of the deflection of light by a gravitational field. There was not, at this stage, an interesting question about the status of the corresponding formal feature of the relevant model since, as such, the formal property of the deflection of light is a derivable consequence of the model provided by the Schwarzschild solution to the Einstein field equations. In this respect the situation is clearly very different from either Monte Carlo simulations or analogue quantum computations.

Two ingredients are necessary in order to provide an explanation of the deflection of light by a gravitational field. The first is an experimental model,\footnote{Our discussion here fits within the broad category of a model based account of measurement. See \cite{Tal:2015} and references therein for details.} which would include both the relevant ingredients from the general theory of relativity \textit{and} the theoretical description of the apparatus and relevant climatic conditions. The second ingredient is the actual experimental source system used to make the observation, in this case the photographic plates attached to a telescope. Together, given the correct result, these provide an explanation of the actual phenomena of light deflection. 

The crucial final ingredient relates to the grasping condition. Should we, in this case, take Eddington to have a suitable `mental model' corresponding to the phenomena in question? It seems difficult not to answer with `Yes!' here. In particular, the mental model of a curved spacetime deflecting light just as a curved two-dimensional surface, say, stretched rubber curved by a weight, deflects the path of a rolling object are precisely the familiar mental pictures deployed in this context for grasping. We thus have a clear case of the deployment of experimental inference with the goal of gaining how-actually understanding of target physical phenomena. It is worth emphasising that our point here is not to try and argue that the goal of all experimental science is to gain how-actually understanding of observable target phenomena.\footnote{For work on the epistemology of experiment see \cite{hacking:1983,galison:1987,franklin:1989,sep-physics-experiment,evans:2019}.} Rather, we take this to be one significant goal of at least some types of experimentation and our focus is on setting up a plausible exemplar to better frame a comparison with analogue quantum emulation. 

\section{Methodological map for analogue quantum simulation}

\begin{figure}[t]
\centering
\includegraphics[width = \textwidth]{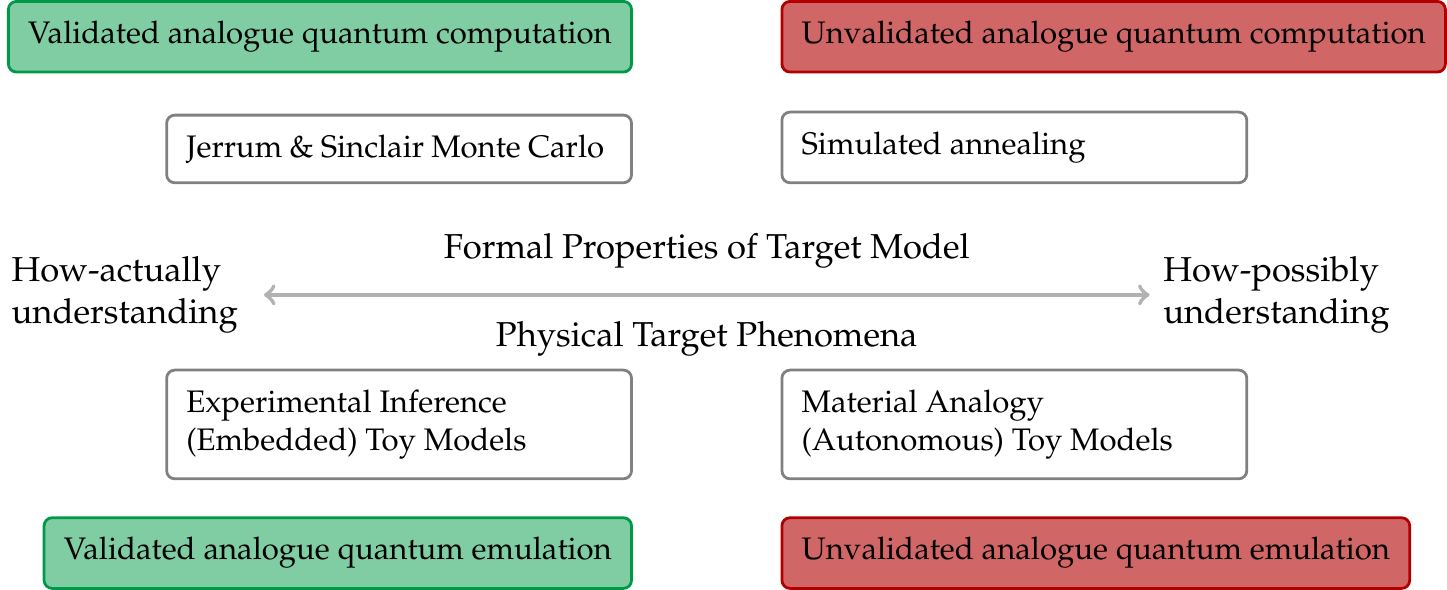}
\caption{Methodological map with respect to the type of understanding aimed at and the object that is to be understood. Whether or not these aims are achievable depends on further features of analogue quantum emulation and computation as discussed in this chapter and the next. 
} 
\label{fig:methodological map2}
\end{figure}

The schema within which we compare analogue quantum computation and emulation to other forms of scientific inference is built around the combination of two distinctions. 
First the distinction between how-actually and how-possibly understanding, introduced in Chapter~\ref{ch:understanding}. 
Second, a distinction based upon the two relevant objects of understanding: formal features of a target model and physical target phenomena. 
We can illustrate the virtue of this schema by situating analogue quantum computation and analogue quantum emulation in comparison with the specific examples of canonical forms of scientific inference just discussed. 
The results of our analysis are summarised in Figure~\ref{fig:methodological map2}.
Let us elaborate. 

The core argument we present in this short book revolves around the distinction between analogue quantum computation and emulation. This distinction is characterised by the respective target; while in analogue quantum computation, we aim to learn about a \emph{formal property of a target model}, in analogue quantum emulation, we aim to learn about a \emph{concrete physical phenomenon}. Within the modified simple view of understanding, we have argued that a central goal of scientists pursuing analogue quantum simulation is to obtain understanding of the respective target. 
The key question when analysing analogue simulations is then whether this understanding can be asserted to be how-actually or how-possibly, that is, whether or not the simulation satisfies the veridicality condition of the modified simple view of understanding. 

We have argued that the veridicality condition can be  established in a validation procedure that involves both mathematical verification of correspondences between different models in certain parameter regimes and experimental certification of models as empirically adequate models of physical phenomena.  
This argument leads us to situate analogue computation and emulation on the methodological map depending on whether or not they are validated. 

A validated analogue quantum computation yields how-actually understanding of its target formal property and in this sense is, rather unsurprisingly, situated on the methodological map alongside exact computer simulations. An unvalidated analogue quantum computation  yields how-possibly understanding. This is because without validation, we cannot assert the veridicality condition of the modified simple view of understanding. Our situation of the methodological map suggest a methodological correspondence with heuristic computer simulations such as simulated annealing. 

A validated analogue quantum emulation yields how-actually understanding of physical target phenomena and is thus methodologically co-situated with conventional experimental inference. 
An unvalidated analogue quantum emulation yields how-possibly understanding of physical target phenomena and is thus methodologically co-situated with material analogies and autonomous toy models. 

The significance of this analysis is that it allows us to isolate the sense in which analogue quantum simulation is both methodologically and epistemically peculiar. It is precisely \textit{because} analogue quantum simulations function in some contexts like material analogies or toy models, in other contexts like computer simulations, and in still other like experiments, that looking for a unified and unequivocal analysis of what we can and cannot learn from them is simply inappropriate. Just as methodological diversity in science in general is an extraordinary recourse, so the diversity of modes of reasoning that can be justified based upon analogue quantum simulation is one of the things that makes it such an exciting new tool for scientific inference. It is by framing this diversity on the methodological map, as we have above, that its full implications can be elicited, and we hope that both scientists and philosophers will profit from our new conceptualisation in the study of analogue quantum simulation, and beyond.

\chapter{Closing remarks}
\label{ch:conclusion}

Drawing the methodological map of modern science as we did in the previous chapter, makes clear why analogue quantum simulations may often seem like something both familiar and revolutionary at the same time. On the one hand, the use of the single term `analogue quantum simulation' might lead one to think that the relevant methodological issues are reducible to those attendant to well-established modes of scientific inference. And indeed, as we have just seen, a number of such relationships could be pointed to. Yet, on the other hand, the complexity and heterogeneity of analogue quantum simulation practice means that the methodological correspondence to other modes of scientific inference are only partial and patchwork: a complicated network of similarities, overlapping and criss-crossing like the various resemblance between members of the same family \cite[66-7]{wittgenstein2009philosophical}. The crucial point is that such family resemblances should not bar us from recognising the novelty of analogue quantum simulation. It is a new instrument for scientific understanding, and requires new philosophical analysis to be properly practiced and understood. We hope that this short book has gone some small way towards that end. 

Much is left to say about the methodological, epistemological, and metaphysical implications of analogue quantum simulation. We hope that our own small effort will serve to spur scientists and philosophers to think and write more about the foundations of this fascinating yet perplexing new area of scientific practice. There are two more conceptual issues we take to be particularly worthy of further attention. 

First, there is the problem of making the grasping condition more concrete by appeal to cognitive science. When writing the preceding chapters, a persistent worry has been to satisfy the sympathetic sceptic with regard to the genuine content of the grasping condition. To what extent has what we have said about grasping genuinely progressed past elaborate hand-waving? Clearly, whilst the relevant philosophical and scientific analysis is still in its infancy, there is a need to provide more concrete and determinate grounds for grasping to be established. This is what we take our norms for observability and manipulability to be the first step towards. In particular, grasping when established via these means is inter-subjectively comparable. As such, in the case of analogue quantum simulation at least, we take there to be a clear pathway towards an analysis of scientific understanding within which the problematic aspects of subjectivity can be constrained. This, of course, is not to suggest that \textit{all} scientific understanding need be of the `active' form in which grasping is established via observability and manipulability. Rather, we take ourselves to have demarcated a species of scientific understanding that is both prevalent in contemporary practice and ripe for further analysis through what might be called the cognitive science of science. 

One particularly fruitful potential further avenue of exploration here is the connection to the newly flourishing literature on the aesthetics of experimentation \cite{murphy:2020,wragge2020,ivanova:2021}. In particular, the aesthetic value of an analogue experiment might be linked to the grasping of the target phenomenon that the experiment on the source system enables. Furthermore, earlier we emphasised that grasping depends upon the dynamics of the simulator being observable in sufficient detail and manipulable to a sufficient degree. It is thus highly noteworthy that in the literature on quantum simulation the level of control over the source system is very often described by scientists in the literature as \textit{exquisite}. Quite possibly, exploration of the cognitive science of scientific understanding may reveal an important cognitive-aesthetic aspect to the `grasping' of target phenomena that analogue quantum simulations allow.  

The second more conceptual issue of particular interest which we would like to point to relates to the effects of errors in analogue quantum simulation. 
We already mentioned that in contrast to digital computation, analogue simulation is intrinsically not error-correctable. 
This may severely limit the scalability of analogue quantum simulations to larger system sizes, an issue that digital simulation does not have because one can (in principle) correct for errors during the computation. 
This raises the important question as to whether and how the nature of errors within analogue quantum simulation limit the strength of inference we are justified in making. 
One might hope that the kinds of errors occurring in target and source system, that is, the deviations of the actual system dynamics from the simulation model are similar \cite{Cubitt:2017ti}. 
Such an argument would require independent evidence, which will typically have to take the form of a characterisation of the error sources. 
But also theoretical arguments based on universality are conceivable.

In Chapter~\ref{ch:norms}, we argued that a heuristic norm for analogue quantum computation is that it outperforms classical computer simulations in terms of the runtime of the algorithm. 
However, we have not said what exactly we mean by such a computational speedup of analogue quantum simulations over classical computer simulations (which are intrinsically digital). 
Indeed, it is far from clear how to compare analogue and digital computations in terms of the scaling of runtime. 
One of the reasons for this is also that the two are incomparable in terms of their tolerance to errors. 
While certain analogue (classical) computations are theoretically able to solve problems which are not efficiently solvable on a classical computer, no actual device will ever achieve that solution because it will be thwarted by errors that are intrinsic to any physical process. 
Moreover, analogue quantum simulators are highly special devices which only allow solving one particular problem (with varying parameters). It is therefore often questionable to what extent such devices might even be considered `computers' in the first place. 

Following up on those issues, the key philosophical question arising from the discussion of verification and validation in Chapter~\ref{ch:norms} is: how do different means of verifying an analogue simulation affect the strength of the inferences we can draw? 
For example, some verification tools will allow statements about fixed problem instances \cite[e.g.][]{flammia_direct_2011,Certification}, while others might only yield statements about the average error when considering a certain set of potential simulation parameters~\cite{wiki_randomized_2020}. 
Depending on the type of verification, we might be warranted to claim different types of inferences. 
A precise philosophical analysis of the different tools available remains outstanding and at the same time seems crucial to the developing technological applications of analogue quantum simulation. 
From a theoretical point of view the techniques of verification which are presently devised for digital computation should be tailored to the specifics of analogue simulation. \\

Science moves quickly, and there is an ever expanding and diversifying range of contemporary experimental and theoretical practice that might be grouped under the heading analogue quantum simulation. We think that there is no more fitting conclusions to our analysis then to pick out a selection of examples not discussed in detail in this book, but of undoubted scientific, and potentially philosophical, significance. 

\begin{itemize}
    \item \textbf{Universality and non-equilibrium fixed points}. A range of recent experiments have shown the ultracold atomic systems far from equilibrium exhibit universality in which measurable experimental properties become independent of microscopic details \cite{prufer:2018,erne:2018,eigen:2018}. Most vividly one of the experimental teams claims that one may use this universality to learn, from experiments with ultracold gases, about fundamental aspects of dynamics studied in cosmology and quantum chromodynamics \cite{prufer:2018}.

    \item \textbf{Analogue simulation of backreaction}. The semi-classical modelling framework applied to analogue and astrophysical black holes makes use of a distinction between a quasi-stationary background and small perturbations on that background. An outstanding and extremely important theoretical challenge (particularly important for black hole evaporation) is to understand backreaction phenomena whereby the perturbations alter the background structure. Analogue classical (and potentially quantum) simulations provide a powerful new tool to understand such phenomena \cite{goodhew:2019}.

    \item \textbf{Digital quantum simulation of analog dynamics}. One of the immediate applications of the intermediate-scale digital quantum computers \cite{arute_quantum_2019} are the simulation of quantum dynamics and static properties of quantum materials \cite{reiher_elucidating_2017}. 
    This approach to quantum simulation is complementary to analogue quantum simulation as it uses a computationally universal digital device rather than a highly tailored analog device. 
    Given our philosophical  analysis of analogue quantum simulations, interesting philosophical questions arise about the epistemological status of such digital quantum simulations. Are they essentially similar to direct analogue simulation, or relevantly different in terms of their potential to obtain understanding of their respective target? 
    And is there even such a thing as a digital quantum emulation?

\end{itemize}

\backmatter
\bibliography{LitAnalogueSim.bib,jc_bib.bib,dumb.bib,mbl_case_study.bib}

\printindex


\end{document}